\documentclass[
  aps,
  physrev,
  reprint,
  onecolumn, 
  superscriptaddress,
  amsmath,
  amssymb,
  longbibliography,
  floatfix
]{revtex4-2}

\usepackage{graphicx}
\usepackage{bm}
\usepackage{mathtools}
\usepackage{braket}     
\usepackage{soul}
\usepackage{makecell}
\usepackage{array}
\usepackage{tikz}
\usepackage{booktabs}
\usepackage[table]{xcolor}
\usepackage[
  colorlinks=true,
  linkcolor=blue,
  citecolor=blue,
  urlcolor=blue
]{hyperref}
\usepackage[capitalise]{cleveref}

\allowdisplaybreaks

\makeatletter
\let\l@subsubsection\@gobbletwo
\makeatother

\newif\ifdraft
\drafttrue

\ifdraft
  \newcommand{\rs}[1]{\textcolor{blue}{[ric: #1]}}
  \newcommand{\ps}[1]{\textcolor{red}{[pasquale: #1]}}
\else
  \newcommand{\rs}[1]{}
  \newcommand{\ps}[1]{}
\fi

\newcommand{\Id}{\mathbb{I}}

\DeclareMathOperator{\Tr}{Tr}

\definecolor{rowblue}{RGB}{190,225,255}

\begin{document}


\title{Dynamics of Rényi entanglement entropies and Eigenstate Thermalization}

\author{Riccardo Senese}
\email{rsenese@sissa.it}
\affiliation{SISSA and INFN, via Bonomea 265, 34136 Trieste, Italy}

\author{Pasquale Calabrese}
\affiliation{SISSA and INFN, via Bonomea 265, 34136 Trieste, Italy}

\begin{abstract}
How is entanglement dynamics encoded in the eigenstates of the time-evolution operator? We address this question by focusing on Rényi entropy dynamics in deterministic chaotic Floquet models in arbitrary dimensions. Inspired by the eigenstate thermalization hypothesis (ETH), we employ spectral representations of reduced density matrix moments (purity and its higher-order generalizations) whose decay through spectral dephasing captures nonequilibrium entanglement growth. These representations closely resemble those governing local-observable dynamics, but involve eigenstates in a replicated Hilbert space and observables are replaced by subsystem permutation operators, such as the swap operator for Rényi-2. 
Focusing on Rényi-2, we combine exact random matrix theory calculations, established ETH results and numerical tests in 1D to formulate ETH-like ansätze for subsystem swap matrix elements in the replicated eigenbasis and their cross-correlations with initial-state overlap statistics. In the thermodynamic limit, purity dynamics is entirely determined by smooth slowly varying functions that emerge from such statistics, paralleling generalized ETH formulations in a single replica. In the limit of large subsystems, these functions exhibit a striking structural simplification: locality of the dynamics together with bulk spectral dephasing reduce their functional dependence to a single fundamental frequency. Our formalism provides a setting where ETH-like statements naturally coexist for both arbitrarily small and large subsystem sizes. We discuss extensions of our results to higher Rényi entropies. Beyond clarifying the spectral foundations of entanglement dynamics, these results lay some of the groundwork for future applications of spectral filter methods to the computation of Rényi evolution in deterministic systems.
\end{abstract}

\maketitle
\tableofcontents

\section{Introduction}
\label{sec:introduction}

The eigenstate thermalization hypothesis (ETH) provides a central framework for understanding dynamics and thermalization in \emph{generic} isolated quantum many-body systems. Its antecedents include von Neumann's quantum ergodic theorem \cite{neumann_beweis_1929,*vonneumann_proof_2010}, Berry's random-wave conjecture \cite{berry_regular_1977}, and early studies of matrix-element statistics in the energy eigenbasis \cite{peres_ergodicity_1984, feingold_ergodicity_1984, jensen_statistical_1985, feingold_distribution_1986}. Its modern formulation was developed in the works of Deutsch and Srednicki \cite{deutsch_quantum_1991, srednicki_chaos_1994, srednicki_approach_1999}. Most notably, ETH provides a simple microscopic justification for the local emergence of thermal behaviour at late times under unitary dynamics \cite{srednicki_chaos_1994, srednicki_approach_1999, rigol_thermalization_2008}. Together with its related generalizations \cite{srednicki_approach_1999, foini_eigenstate_2019, chan_eigenstate_2019, foini_outofequilibrium_2025}, it also provides the information necessary to reconstruct the dynamics of correlation functions, even in the transient regime before local equilibrium is established.

The simplest ETH statements are those related to chaotic many-body Floquet dynamics $U(t)=U_F^t$, $t \in \mathbb Z$, where for simplicity we assume that $U_F$ does not possess symmetries and that it has a local nature (e.g.~it is generated by a time-dependent local Hamiltonian). We refer the reader to the pedagogical reviews on ETH in Refs.~\cite{dalessio_quantum_2016, deutsch_eigenstate_2018, mori_thermalization_2018, patil2026quantum} for the important related case of static Hamiltonians. 
Unlike the latter, the absence of energy conservation (or any other (quasi)local conservation) reduces the ETH ansatz to its simplest form \cite{chan_eigenstate_2019, fritzsch_eigenstate_2021}
\begin{equation}
\label{eq:FloquetStETH}
    O_{mn} \equiv \braket{m|O|n}= \frac{\Tr[O]}{d} \, \delta_{mn} + \frac{1}{\sqrt d}f_O(\omega_{mn})R_{mn} \ .
\end{equation}
Here $O$ is a simple local observable, $\ket{m}$ denotes an eigenstate of $U_F$ such that $U_F\ket{m}=e^{-i\varphi_m}\ket{m}$, $d=\dim \mathcal H$ is the Hilbert space dimension and $\omega_{mn}=\varphi_m-\varphi_n ({\rm mod} \ 2 \pi)$. $R_{mn}$ denote erratically varying numbers with order one variance $\overline{|R_{mn}|^2}$, where the bar symbol indicates averages over eigenstates in a neighbourhood of $m$ and $n$ (note that for arbitrarily small but fixed averaging windows, these averages still involve exponentially many terms in system size). Other interpretations of the bar symbol are possible \cite{foini_eigenstate_2019}, e.g.~for fixed indices $m$ and $n$ those based on weak (but sufficient to mix nearby eigenstates) random unitary-preserving perturbations of $U_F$ \cite{deutsch_quantum_1991, reimann_eigenstate_2015}. The crucial feature in \eqref{eq:FloquetStETH}, like in all ETH ansätze, is that $f_O(\omega)$ (which is $2\pi$-periodic in the present Floquet case) is a continuous order-one function that varies \emph{very slowly} with $\omega$ when compared to the level-spacing scale $\delta \omega=\varphi_{m+1}-\varphi_{m}\simeq 2\pi/d$. 

The ansatz \eqref{eq:FloquetStETH} is sufficient to understand relaxation of out-of-equilibrium 1-point functions to their infinite-temperature values: due to spectral dephasing, $\braket{\psi(t)|O|\psi(t)}-\Tr[O]/d$ is exponentially (in system size) close to zero at almost all times during its infinite-time evolution (here $\ket{\psi(t)}=U(t)\ket{\psi(0)}$) \cite{srednicki_approach_1999}. \cref{eq:FloquetStETH} also enables one to link $f_O(\omega)$ to connected dynamical 2-point functions at infinite temperature: $|f_O(\omega)|^2$ coincides for $d \to \infty$ with the Fourier transform of $\braket{O(t)O}_c$, where $O(t)=U^\dag(t)OU(t)$ and $\braket{\ldots}=\Tr[\ldots]/d$ \cite{feingold_distribution_1986, srednicki_approach_1999}. In other words, $f_O(\omega)$ is the spectral function controlling the degree of time-dependent dephasing which underlies dynamics \cite{schonle_eigenstate_2021}. Exactly as in the static-Hamiltonian case, also in Floquet systems \cref{eq:FloquetStETH} encodes only the first two orders of a full-ETH hierarchy \cite{foini_eigenstate_2019, chan_eigenstate_2019} (see also Refs.~\cite{pappalardi_eigenstate_2022, pappalardi_full_2025, vallini_refinements_2026})
\begin{equation}
\label{eq:nETHFloquet}
    \overline{O_{i_1i_2}O_{i_2i_3}\ldots O_{i_{n-1}i_n}O_{i_ni_1}} = \frac{1}{d^{n-1}}h^{(n)}_O(\omega_1,\ldots,\omega_{n-1}) \ , \qquad i_j \neq i_k  \ \forall  \ j \neq k, \   n = 1, 2 ,\ldots \ ,
\end{equation}
where $\omega_j=\varphi_{i_j} - \varphi_{i_{j+1}} ({\rm mod} \ 2\pi )$, $h_O^{(n)}$ is a continuous order-one function and the bar symbol has the same meaning as above. In the following we will refer to \eqref{eq:nETHFloquet} as “$n$-ETH ansatz”. A similar formula also holds when the $n$ operators are all different, i.e.~$O \to O^{(1)}, O^{(2)}, \ldots, O^{(n)}$. The $n$-ETH ansatz \eqref{eq:nETHFloquet}, together with factorization rules for repeated indices
\begin{equation}
\label{eq:factorizationFullETH}
    \overline{O_{i_1i_2}\ldots O_{i_\ell i_1} O_{i_{1}i_{\ell+1}}\ldots O_{i_ni_1}} \sim \overline{O_{i_1i_2}\ldots O_{i_\ell i_1}}\ \cdot \overline{O_{i_{1}i_{\ell+1}}\ldots O_{i_ni_1}} \ , \qquad i_j \neq i_k \ \forall \ j \neq k \ ,
\end{equation}
and the entropic suppression in the spectral sums of index-patterns corresponding to non-cactii diagrams \cite{foini_eigenstate_2019} (crossing partitions in the language of \cite{pappalardi_eigenstate_2022}), encodes all the information needed to reconstruct higher-order dynamical functions $\braket{O(t_1)\ldots O(t_n)}$. The information needed to reconstruct the dynamics of observables starting from an initial state $\ket{\psi(0)}$ is instead encoded in the cross-correlations between the matrix elements of $\rho(0)=\ket{\psi(0)}\bra{\psi(0)}$ and those of the observables \cite{foini_outofequilibrium_2025}, e.g. at the lowest order
\begin{equation}
\label{eq:outETHsinglerep}
    \overline{O_{mn}\,\rho(0)_{nm}} = \frac{1}{d^2} g_{O,\psi(0)}(\omega_{mn}) \ , \qquad m \neq n \ .
\end{equation}
Also here $g_{O,\psi(0)}$ is a continuous order-one function. Importantly, the scaling with $d$ in both \cref{eq:nETHFloquet} (for $n\ge 3$) and \cref{eq:outETHsinglerep} emerge from delicate cross-correlations surviving strong cancellations between the erratically varying phases of the terms in the local-spectrum averages \cite{foini_eigenstate_2019, foini_outofequilibrium_2025, senese2026nonequilibrium}. Indeed, such cancellations are responsible for the decays $d^{-(n-1)}$ and $d^{-2}$ in each equation, which are significantly faster than those predicted by multiplying the typical scales $|O_{mn}|\simeq d^{-1/2}$ and $|\rho(0)_{nm}|\simeq d^{-1}$ of each individual term.\\

Despite the enormous impact that the simple ideas above have had in defining a modern spectral understanding of thermalization, observable dynamics and time-dependent dephasing, it remains largely unexplored whether it is possible to generalize them in a simple way to describe quantum complexity measures, and in particular \emph{entanglement} and its \emph{evolution}. As the fundamental witness of the departure of quantum theory from the classical world, entanglement quantifies the amount of nonclassical correlations between subregions of a quantum system \cite{bengtsson_geometry_2006, amico_entanglement_2008, horodecki_quantum_2009}. It is a key resource for quantum computation \cite{nielsen_quantum_2010}, and its increase over time is strongly tied to the emergence of local thermalization in the many-body context. In systems whose time-evolution generator has a local nature, entanglement dynamics is known to possess a high degree of universality, famously manifested by the onset of ballistic growth of subsystem von Neumann entropy irrespective of microscopic details \cite{calabrese_evolution_2005, lauchli_spreading_2008, fagotti_evolution_2008, schuch_entropy_2008, calabrese_entanglement_2009, kim_ballistic_2013, hauke_spread_2013, schachenmayer_entanglement_2013, buyskikh_entanglement_2016, alba_entanglement_2017,nahum_quantum_2017, alba_entanglement_2018, alba_quantum_2019, bertini_entanglement_2019, calabrese_entanglement_2020, bertini_growth_2022}. The ballistic growth has been shown or argued to apply also to generic Rényi entanglement entropies in several setups \cite{calabrese_evolution_2005, fagotti_evolution_2008, nahum_quantum_2017, ho_entanglement_2017, mezei_entanglement_2017, chan_solution_2018, zhou_emergent_2019, bertini_entanglement_2019, calabrese_entanglement_2020,bertini_growth_2022}. However, in chaotic models featuring local conservation laws (including energy conservation alone), which induce sub-ballistic transport of conserved charges, recent works have provided analytical and numerical evidence that Rényi growth can be sub-ballistic for Rényi index $k>1$ \cite{rakovszky_subballistic_2019, huang_dynamics_2020, zhou_diffusive_2020, znidaric_entanglement_2020, rakovszky_entanglement_2021, znidaric_reply_2021, lastres2026geometry}. Remarkably, entanglement and its dynamics can be reconstructed \cite{daley_measuring_2012, elben_renyi_2018, huang_predicting_2020, elben_randomized_2022} from experimental measurements both at the level of von Neumann \cite{lanyon_efficient_2017} and most commonly Rényi \cite{islam_measuring_2015, kaufman_quantum_2016, brydges_probing_2019, rath_entanglement_2023, joshi_observing_2024, hu_demonstration_2025, yang2026probing} entropies.\\

In this work, we generalize the ETH ideas above to the multireplica setting needed to describe entanglement evolution. Focusing on integer-index $k$-Rényi entropies with $k>1$, and especially $k=2$, we show that ansätze \emph{structurally identical} to \cref{eq:FloquetStETH,eq:outETHsinglerep} underlie the dynamics and late-time saturation of Rényi entanglement entropies under deterministic unitary evolution. Conceptually, all that is needed is to replace single-replica eigenstates in  \cref{eq:FloquetStETH,eq:outETHsinglerep} with multireplica ones and to consider the statistics of subsystem cyclic permutation operators (like the subsystem swap for $k=2$) in place of local single-replica observables $O$. Our strategy is threefold: \textbf{(i)} we obtain the skeleton of the relevant ansätze by means of exact random matrix theory (RMT) \cite{mehta_random_matrices} calculations using global ensembles; indeed, also in the single-replica case such ensembles provide the correct skeleton behind  \cref{eq:FloquetStETH,eq:nETHFloquet,eq:factorizationFullETH,eq:outETHsinglerep}, including the delicate cross-correlations that survive strong phase cancellations; \textbf{(ii)} to obtain the full ansätze we dress the RMT skeletons by slowly varying continuous functions of the relevant frequencies, and in simple limits we connect them to the $n$-ETH statements in \eqref{eq:nETHFloquet}; our perspective also enables us to identify an important structural simplification in the frequency dependence of spectral functions for large subsystems; \textbf{(iii)} we carefully verify the conjectured ansätze in a 1D chain via exact diagonalization and exact time-evolution numerics. Our focus here is on deterministic Floquet dynamics without symmetries, as this provides the simplest setting. Generalizations to Floquet models with discrete symmetries and local conservation laws, or to static-Hamiltonian dynamics, are important directions to address in future work. They will differ in some of the details (e.g.~the need to resolve symmetry sectors or the emergence of different scaling laws for Rényi growth), but will still obey generalized ETH statements very similar to the ones we introduce here.

We note that our results significantly advance or largely complement earlier contributions to this replica-ETH program (or closely related programs) from Refs.~\cite{kaneko_characterizing_2020, shi_local_2023, hahn_eigenstate_2024, jindal_generalized_2024, pain_entanglement_2024, pain_imprints_2026}. We highlight differences and similarities with these earlier works, whenever relevant, directly in the overview of our main results in \cref{sec:overview} and in the main text.

We would like to make apparent another fundamental motivation of our work. While ETH statements have historically been a cornerstone for \emph{conceptual} understanding, recent works \cite{banuls_entanglement_2020, lu_algorithms_2021, cakan_approximating_2021, yang_classical_2022, schuckert_probing_2023, ghanem_robust_2023, hemery_measuring_2024, morettini_energyfiltered_2024, luo_probing_2024, pappalardi_microcanonical_2024, yang2026thermal} have started exploring whether it is possible to \emph{make use} of ETH (and related statements) to more efficiently simulate features of quantum systems in and out of equilibrium. In such spectral filter methods, the core simplification relies on the fact that spectral ETH functions are slowly varying compared to level-spacing scales. In establishing here the existence of slowly varying ETH functions directly related to entanglement growth, we hope to ease extensions of these methods towards the characterization of entanglement dynamics itself. \\

The paper is organized as follows. In \cref{sec:overview} we introduce the relevant quantities and their associated Lehmann representations for the Rényi-2 case, and provide an overview of all our main results. In \cref{sec:RMTbriefresults} we briefly outline the derivation of the RMT replica-ETH skeletons, with full details on the calculations reported in Appendix \ref{appendix:RMT}. In \cref{sec:fullansatzsection} we complete the ETH ansätze by dressing the RMT skeletons. We also discuss their relation to dynamics, connect them with established ETH results in simple limits, and analytically demonstrate the emergence of an important frequency-dependence simplification for the spectral functions in the limit of large subsystems. In \cref{sec:numver} we numerically verify all our conclusions in a 1D spin-1/2 chaotic Floquet chain. In \cref{sec:higherRenyi} we discuss extensions of the results to arbitrary Rényi entropies with integer-index $k>2$. We conclude in \cref{sec:discussion} with an outlook on future interesting directions. A few appendices follow presenting additional results and derivations.

\section{Overview and main results}
\label{sec:overview}

\subsection{Purity and spectral representations}

Here, and for most of the work, we restrict our attention to Rényi-2 entropy because it represents the simplest case and yet perfectly exemplifies the more general treatment. Indeed, the generalization to higher Rényi entropies is straightforward in several (but not all) aspects, as we discuss in \cref{sec:higherRenyi}.

Consider a pure quantum state $\ket{\psi} \in \mathcal H$ on a bipartite $D$-dimensional lattice $A \cup B$ with $V=|A|+|B|$ sites, and its reduced density matrix 
\begin{equation}
    \rho_A = \Tr_B[\ket \psi \bra \psi] \ .
\end{equation}
With a slight abuse of notation we indicate with $A$ and $B$ both the physical subsystems and their respective Hilbert space. We denote Hilbert space dimensions as $d_A = {\rm dim}(A), \ d_B = {\rm dim}(B)$ and  $d = d_A d_B = {\rm dim}(\mathcal H)$. Assuming a local Hilbert space $d_{\rm loc}\ge 2$ at each site, we have $d = d_{\rm loc}^V$. 
The Rényi-2 entropy $\mathbb S^{(2)}_A$ associated with the reduced density matrix $\rho_A$ is
\begin{equation}
    \mathbb S^{(2)}_A = -\ln \mathbb P_A \ , \qquad \quad \mathbb P_A = \Tr_A[\rho_A^2 ] \ ,
\end{equation}
where $\mathbb P_A$ is the subsystem purity. Because $\ket \psi$ is pure we have $\mathbb S_B^{(2)}=\mathbb S_A^{(2)}$, and the Rényi-2 entropy is a measure of quantum entanglement between $A$ and $B$ \cite{bengtsson_geometry_2006}. The knowledge of $\mathbb S_A^{(2)}$ as $A$ varies captures aspects of the quantum complexity of $\ket \psi$. 

Our goal is to explore the dynamics of $\mathbb S_A^{(2)}$ during quantum evolution far from equilibrium, from the perspective of Lehmann representations \cite{mahan_manyparticle_1990, fetter_quantum_2003}. In particular, starting from pure initial states $\ket{\psi(0)}$, we consider global quantum quenches \cite{essler_quench_2016, eisert_quantum_2015} induced by a deterministic unitary evolution $U(t)$
\begin{equation}
    \ket{\psi(t)} = U(t) \ket{\psi(0)} \ .
\end{equation}
The primary question we have in mind is: how do we relate entanglement evolution to precise features of the initial state $\ket{\psi(0)}$ and of eigenstates of the time-evolution operator?
Here $\ket{\psi(0)}$ is assumed to have non-vanishing overlap with exponentially many (in the volume $V$) eigenstates of $U(t)$. This condition is typically satisfied in the experimentally relevant setup where $\ket{\psi(0)}$ is some simple sub-volume-law entangled state (e.g.~a product state or an area-law ground state), and it is expected more in general when the preparation of $\ket{\psi(0)}$ does not involve processes induced by $U(t)$. Indeed, in the following we also consider situations in which $\ket{\psi(0)}$ has a product state form $\ket{\psi (0)} = \ket{a}_A \otimes \ket{b}_B$, where $\ket{a}_A$ and $\ket{b}_B$ are highly entangled respectively within $A$ and $B$. We will furthermore assume that the dynamics induced by $U(t)$ is local, in the sense of being generated by a local (time-dependent) Hamiltonian or a deterministic local quantum circuit.

The core of our treatment is based on the following standard identities (replica trick)
\begin{align}
\label{eq:mainidentity}
    \mathbb P_A(t) &= \Tr_A[\rho_A(t)^2] =\Tr_{A^{\otimes 2}}\big[S_A \big( \rho_A(t) \otimes \rho_A(t)\big)\big]  \\ \nonumber
    & = \Tr_{\mathcal H^{\otimes 2}}\big[S_A \big( \rho(t) \otimes \rho(t)\big)\big] =\braket{\psi(t)|^{\otimes 2}S_A|\psi(t)}^{\otimes 2} \ ,
\end{align}
where $\rho(t) = \ket{\psi(t)}\bra{\psi(t)}$, $\rho_A(t)=\Tr_B[\rho(t)]$ and $S_A=S_A^\dag = S_A^{-1}$ denotes the subsystem swap operator acting only within $A$ across 2 replicas, i.e.
\begin{equation}
    S_A \big(\ket{a_1}_A\ket{b_1}_B \otimes \ket{a_2}_A\ket{b_2}_B\big) = \ket{a_2}_A\ket{b_1}_B \otimes \ket{a_1}_A\ket{b_2}_B \ .
\end{equation}
\cref{eq:mainidentity} follows from the following standard property of the swap operator
\begin{equation}
\label{eq:intertwofS}
    \Tr_{\mathcal V}[O_1 O_2] = \Tr_{\mathcal V^{\otimes 2}}\Big[S_{\mathcal V} \big( O_1 \otimes O_2 \big)\Big] \ ,
\end{equation}
where $O_1, O_2 \in {\rm End}(\mathcal V)$, and from the fact that $S_A$ acts as the identity on $B^{\otimes 2}$. The crucial point in the last 2 equalities of \eqref{eq:mainidentity} is that everything is defined on the global $\mathcal H^{\otimes 2}$. This enables us to express $\mathbb P_A(t)$ in a Lehmann representation, i.e.~via the resolution of the identity $\Id^{\otimes 2}$ in terms of the eigenstates of $U(t)^{\otimes 2}$
\begin{align}
\label{eq:eigenstates2replica}
    U(t)^{\otimes 2} \ket{\bm m}&=e^{-i (\varphi_m + \varphi_{m'})t}\ket{\bm m} \ , \qquad   \ket{\bm m} \in \mathcal H_+^{\otimes 2}\ , \\ 
    \label{eq:symmetricEigen}
    \ket{\bm m}&=\frac{\ket{m,m'}+\ket{m',m}}{\sqrt{2(1+\delta_{m,m'})}} \ , \qquad  m\le m' \ .
\end{align}
We have restricted the set of relevant multireplica eigenstates to the symmetric sector $\mathcal H_+^{\otimes 2}$ of the global replica symmetry induced by $[S_{\mathcal H},U(t)^{\otimes 2}]=0$, where $S_{\mathcal H}$ is the \emph{global} swap operator and $S_{\mathcal H}\ket{\bm \phi} = \ket{\bm \phi}$ \ for all \ $\ket{\bm \phi} \in  \mathcal H_+^{\otimes 2}$. Indeed, these are the only eigenstates with non-vanishing overlap with $\ket{\psi(0)}^{\otimes 2}$ in \eqref{eq:mainidentity}. In general, the eigenvalues $\varphi_m$ denote energies in quantum dynamics induced by time-independent Hamiltonians with $t \in \mathbb R$, or the eigenphases of a Floquet unitary $U_F$ for which $U(t) = U_F^{t}$, $t \in \mathbb Z$. In the following sections we will restrict our attention to the Floquet case, but for the time being we can keep the setup general.

Given \eqref{eq:eigenstates2replica}, the Lehmann sum for $\mathbb P_A(t)$ reads
\begin{equation}
\label{eq:mainLehmanSum}
    \mathbb P_A(t) = \sum_{m\le m', n \le n'}e^{i(\varphi_m + \varphi_{m'} - \varphi_n - \varphi_{n'})t} \braket{\bm m|S_A|\bm n}\braket{\bm n|\rho(0)^{\otimes 2}|\bm m} \ .
\end{equation}
\cref{eq:mainidentity,eq:mainLehmanSum} make apparent that $\mathbb P_A(t)$, aside for the promotion of the global system to a 2-replica and the introduction of the swap $S_A$ in place of a physical observable, shares the same structural form as the dynamics of expectation values of ordinary 1-replica local observables $O$ after a quantum quench
\begin{equation}
\label{eq:dyn1replicaObs}
    \braket{\psi(t)|O|\psi(t)} = \sum_{m,n}e^{i(\varphi_m - \varphi_n)t}\braket{m|O|n}\braket{n|\rho(0)|m} \ .
\end{equation}
We stress that $S_A$, like $O$, acts nontrivially only in a spatial subregion (although it couples different replicas).
The analogy between \eqref{eq:mainLehmanSum} and \eqref{eq:dyn1replicaObs} motivates several questions inspired by the standard ETH treatments of dynamics:
\begin{itemize}
    \item Do the matrix elements $\braket{\bm m|S_A|\bm n}$ possess an ETH-like structure?
    \item What are the cross-correlations between these matrix elements and the overlaps $\braket{\bm n|\rho(0)^{\otimes 2}|\bm m}$ that are relevant for reconstructing purity dynamics (\emph{cf.}~Refs.~\cite{foini_outofequilibrium_2025, senese2026nonequilibrium})?
    \item Can we interpret the growth of entanglement after global quantum quenches as due to a fast decay of $\mathbb P_A(t)$ induced by dephasing in \cref{eq:mainLehmanSum}? Is the latter controlled by generalized ETH spectral functions? 
\end{itemize}

There is also another point that emerges naturally from the standard ETH perspective on dynamics. Working for simplicity at infinite temperature, as mentioned in the introduction the simplest dynamical correlator from the point of view of standard ETH are equilibrium dynamical 2-point functions
\begin{equation}
\label{eq:standard2pointdynam}
  \braket{O(t)O} \ , \quad \qquad \braket{\ldots} \equiv \frac{1}{d}\Tr_{\mathcal H}[\ldots] \ ,
\end{equation}
because their Fourier transforms map directly to the slowly varying functions $f_O(\omega)$ entering the ansatz  \eqref{eq:FloquetStETH}. Therefore, we are naturally led to consider
\begin{align}
\label{eq:Cplus}
    C_A^{(+)}(t) \equiv \braket{S_A(t)S_A}_2 &= \frac{1}{d_+}\sum_{m\le m', n \le n'}e^{i(\varphi_m + \varphi_{m'} - \varphi_n - \varphi_{n'})t} |\braket{\bm m|S_A|\bm n}|^2 \ , \\
   \braket{\ldots}_2 &\equiv \frac{1}{d_+}\Tr_{\mathcal H^{\otimes 2}_+}[\ldots] \ , \qquad  d_+ \equiv \dim(\mathcal H^{\otimes 2}_+) = \frac{d(d+1)}{2} \ ,
\end{align}
where $S_A(t)=(U^\dag(t))^{\otimes 2} S_A U(t)^{\otimes 2}$. For dynamics induced by time-independent Hamiltonians, analogous correlators could be considered at any finite temperature, but the choice \eqref{eq:Cplus} is the natural one given our subsequent focus on Floquet dynamics.
Although at first sight $C_A^{(+)}(t)$ does not appear to have any obvious relation with purity dynamics, it turns out to admit a very direct connection to it \cite{zanardi_entangling_2000}. Consider the following ensemble of initial states $\ket{\psi}=\ket{\psi(0)}$
\begin{equation}
\label{eq:prodEnsemble}
    \ket{\psi} = \ket{a}_A \ket{b}_B  \ , \qquad \ket{a}_A, \ket{b}_B  \ \text{Haar distributed in } A, B \ ,
\end{equation}
which have zero entanglement between $A$ and $B$ but significant entanglement, typically, within each subsystem. Then one can prove that \cite{zanardi_entangling_2000} (a short proof is provided in Appendix \ref{sec:appRMTproduct})
\begin{equation}
\label{eq:Eprod}
    \mathbb E_{\ket{\psi}}[\mathbb P_A(t)] = \frac{d+1}{(d_A+1)(d_B+1)}C_A^{(+)}(t) + \frac{d_A + d_B}{(d_A+1)(d_B+1)} \ ,
\end{equation}
where $\mathbb E_{\ket{\psi}}$ denotes average over the ensemble of initial states \eqref{eq:prodEnsemble}. 
Therefore, up to an offset and a constant of proportionality, $C_A^{(+)}(t)$ in \cref{eq:Cplus} exactly coincides with the time-dependent purity averaged over the ensemble of initial states \eqref{eq:prodEnsemble}. We note that identical product-state averages of subsystem purity in the context of spectral representations of entanglement dynamics (\emph{cf.}~\cref{eq:mainLehmanSum}) have been employed also in Ref.~\cite{pain_entanglement_2024}. Importantly, as we prove in Appendix \ref{app:concentrationHaar} using results from Ref.~\cite{hayden_aspects_2006}, in the limit of $d_A,d_B\to \infty$  the measure \eqref{eq:prodEnsemble} induces concentration around $\mathbb E_{\ket{\psi}}[\mathbb P_A(t)]$. In particular, we prove the following inequality
\begin{equation}
\label{eq:ncxmbvmxcx}
    {\rm Pr}\left( \big|\mathbb P_A(t) - \mathbb E_{\ket{\psi}}[\mathbb P_A(t)]\big|>\varepsilon \right) \le 8 \exp\left(-c d_{\min}\varepsilon^2\right) \ , \qquad d_{\min} = \min(d_A,d_B) = e^{V_{\rm min}\ln d_{\rm loc}} \ ,
\end{equation}
where $\varepsilon>0$ is an arbitrarily small constant, $c>0$ is a universal number, ${\rm Pr}$ denotes probability with respect to the distribution \eqref{eq:prodEnsemble}, and $V_{\rm min}=\min(|A|,|B|)$. This bound implies that in the limit of large subsystems, up to exponentially rare fluctuations, individual purities $\mathbb P_A(t)$ from randomly sampled initial states $\ket \psi$ coincide with $\mathbb E_{\ket{\psi}}[\mathbb P_A(t)]$ at arbitrary fixed times. This implies that $C_A^{(+)}(t)$ in the limit of large subsystems describes purity evolution (from typical product states \eqref{eq:prodEnsemble}) without the need of any average.

An important point must be kept in mind, here and in the following, regarding the limit $V_{\rm min}\to \infty$ (as opposed to simply considering $d\to\infty$ with either $A$ or $B$ finite). For systems with (entangling) local interactions all along the boundary $\partial$ between $A$ and $B$ and for any \emph{fixed} time $t$ the purity is lower bounded by \cite{bravyi_liebrobinson_2006, unanyan_shorttime_2010, vanacoleyen_entanglement_2013, shi_bounds_2024, toniolo_dynamical_2025}
\begin{equation}
\label{eq:purityboundarea}
    \mathbb P_A(t) \ge \mathbb P_A(0)\exp(-\alpha(t) |\partial|) \ ,
\end{equation}
where $|\partial|$ is the size of the boundary $\partial$ and $\alpha(t)>0$ is a system-size independent constant. In this work we always assume that $|\partial|$ is \emph{subextensive} with respect to $V_{\rm min}$, i.e.~$|\partial|=o(V_{\rm min})$. In practice, for generic interacting dynamics at any fixed time $|t|>0$, the purity $\mathbb P_A(t)$ is expected to saturate the bound \eqref{eq:purityboundarea} for some $\alpha$. This implies that $\mathbb P_A(t)$ for $|t|>0$ trivially vanishes when $V_{\rm min}\to \infty$ in dimension $D>1$ because $|\partial| \to \infty$ (in $D=1$ the boundary can diverge subextensively only in the presence of many entanglement cuts). However, the bound in \eqref{eq:ncxmbvmxcx} remains meaningful because, while at fixed $t$ the purities individually decay exponentially with $|\partial|$, it predicts a much faster exponential decay with $V_{\rm min}$ for their absolute difference (with probability approaching one) as seen by setting $\varepsilon = e^{-\gamma V_{\rm min}}$ for $0<\gamma<\ln d_{\rm loc}/2$. Furthermore, for systems in $D>1$ a natural intensive quantity to consider (in place of the purity) is the Rényi-2 entropy per unit boundary area
\begin{equation}
\label{eq:entropyperarea}
     s^{(2)}_A(t) = \lim_{V_{\rm min}\to \infty} \frac{\mathbb S_A^{(2)}(t)}{|\partial|} \ .
\end{equation}
This is the quantity one is typically interested in for $V_{\rm min}\to \infty$ and $D>1$.

In all of the following, we employ the symbol “$\sim$” to express asymptotic equality in the limit of $d \to \infty$ (with no additional assumption on $V_{\rm min}$, i.e.~on $d_A$ and $d_B$ individually).

\subsection{Summary of main results}

Although the formalism above is adequate for both time-independent and time-dependent quench protocols, from now on we focus mostly on the latter. Indeed, as discussed in the introduction, Floquet systems feature the simplest form of ETH-related statements, and hence represent the natural starting point for investigating ETH behaviour relevant for entanglement measures. RMT results that concern also systems evolving according to static Hamiltonians will be explicitly highlighted in the next sections. 

We consider local deterministic Floquet operators $U_F$ that possess no symmetry (not even time-reversal invariance), and hence whose spectra follow the circular unitary ensemble (CUE) statistics \cite{dyson_statistical_1962, mehta_random_matrices}. $U_F$ could represent the evolution in a time interval $\tau$ induced by a time-dependent periodic Hamiltonian $H(t+\tau) = H(t)$, or a deterministic quantum circuit. As in \eqref{eq:eigenstates2replica}, our convention for the eigenphases is $U_F \ket{m} = e^{-i\varphi_m}\ket{m}$. Based on the fact that standard single-replica ETH ansätze apply equally well (with minor modifications) to systems satisfying CUE (GUE) or COE (GOE) \cite{dyson_statistical_1962, mehta_random_matrices} statistics (see, e.g., Refs.~\cite{dalessio_quantum_2016, hamazaki_randommatrix_2019, patil2026quantum}), we expect that the overall functional form of the replica-ETH ansätze reported below applies also to Floquet systems with time reversal invariance (COE).

In the numerical checks presented later on, we will assume for simplicity that subsystem $A$ is a contiguous set of sites. However, such condition plays no role in our RMT calculations, nor in the rest of our arguments as long as the boundary $\partial$ between $A$ and $B$ scales subextensively with respect to $V_{\rm min}$ (\emph{cf.}~\cref{eq:purityboundarea,eq:entropyperarea} and \cref{sec:simplification}). We therefore believe all our results to hold for quite general bipartition geometries. 

Our main results and conjectures, obtained by a combination of exact calculations within RMT (\cref{sec:RMTbriefresults}), educated guesses based on established ETH results (\cref{sec:fullansatzsection}), and numerical checks via ED in 1D chains (\cref{sec:numver}), are summarized below.

\subsubsection{ETH ansatz for the subsystem swap operator}
\label{sec:ETHforsubswap}

In the limit $d = d_A d_B \gg 1$ (with no further individual assumptions on $d_A$ and $d_B$), the matrix elements of the subsystem swap appearing in \cref{eq:Cplus} follow an ETH-like ansatz of the form
\begin{align}
\label{eq:ETHansatz1}
\braket{\bm m|S_A|\bm m} & = \left( \frac{1}{d_A} + \frac{1}{d_B}\right) + \mathcal O(d^{-1})   \ ,\\
\label{eq:ETHansatz2}
    |\braket{\bm m|S_A|\bm n}|^2 & =  \frac{2}{d^2}f_A^2(\Omega, \omega_1, \omega_2) \big|R_{\bm m, \bm n}^{(A)}\big|^2 + \mathcal O(d^{-3})  \ , \qquad \quad  \ket{\bm m} \neq \ket{\bm n}  \ .
\end{align}
The leading order of \cref{eq:ETHansatz1} equals the leading order of the purity of global Haar random states, and is hence consistent with local relaxation to infinite temperature ensembles at late times \cite{dalessio_longtime_2014, lazarides_equilibrium_2014, ponte_periodically_2015} (see \cref{eq:mainLehmanSum} or further discussions below). It also matches the results of Ref.~\cite{kaneko_characterizing_2020}, which also studied matrix elements of $S_A$ in a basis of eigenstates but restricted the analysis to diagonal elements only (i.e.~those relevant for saturation at late times, but not for dynamics). 

The function $f_A$  in \cref{eq:ETHansatz2} denotes a continuous, positive, order one function which is very slowly varying (with respect to level separation) in the frequencies
\begin{equation}
    \Omega = \varphi_m + \varphi_{m'} - \varphi_n - \varphi_{n'},  \quad \omega_1 = \varphi_m - \varphi_n,  \quad \omega_2= \varphi_m - \varphi_{n'} \ .
\end{equation}
Here $\Omega$ is the fundamental dephasing frequency appearing in \cref{eq:mainLehmanSum,eq:Cplus}, while $\omega_1$, $\omega_2$ are convenient auxiliary frequencies. Consistently with the definition of each $\varphi$ as a complex phase, $f_A$ is periodic with period $2\pi$ in each argument. As in the standard ETH, the off-diagonal terms \eqref{eq:ETHansatz2} are exponentially suppressed with respect to the diagonal ones \eqref{eq:ETHansatz1} (note also that the off-diagonal prefactor $2/d^2$ is nothing but the leading order of $d_+^{-1}$, in analogy to the standard off-diagonal ETH). $R_{\bm m, \bm n}^{(A)}$ denote erratically varying complex variables such that
\begin{equation}
\label{eq:Rmn}
    \overline{\big|R_{\bm m, \bm n}^{(A)}\big|^2} = 1 \ ,
\end{equation}
where, similarly to the introduction, the bar indicates average over neighbouring pairs of 2-replica eigenstates $\bm m,\bm n$. Importantly, in the limit of $d \gg1$ the averages include a very large number of points within spectral regions where $f_A(\Omega, \omega_1, \omega_2)$ stays effectively constant. Quantities related to our off-diagonal matrix elements \eqref{eq:ETHansatz2} have been investigated analytically and numerically in Ref.~\cite{hahn_eigenstate_2024}, and their results are consistent with our ansatz (see further comments below). From our numerics we find that, for sufficiently large subsystem sizes (e.g.~$|A| \gtrsim4$) the statistics of $R_{\bm m, \bm n}^{(A)}$ is \emph{independent} of $A$ in the sense that the higher moments $\overline{|R|^{\ell}}$, $\ell > 2$, are compatible with a complex Gaussian distribution of unit variance, i.e.~$P(R)=\exp(-|R|^2)/\pi$. For very small subsystems, on the other hand, the higher moments retain a dependence on $A$ and hence deviate from the Gaussian prediction. Related observations were made in Ref.~\cite{hahn_eigenstate_2024}. For completeness, we discuss in \cref{sec:RMTbriefresults,sec:fullansatzsection} results analogous to \cref{eq:ETHansatz1,eq:ETHansatz2} also for eigenstates in the antisymmetric sector $\mathcal{H}_-^{\otimes 2}$
\begin{equation}
\label{eq:antisymmEigen}
    \ket{\widetilde{\bm m}} = \frac{\ket{m,m'}-\ket{m',m}}{\sqrt{2}} \ , \qquad m< m' \ .
\end{equation}
The changes are minimal: in \eqref{eq:ETHansatz1} the plus sign is replaced by a minus sign, while the off-diagonal part stays identical. 
We stress that, consistently with $\mathbb S_B^{(2)}=\mathbb S_A^{(2)}$ for pure states, the matrix elements $\braket{\bm m |S_A|\bm n}$ coincide with those where $S_A \to S_B$. Indeed, using $S_{\mathcal H} = S_A S_B$, $S_A^2=\Id$ and $S_{\mathcal H} \ket{\bm n} = \ket{\bm n}$, we get
\begin{equation}
\label{eq:excSAandSB}
    \braket{\bm m |S_A|\bm n} = \braket{\bm m |S_A S|\bm n} = \braket{\bm m |S_B|\bm n} \ .
\end{equation}
This proves that the statistics in \cref{eq:ETHansatz1,eq:ETHansatz2} must be invariant under $d_A \leftrightarrow d_B$, as already apparent in the diagonal part. 

It is important to remark that although we prefer to work with (anti)symmetrized eigenstates, because they resolve the replica symmetry $[S_{\mathcal H},U(t)^{\otimes 2}]=0$ and lead to further simplifications (\emph{cf.}~\cref{sec:RMTbriefresults} and Appendix \ref{appendix:RMT}), an ETH-like ansatz is present also for the unsymmetrized matrix elements $\braket{m,m'|S_A|n,n'}$. These, and hence their ETH ansatz, can be related to the (anti)symmetrized ones by combining symmetric and antisymmetric statistics together. Using \cref{eq:intertwofS} it is also elementary to express them as
\begin{align}
\label{eq:redtranmatrix1}
    \braket{m,m'|S_A|n,n'} &= \Tr_A\left[\rho_A^{(nm)}\rho_A^{(n'm')}\right] \ , \\
\label{eq:redtranmatrix2}
    \rho_A^{(nm)} &\equiv \Tr_B\big[\ket n\bra m\big] \ ,
\end{align}
in terms of reduced eigenstate-transition matrices. In the context of ETH, both $\rho_A^{(mm)}$ as $m$ varies and $\rho_A^{(nm)}$ with $n \neq m$ have been considered in Ref.~\cite{dymarsky_subsystem_2018}. Based on their diagonal results (naturally generalized here to the Floquet context), \cref{eq:ETHansatz1} can be understood to hold when $m=n\neq m'=n'$ (i.e.~not only when all indices are equal) and $d_A \ll d_B$ because $\braket{m,m'|S_A|m,m'} = \Tr_A\big[\rho_A^{(mm)}\rho_A^{(m'm')}\big] $ and $\|\rho_A^{(mm)}-\rho_A^{(m'm')}\|\to 0$ in the limit of $d \to \infty$ \cite{dymarsky_subsystem_2018}. This shows that such diagonal matrix elements coincide at leading order with the purity of a single eigenstate $\ket m$ (see also related statements in Ref.~\cite{jindal_generalized_2024}). The other unsymmetrized matrix element entering \eqref{eq:ETHansatz1}, i.e.~$\braket{m,m'|S_A|m',m} = \Tr_A\big[\rho_A^{(m'm)}\rho_A^{(mm')}\big]$, is instead subleading in the fixed-$d_A$ limit due to the claim from Ref.~\cite{dymarsky_subsystem_2018} that $\|\rho_A^{(mn)}\| \propto d^{-1/2}$ at fixed $d_A$ for $m\neq n$. This off-diagonal scaling is also consistent with the validity of \eqref{eq:ETHansatz2}. For example, when all four eigenstate indices are distinct the $\mathcal O(d^{-2})$ upper bound on the squared matrix elements it provides is saturated by our ansatz. In presence of some coincident indices the bound can be less tight but still consistent.\\

\cref{eq:redtranmatrix1,eq:redtranmatrix2} also allow us to make more explicit the connections with Ref.~\cite{hahn_eigenstate_2024}. Indeed, expanding each eigenstate $\ket m$ into a product state basis $\ket{\Psi_a}_A \ket{\Psi_b}_B$ with respect to our bipartition (see \cref{eq:SchmidtIdentity1,eq:SchmidtIdentity2,eq:SchmidtIdentity3} in \cref{sec:numver}) one obtains that the fundamental quantity $M_A(nmn'm')$ in Ref.~\cite{hahn_eigenstate_2024} (in their notation) is nothing but $\braket{m,m'|S_A|n,n'}$. Most of the focus of Ref.~\cite{hahn_eigenstate_2024} is then on quantities dependent on two distinct bipartitions $A\cup B$ and $\widetilde{A} \cup \widetilde B$, that they relate to OTOC dynamics and evolution-operator purity, mostly within a framework where the correlators are averaged over an ensemble of local Floquet circuits. However, it is possible to connect some of their quantities \emph{before ensemble average} to our $C_A^{(+)}(t)$ from \cref{eq:Cplus,eq:Eprod}. The distinction is in the respective interpretations: while Ref.~\cite{hahn_eigenstate_2024} recognizes their importance in the description of \emph{operator} purity, we use \cref{eq:Eprod} and the concentration-of-measure results of Appendix \ref{app:concentrationHaar} to relate $C_A^{(+)}(t)$ to evolving \emph{state} purity from typical initial states from the ensemble \eqref{eq:prodEnsemble}. This provides us with a simple warm up before our general treatment of deterministic quench dynamics from more general \emph{individual} initial states (see \cref{eq:mainLehmanSum} and \cref{sec:summaryPurityForSinglePsi} below). We also remark that our treatment is explicitly constructed around the \emph{absence} of averages over realizations of dynamics: our statistics and coarse-grained dynamical predictions emerge entirely from local-in-the-spectrum averages of multireplica eigenstates of a \emph{fixed} deterministic $U(t)$. This marks an additional conceptual difference with respect to Ref.~\cite{hahn_eigenstate_2024} (see, however, their Appendix C for a discussion of single-realization results in a specific class of circuits). The four-eigenstate correlations considered in Ref.~\cite{pain_imprints_2026}, closely related to those studied in Ref.~\cite{hahn_eigenstate_2024}, admit the same interpretation in terms of subsystem swap matrix elements.

\subsubsection{Dynamics of purity averaged over initial product states} 
The Lehmann sum \eqref{eq:Cplus} for $C_A^{(+)}(t)$, which gives direct access to the averaged purity $\mathbb E_{\ket{\psi}}[\mathbb P_A(t)]$ via \cref{eq:Eprod}, comprises seven classes of pairs of 2-replica eigenstates, schematically represented in \cref{tab:classesOfEigenstates}. 
\begin{table}[t]
\centering
\begingroup
\normalsize
\renewcommand{\arraystretch}{0.9}
\setlength{\extrarowheight}{3pt}
\setlength{\tabcolsep}{5pt}

\begin{tabular}{ccc}
\toprule
\rowcolor{white}[\tabcolsep]
Class & $(m,m')$ & $(n,n')$ \\
\midrule
\rowcolor{white}[\tabcolsep]
$\mathrm{I}$   & $(a,a)$ & $(a,a)$ \\
\rowcolor{white}[\tabcolsep]
$\mathrm{II}$  & $(a,a)$ & $(a,b)$ \\
\rowcolor{white}[\tabcolsep]
$\mathrm{III}$ & $(a,a)$ & $(b,b)$ \\
\rowcolor{rowblue}[\tabcolsep]
$\mathrm{IV}$  & $(a,b)$ & $(a,b)$ \\
\rowcolor{white}[\tabcolsep]
$\mathrm{V}$   & $(a,a)$ & $(b,c)$ \\
\rowcolor{white}[\tabcolsep]
$\mathrm{VI}$  & $(a,b)$ & $(a,c)$ \\
\rowcolor{rowblue}[\tabcolsep]
$\mathrm{VII}$ & $(a,b)$ & $(c,d)$ \\
\bottomrule
\end{tabular}
\endgroup
\makeatletter
\global\let\CT@row@color\relax
\makeatother
    \caption{The seven possible classes of pairs of 2-replica symmetric eigenstates $\ket{\bm m}$ and $\ket{\bm n}$, \emph{cf.}~\cref{eq:symmetricEigen}, appearing in the Lehmann sums in \cref{eq:mainLehmanSum,eq:Cplus}.  The indices $a,b,c,d$ are all distinct. Even within the convention $m\le m'$, $n\le n'$ from \eqref{eq:symmetricEigen}, some of the classes contain other representative states that can be obtained by a permutation of their characteristic indices. The blue-shaded classes $\mathrm{IV}$ and $\mathrm{VII}$ are the only ones relevant in the thermodynamic limit.}
    \label{tab:classesOfEigenstates}
\end{table}
Of these seven classes, only classes IV $\rightarrow$ $\{(a,b),(a,b)\}$
and VII $\rightarrow$ $\{(a,b),(c,d)\}$ (with $a \neq b \neq c \neq d$) can contribute at $\mathcal O(1)$ to $C_A^{(+)}(t)$, while the others contribute only at $\mathcal{O}(d^{-1})$ and are therefore subleading with respect to the saturation value of $C_A^{(+)}(t)$ at late times. This is very similar to the entropic suppression of crossing partitions in full-ETH \cite{foini_eigenstate_2019, pappalardi_eigenstate_2022}. To understand the suppression, we can split $C_A^{(+)}(t)$ into a time-independent component $C_A^{(\rm diag)}$ where we restrict the Lehmann sum to $\ket{\bm m }=\ket{\bm n}$, and a time-dependent component $C_A^{(\rm off)}(t)$ where only off-diagonal elements $\ket{\bm m} \neq \ket{\bm n}$ appear. At leading order in $d \gg 1$, we get from \cref{eq:ETHansatz1}
\begin{align}
\label{eq:Cdiag}
    C_A^{(\rm diag)} \sim \left( \frac{1}{d_A} + \frac{1}{d_B}\right)^2 \sim \braket{S_A}_2^2 \ .
\end{align}
We see that of the 2 diagonal classes in \cref{tab:classesOfEigenstates}, i.e.~classes I and IV, only IV has an adequate entropic factor to counterbalance the $d_+^{-1}=2d^{-1}(d+1)^{-1}$ prefactor in \eqref{eq:Cplus}, while class I only contributes with a subleading correction to \eqref{eq:Cdiag}. In the limit of $d \to \infty$, only diagonal terms survive dephasing at late times, and hence $C_A^{(\rm diag)}$ captures the late time relaxation of $C_A^{(+)}(t)$. Combining \eqref{eq:Cdiag} with \cref{eq:Eprod}, we obtain that $\mathbb E_{\ket{\psi}}[\mathbb P_A(t)]$ relaxes at late times to
\begin{equation}
\label{eq:PurProdRelax}
    \mathbb P_A^{(\rm diag)}\sim  \frac{1}{d_A} + \frac{1}{d_B}\ .
\end{equation}
This coincides, at leading order, with the purity of a global Haar random state and is hence consistent with the standard picture of local relaxation of Floquet dynamics to an infinite-temperature ensemble.

By analogous reasoning we see that, in the time-dependent component $C_A^{(\rm off)}(t)$, classes II, III, V and VI from \cref{tab:classesOfEigenstates} lack an entropic factor that counterbalances the $1/d_+$ prefactor in \eqref{eq:Cplus} and the $2/d^2$ prefactor in the off-diagonal ETH ansatz \eqref{eq:ETHansatz2}. Therefore, all these classes contribute only at $\mathcal{O}(d^{-1})$ to $C_A^{(+)}(t)$, and hence the class VII $\rightarrow$ $\{(a,b),(c,d)\}$ drives alone the decay of $C_A^{(+)}(t)$ from $C_A^{(+)}(0)=1$ to the relaxation at $C_A^{(\rm diag)}$ \footnote{We note that in the specific case of $d_A, d_B \propto \sqrt{d}$, one gets a saturation value $C_A^{(\rm diag)}\propto d^{-1}$, which coincides with the order of contribution of off-diagonal classes V and VI at $t=0$. However, at the late times where saturation emerges, classes V and VI have dephased, and so $C_A^{(\rm diag)}$ is dominant.}. By means of \cref{eq:Eprod} we conclude that also the decay and saturation of $\mathbb E_{\ket{\psi}}[\mathbb P_A(t)]$ is governed, at leading order, only by classes IV and VII. 

Using the ETH ansatz \eqref{eq:ETHansatz2}, and the fact that the level spacing in the eigenphases $\varphi_j$ becomes uniformly suppressed to the average value of $2\pi/d$, we obtain for large $d$
\begin{align}
\label{eq:CoffFourier}
    C_A^{(\rm off)}(t) \sim \int_0^{2\pi}\frac{d\Omega}{2\pi} e^{i\Omega t}F_A(\Omega)  \ , \quad \qquad F_A(\Omega) =  \int_0^{2\pi} \frac{d\omega_1\omega_2}{(2\pi)^2}  f_A^2(\Omega,\omega_1,\omega_2)  \ .
\end{align}
The asymptotic equality in the first equation must be intended for $t$ arbitrary but fixed while $d \to \infty$.
This is a crucial result, because it relates the decay of the average purity $\mathbb E_{\ket \psi}[\mathbb P_A(t)]$ to a slowly varying function that can be extracted directly from the 2-replica matrix elements \eqref{eq:ETHansatz2}. In the present setting (no symmetries) in general we expect that $\mathbb E_{\ket \psi}[\mathbb P_A(t)]$ decays to zero exponentially with $t$ until saturation. A similar asymptotic decay can characterize also $\mathbb E_{\ket \psi}[\mathbb P_A(t)]- \mathbb P_A^{(\rm diag)}$ \cite{znidaric_twostep_2023, jonay_physical_2024}. This would imply that $\lim_{d\to\infty }F_A(\Omega)$ is smooth ($C^\infty$) and analytic in a horizontal strip of the complex plane which includes the whole real axis. The same remains true when $|A|,|B|\to\infty$ ($\mathbb P_A^{(\rm diag)}\to0$), for which the exponential decay is generally strongly believed to indeed hold asymptotically (keep also in mind \cref{eq:entropyperarea} in this limit). We note that $C_A^{(+)}(0)=1$ provides an exact sum-rule for $f_A^2$. Indeed, for $d$ large (and no assumptions on $d_A$ and $d_B$ singularly)
\begin{align}
\label{eq:sumrulfA}
    1=C_A^{(+)}(0) = C_A^{(\rm diag)}+C_A^{(\rm off)}(0) \quad \longrightarrow  \quad \int_0^{2\pi} \frac{d\Omega \, d\omega_1 d\omega_2}{(2\pi)^{3}}f_A^2(\Omega, \omega_1, \omega_2) \sim 1- \left( \frac{1}{d_A} + \frac{1}{d_B}\right)^2  \ .
\end{align}
Importantly, \cref{eq:CoffFourier} implies that being able to estimate $f_A^2(\Omega,\omega_1, \omega_2)$ via spectral filter methods \cite{luo_probing_2024} for large (sub)system sizes would give direct access to the typical growth of Rényi-2 entropy for a fixed deterministic $U_F$ and initial states drawn from \eqref{eq:prodEnsemble}.

\subsubsection{Connections with full-ETH}
\label{sec:connFeth}

In the limit of $d_A$ arbitrarily large but finite while $d \to \infty$, unlike other probes of quantum complexity and information scrambling \cite{foini_eigenstate_2019, hahn_eigenstate_2024}, the spectral function $f_A^2(\Omega,\omega_1,\omega_2)$ associated with the decay of $C_A^{(+)}(t)$ is at leading order determined only by sums of products of 2-ETH terms (\emph{cf.}~\cref{eq:FloquetStETH,eq:nETHFloquet}). 4-ETH terms, although present, contribute only at subleading orders. This becomes apparent from the following standard identity for the subsystem swap operator
\begin{equation}
\label{eq:SAexpanded}
    S_A = \sum_{\mu=1}^{d_A^2}O_\mu \otimes O_\mu \ ,
\end{equation}
where $\{O_\mu\}$ denotes a Hermitian orthonormal basis of single-replica operators in $A$ (with respect to the Hilbert-Schmidt product $\Tr_A[T_1^\dag T_2]$ in $A$). This identity has been implicitly used in the past \cite{ho_entanglement_2017} to connect the dynamics of Rényi entropies to operator spreading. From \eqref{eq:SAexpanded} we see that $f_A^2$ can be related at leading order to the spectral functions associated with 2-ETH matrix elements of the form $(O_\mu)_{mn}(O_\nu)_{nm}$. Similar conclusions remain true for higher Rényi entropies, see \cref{sec:higherRenyi}. We find this a simple yet interesting observation. However, we stress that  \cref{eq:SAexpanded} does \emph{not} imply that the spectral function $f_A^2$ is similar to any 2-ETH function. Indeed, using  \eqref{eq:SAexpanded} in the definition \eqref{eq:Cplus} of $C_A^{(+)}(t)$ leads to a sum involving $d_A^4$ products of 2-ETH terms, many of them involving operators $O_\mu$ that act nontrivially over quite extended subregions of $A$. This suggests that, as soon as $|A|\gg1$, the defining features of $f_A$ can be quite distinct from those of any individual 2-ETH function that contributes to it. The reason why we find \cref{eq:SAexpanded} useful is that it enables us to advocate, in the limit of $d_A$ finite while $d \to \infty$, for the existence of ETH statements related to Rényi entropies on the basis of more established single-replica ETH results.

To further strengthen the connections with Ref.~\cite{hahn_eigenstate_2024}, we remark that all the identities relating their fundamental building block $M_A(nmn'm')$ to physical quantities defined by sums over basis of operators in $A$ are simply a consequence of \cref{eq:SAexpanded}. We also highlight that some explicit connections between spectral representations for Rényi entropies and full-ETH have been worked out in Ref.~\cite{jindal_generalized_2024}, although within a framework that differs substantially from the one we employ here. From their perspective, it is not immediately evident that only 2-ETH terms dominate in the present limit, while it is trivial to verify it using \eqref{eq:SAexpanded}.

\subsubsection{ETH ansatz for purity dynamics without any average}
\label{sec:summaryPurityForSinglePsi}
In the Lehmann sum \eqref{eq:mainLehmanSum} for the purity $\mathbb P_A(t)$ in a quench from an initial state $\ket{\psi(0)}$ (with the characteristics specified above, e.g.~$\ket{\psi(0)}$ lowly entangled in the simplest case), the additional spectral ingredients are the overlaps $\braket{m|\psi(0)}$. In a Floquet system belonging to the CUE class one expects for large $d$ \cite{haake_quantum_2010, herrmann_characterizing_2023, schmid2026genuine} (\emph{cf.}~also Appendix \ref{sec:appRMTsingle})
\begin{equation}
\label{eq:zmDef}
    \braket{m|\psi(0)} = \frac{z_m}{\sqrt{d}} \ ,
\end{equation}
where $z_m$ is an erratically varying order-one complex amplitude that can be interpreted (at the level of local-spectrum averages over $m$) as a complex Gaussian random variable of unit variance for what concerns its moments $|z_m|^{q}$. Similar relations also hold for Floquet systems in the COE class. We have
\begin{equation}
\label{eq:Rtilde}
    \braket{\bm n |\rho(0)^{\otimes 2}|\bm m} = \frac{2}{\sqrt{(1+\delta_{m,m'})(1+\delta_{n,n'})}} \frac{\widetilde{R}_{\bm n,\bm m}}{d^2} \ , \qquad \quad \widetilde{R}_{\bm n, \bm m}=z_n z_{n'} z_m^* z_{m'}^*  \ .
\end{equation}
Combining \cref{eq:Rtilde} with \cref{eq:ETHansatz1,eq:ETHansatz2}, and given the RMT results of \cref{sec:RMTbriefresults} together with the considerations in \cref{sec:fullansatzsection}, we arrive at the following generalized “out-of-equilibrium ETH” ansatz \cite{foini_outofequilibrium_2025}
\begin{align}
\label{eq:outEqETH1}
    \overline{\braket{\bm m|S_A|\bm m}\braket{\bm m |\rho(0)^{\otimes 2}|\bm m}} &= \frac{2}{d^2}\left( \frac{1}{d_A} + \frac{1}{d_B} \right)\big(1 + \mathcal{O}(d^{-1})\big) \ , \\
\label{eq:outEqETH2}
    \overline{\braket{\bm m|S_A|\bm n}\braket{\bm n |\rho(0)^{\otimes 2}|\bm m}}  &= \frac{4}{d^4} g_A(\Omega, \omega_1, \omega_2) + \mathcal{O}(d^{-5})  \ ,
\end{align}
where $g_A$ is an order-one function that depends both on $A$ and $\ket{\psi(0)}$, and that is very slowly varying compared to level-spacing scales. As before the bar indicates average over neighbouring 2-replica eigenstates, which happens on eigenphase scales where $g_A$ stays constant (aside for the subleading classes II and III from \cref{tab:classesOfEigenstates}, see below). The previous two ansätze imply that at leading order in $d^{-1}$
\begin{align}
\label{eq:jagkjgV1}
&\overline{\widetilde{R}_{\bm m , \bm m}} =\overline{|z_m|^4} \sim 2 \ ,  \qquad m = m' \ ,\\
\label{eq:jagkjgV2}
    &\overline{\widetilde{R}_{\bm m , \bm m}} =\overline{|z_m|^2 |z_{m'}|^2} \sim 1 \ ,  \qquad m \neq m' \ ,\\
    \label{eq:crossCorrRRtilde} 
    &\overline{R^{(A)}_{\bm m, \bm n}\widetilde{R}_{\bm n , \bm m}} \sim \sqrt{(1+\delta_{m,m'})(1+\delta_{n,n'})}\frac{\sqrt{2}}{d}\frac{ g_A(\Omega,\omega_1,\omega_2)}{f_A(\Omega,\omega_1,\omega_2)} \ , \qquad \ket{\bm m} \neq \ket{\bm n} \ .
\end{align}
The first two of the previous equations are consistent with the interpretation of $z_m$ as a complex Gaussian variable of unit variance for what concerns its moments.
From the last equation we see that, similarly to what happens in spectral representations for the dynamics of 1-replica observables \cite{foini_outofequilibrium_2025} (\emph{cf}.~\cref{eq:outETHsinglerep,eq:dyn1replicaObs}), the erratically varying complex variables $R^{(A)}\widetilde{R}$ must possess strongly fluctuating phases. Indeed, while the terms $R^{(A)}\widetilde R$ are individually of \emph{order one}, their local-spectrum average decays as $d^{-1}$. This spectral-interference mechanism is generic for quench dynamics in chaotic models \cite{foini_outofequilibrium_2025, senese2026nonequilibrium}, and it is at the core of the entropic counterbalancing that ensures Lehmann sums contribute correctly at $\mathcal O(1)$. In particular, \cref{eq:crossCorrRRtilde} shows that the continuous spectral function $g_A$ governing the dynamics of $\mathbb P_A(t)$ emerges from delicate cross-correlations between $R^{(A)}$ and $\widetilde{R}$. 

We note that to resolve an $\mathcal O(d^{-1})$ average of approximately independent terms each with $\mathcal O (1)$ magnitude, the average must include $\simeq d^2$ terms. For class VII from \cref{tab:classesOfEigenstates} (four distinct eigenstate indices) this implies that each average involves a fraction $d^{-1/2}$ of all eigenstates, which is perfectly local in the spectrum for $d \to \infty$. For classes V and VI (three distinct indices) the fraction is $d^{-1/3}$, which still decays quickly to zero for $d \to \infty$. On the other hand, for the off-diagonal classes II and III (two distinct indices) the fraction required to resolve the suppression in \eqref{eq:crossCorrRRtilde} is order one, and so the bar symbol cannot be associated with a local-spectrum average. However, these classes are subleading in the Lehmann representation (see below) and so this has no implications for the reconstruction of purity dynamics.
We remark that the role of higher-order eigenstate correlations in purity dynamics starting from a specified $\ket{\psi(0)}$ has been investigated in Refs.~\cite{shi_local_2023, jindal_generalized_2024}. In particular, the cross-correlations behind \eqref{eq:crossCorrRRtilde}, including the expected $d^{-4}$ scaling in \cref{eq:outEqETH2}, are implicitly captured by the formalism of Ref.~\cite{jindal_generalized_2024}. The explicit purity-dynamics calculations in these works, however, focus on initial states of the form $\ket{\psi(0)}=\ket{a}_{H_A}\ket{b}_{H_B}$, where $\ket{s}_{H_S}$ is an eigenstate of the subsystem Hamiltonian $H_S$ with $S=A,B$.

As in the case of $C_A^{(+)}(t)$, the scalings with $d^{-1}$ in \cref{eq:outEqETH1,eq:outEqETH2} imply that only the diagonal class IV $\rightarrow$ $\{(a,b),(a,b)\}$ and the off-diagonal class VII $\rightarrow$ $\{(a,b),(c,d)\}$ of \cref{tab:classesOfEigenstates} can contribute at $\mathcal O(1)$ to $\mathbb P_A(t)$, while all the remaining classes contribute at most at $\mathcal O(d^{-1})$ due to absence of entropic counterbalancing. Similarly to what done for the spectral representation for $C_A^{(+)}(t)$, we can split $\mathbb P_A(t)$ in a time-independent diagonal part $\mathbb P_A^{(\rm diag)}$ and a time-dependent off-diagonal one $\mathbb P_A^{(\rm off)}(t)$. We immediately obtain from \cref{eq:outEqETH1} that $\mathbb P_A^{(\rm diag)}$ matches the Haar-random-purity value in \cref{eq:PurProdRelax}. This also demonstrates that the $\mathcal O(d^{-1})$ contributions of all the subleading classes from \cref{tab:classesOfEigenstates} are exponentially smaller than the value $\mathbb P_A(t)$ relaxes to at late times.  
For the off-diagonal part, similarly to \cref{eq:CoffFourier}, we derive the following spectral representation in the limit of large $d$
\begin{align}
\label{eq:PAFourier}
    \mathbb P^{\rm (off)}_A(t) \sim \int_0^{2\pi}\frac{d\Omega}{2\pi} e^{i\Omega t}G_A(\Omega)  \ , \quad \qquad G_A(\Omega) =  \int_0^{2\pi} \frac{d\omega_1\omega_2}{(2\pi)^2}  g_A(\Omega,\omega_1,\omega_2)  \ .
\end{align}
 We can obtain from this and from \eqref{eq:PurProdRelax} a sum rule for $g_A$
\begin{align}
\label{eq:sumrulgA}
    \mathbb P_A(0)& = \mathbb P_A^{(\rm diag)}+\mathbb P_A^{(\rm off)}(0)\quad \longrightarrow  \quad \int_0^{2\pi} \frac{d\Omega \, d\omega_1 d\omega_2}{(2\pi)^{3}}g_A(\Omega, \omega_1, \omega_2) \sim \mathbb P_A(0)- \frac{1}{d_A} - \frac{1}{d_B} \ .
\end{align}
As before, under the assumption that in the present setup (no symmetries) $\mathbb P_A^{(\rm off)}(t)$ decays exponentially with $t$ \cite{znidaric_twostep_2023, jonay_physical_2024}, then $\lim_{d\to\infty}G_A(\Omega)$ is a smooth function that is analytic in a strip of the complex plane that includes the real axis. In the large-subsystems limit $V_{\rm min}\to \infty$ (where $\mathbb P_A^{(\rm off)}(t)\to\mathbb P_A(t)$) the leading asymptotic decay in $t$ is indeed generally expected to be exponential (ballistic growth of Rényi-2 entanglement), \emph{cf.}~also \cref{eq:entropyperarea}.\\

Using the identity \eqref{eq:SAexpanded} inside \cref{eq:outEqETH1,eq:outEqETH2} it is possible to relate the expression \eqref{eq:mainidentity} for the purity $\mathbb P_A(t)$ and its associated spectral weights to sums of $d_A^2$-many out-of-equilibrium-ETH single-replica terms \cite{foini_outofequilibrium_2025}. This is the out-of-equilibrium analogue of the connection with full-ETH discussed in \cref{sec:connFeth}.

In establishing the existence of a slowly-varying smooth spectral function governing entanglement growth, and that can be extracted from the spectral statistics in \eqref{eq:outEqETH2}, we hope to motivate future work on the construction of spectral filter methods \cite{luo_probing_2024} that can probe $g_A$ for large (sub)systems.

\subsubsection{Simplified spectral function for large subsystems}

Using only locality of the dynamics, together with standard claims related to bulk spectral dephasing, we analytically demonstrate the following simplification in the frequency dependence of the off-diagonal function $f_A$ from \cref{eq:ETHansatz2}
\begin{align}
\label{eq:adjkdafjh1}
    f_A^2(\Omega,\omega_1,\omega_2) =  \mathsf F_A(\Omega) + r_A^{(f)}(\Omega, \omega_1, \omega_2)\ , 
\end{align}
where in the limit of $d_A,d_B \to \infty$ (i.e.~$V_{\rm min}\to \infty$) we have
\begin{equation}
\label{eq:adjkdafjh1bis}
|1-\mathsf F_A(\Omega)|= \mathcal O \big(e^{-\alpha |\partial|}\big) \ , \qquad |r_A^{(f)}(\Omega, \omega_1, \omega_2)|=\mathcal O\big( e^{-\alpha' V_{\rm min}}\big) \ ,
\end{equation}
with $\alpha, \alpha'>0$ system-size independent constants. The previous two equations state that the spectral function $f_A$, which depends a priori on a set of three frequencies, in the limit of large subsystems $A$ and $B$ becomes independent of $\omega_1$ and $\omega_2$. Indeed, the remainder $r_A^{(f)}$ is exponentially suppressed in $V_{\rm min}$, and it is hence negligible with respect to $\mathsf F_A(\Omega)$ under our assumption that boundary size $|\partial|$ is subextensive with respect to $V_{\rm min}$. Therefore for large $V_{\rm min}$ the function $\mathsf F_A(\Omega)$ encodes alone the time dependence of the (typical, see \cref{eq:ncxmbvmxcx}) purity. In the scaling limit $V_{\rm min}\to\infty$ (for which $|\partial|\to\infty$ in $D>1$) it encodes the Rényi-2 entropy per boundary area of \cref{eq:entropyperarea}. Indeed $\mathsf F_A$ approaches $1$ exponentially with $|\partial|$ as a consequence of \cref{eq:sumrulfA} and of the expected saturation of the in bound \cref{eq:purityboundarea}. From \eqref{eq:CoffFourier} we also see that $\mathsf F_A(\Omega)$ approaches the marginal $F_A(\Omega)$ exponentially fast with $V_{\rm min}$. 

Under the assumption that $\ket{\psi(0)}$ is a product state, or that it has subextensive entanglement with respect to the bipartition into $A$ and $B$ (this latter case requiring a few extra technical assumptions described in \cref{sec:simplification}), we obtain a similar result also for $g_A$ from \cref{eq:outEqETH2}
\begin{align}
\label{eq:adjkdafjh2}
   & g_A(\Omega,\omega_1,\omega_2) =  \mathsf G_A(\Omega) + r_A^{(g)}(\Omega, \omega_1, \omega_2)\ , \\
\label{eq:adjkdafjh2bis}
&|\mathbb P_A(0)-\mathsf G_A(\Omega)|=\mathcal O \big(e^{-\beta |\partial|}\big) \ , \qquad |r_A^{(g)}(\Omega, \omega_1, \omega_2)|=\mathcal O \big( e^{-\beta' V_{\rm min}}\big) \ ,
\end{align}
where $\beta,\beta'>0$ are system-size independent constants and $\mathsf G_A(\Omega)$ approaches $\mathbb P_A(0)$ as a consequence of \eqref{eq:sumrulgA}. Therefore in the limit of large $V_{\rm min}$ the the function $\mathsf G_A(\Omega)$ determines alone the time-dependence the purity $\mathbb P_A(t)$ (and for $V_{\rm min}\to \infty$ of the Rényi-2 entropy per boundary area \eqref{eq:entropyperarea}).

In \cref{sec:numver} we verify the emergence of such simplifications from our ED numerics. We stress that the emergent independence from some fundamental frequencies, which can greatly aid the numerical reconstruction of the spectral functions, is generally not expected to arise in standard full-ETH spectral functions that depend on multiple frequencies \cite{foini_eigenstate_2019, chan_eigenstate_2019}. It is an emergent property tied to the structure of the subsystem swap operator. Related observations, although based on different large-scale limits and physical arguments, have been made in Ref.~\cite{hahn_eigenstate_2024} for quantities that can be related to our $f_A^2$ (\emph{cf.}~also earlier statements in \cite{chan_eigenstate_2019} and related locality-based consistency requirements in the spectral analysis of \cite{jindal_generalized_2024}). To the best of our knowledge, the extension of such auxiliary-frequency independence to out-of-equilibrium-ETH spectral functions like $g_A$ has instead no antecedent.

\section{ETH skeleton from RMT}
\label{sec:RMTbriefresults}

The simplest way to analytically motivate the structural form of the ETH ansätze in \cref{eq:FloquetStETH,eq:nETHFloquet,eq:factorizationFullETH,eq:outETHsinglerep} is to employ RMT \cite{mehta_random_matrices}. This traces back to Wigner's highly influential idea \cite{wigner_statistical_1951, wigner_characteristic_1955, wigner_random_1967} that, locally in the spectrum, the statistics of eigenvalues and eigenstates of a \emph{structured} large matrix (representing for example a physical Hamiltonian or a Floquet unitary) might behave very similarly to the statistics one would obtain from a large \emph{random} matrix. This applies to physical systems that are sufficiently chaotic \cite{bohigas_characterization_1984}, e.g.~that lack physically relevant symmetries or whose few symmetries can be resolved into large independent sectors. Although global random matrices do not possess key features of physical unitaries, e.g.~the \emph{spatial locality} or the \emph{few-body} nature of their generators, the matrix-element  statistics calculated using RMT perfectly captures the “skeleton” of its physical counterpart, and it “only” misses the continuous slowly varying functions that appear in ETH-like ansätze. Obviously, while RMT results hold for arbitrary operators, they are expected to describe the skeleton of physical matrix-element statistics only when the operator is “generic” in relation to the Hamiltonian or unitary providing the physical eigenstates (\emph{cf.}~\cite{mierzejewski_quantitative_2020} for related ideas).

Therefore, our strategy is to use RMT to derive the skeleton of the ETH ansatz \eqref{eq:ETHansatz1} and \eqref{eq:ETHansatz2} for the subsystem swap $S_A$ (which can be regarded as a generic operator), and its out-of-equilibrium analogue from \cref{eq:outEqETH1,eq:outEqETH2}. In \cref{sec:fullansatzsection} we then “dress up” the skeleton by the expected continuous functions that encode the leading coarse-grained dynamics on timescales that diverge with system size. We note that RMT has already been extensively used to characterize the entanglement properties of random pure states and their connection to many-body eigenstates (\emph{cf.}~Ref.~\cite{bianchi_volumelaw_2022}). Here, instead, our focus is on the matrix-element statistics underlying the \emph{dynamics} of Rényi entropies.

\subsection{Haar averages from GUE and CUE}
\label{sec:GUEandCUE}

We restrict our attention to unitary dynamics that lacks time-reversal symmetry (as well as any other symmetry) and derive RMT results that capture \emph{both} time-independent Hamiltonians and Floquet unitaries. Therefore, although the dressing of the next section is restricted to the latter case, the skeletons we obtain will be relevant for future extensions of our ansätze to static Hamiltonians.

For static Hamiltonians $H$, the lack of time-reversal symmetry implies that $H$ is a $d \times d$ Hermitian matrix with complex entries, and hence the relevant RMT ensemble is the GUE \cite{mehta_random_matrices}. This is defined by a probability distribution $P(H)$ in which the entries of $H$ are drawn from a Gaussian distribution
\begin{equation}
    P(H) dH = \exp\big[-a\Tr(H^2)+b\Tr(H)+c\big]dH \ , \qquad a >0, \{b,c\}\in\mathbb R^{2} \ ,
\end{equation}
where $dH \equiv \prod_{j\le k}dH_{jk}^{\rm (r)}\prod_{j< k}dH_{jk}^{\rm (i)}$ and $H_{jk}=H_{jk}^{\rm (r)}+ i H_{jk}^{\rm (i)}$. Importantly, such $P(H)$ is invariant under arbitrary unitary rotations $H' = VHV^{-1}$, $V \in \mathrm U(d)$, that is $P(H)=P(H')$ and $dH = dH'$. A simple change of basis $H = U D U^\dag$, with $D={\rm diag}(E_1, \ldots, E_d)$ and a chosen ordering, e.g.~$E_m \le E_{m+1}$, leads to
\begin{equation}
\label{eq:PGUE}
    P(H)dH = P_E(E_1,\ldots,E_d)dE_1 \ldots dE_d \times d\tilde\mu(U) \ ,
\end{equation}
where $E_m$ denote eigenenergies and $d\tilde\mu(U)$ denotes the normalized invariant measure on the quotient space $\mathrm U(d)/\mathrm U(1)^d$. The presence of $\mathrm U(1)^d$ is due to the additional freedom of multiplying columns of $U$ by arbitrary phases. For averages of physical quantities, which are phase-independent by construction, we can replace $d\tilde \mu(U)\to d\mu(U)$, where $d\mu(U)$ is the normalized Haar measure on the full unitary group $\mathrm U(d)$ (with this substitution all unphysical quantities average trivially to zero). $P_E$ in \cref{eq:PGUE} is the term responsible for level repulsion and the emerge of a Wigner-Dyson distribution \cite{mehta_random_matrices}, and importantly, it does not depend on eigenstates but only on eigenenergies. The eigenstate $\ket{m}$ of $H$ appears solely as the $m$-th column of $U$, i.e.~$U\ket{e_m}=\ket{m}$, with $\{\ket{e_m}\}$ denoting the computational basis. As a consequence, GUE averages of matrix elements of fixed operators in the basis of eigenstates amount to simple Haar averages
\begin{equation}
\label{eq:HaarAvME}
    \mathbb E_{{\rm Haar}}\left[\braket{m_1, \ldots , m_k|O^{(k)}|n_1, \ldots, n_k}\right] = \int d\mu (U) \,\left[ (U^\dag)^{\otimes k}O^{(k)}U^{\otimes k}\right]_{m_1\ldots m_k,n_1\ldots n_k} \ ,
\end{equation}
where $O^{(k)}$ is a generic fixed operator in the $k$-replica space and the indices $m_j$, $n_j$ on the right-hand side refer the computational-basis states $\ket{e_{m_j}}$, $\ket{e_{n_j}}$. 

For chaotic Floquet systems without time-reversal symmetry, one models the Floquet operator $U_F$ by the CUE \cite{dyson_statistical_1962}, meaning that $U_F$ is drawn from the normalized Haar measure $d\mu(U_F)$ on $\mathrm{U}(d)$. We write 
\begin{equation} 
U_F=UD_FU^\dag \ , \qquad D_F={\rm diag}(e^{-i\varphi_1},\ldots,e^{-i\varphi_d}) \ , 
\end{equation} 
where we fix, e.g., $0\le \varphi_1 \le \ldots \le \varphi_d < 2\pi$ and the columns of $U$ are the eigenvectors of $U_F$. This implies
\begin{equation}
    d\mu(U_F) = P_\varphi(\varphi_1, \ldots, \varphi_d)d \varphi_1 \ldots d \varphi_d \times d\tilde \mu(U) \ .
\end{equation}
With the same qualifications as above for the arbitrary phases of the eigenvectors,  we can replace $d\tilde \mu(U) \to d\mu(U)$. Consequently, CUE averages of matrix elements of fixed operators in the eigenbasis of $U_F$ are again given by \cref{eq:HaarAvME}. The eigenvector averages are therefore treated identically for GUE Hamiltonians and CUE Floquet operators.

In the following subsections we report only the main equations used in the RMT calculations and the main results. For all the details and intermediate steps, see Appendix \ref{appendix:RMT}.

\subsection{Matrix elements of (anti)symmetrized eigenstates}
\label{sec:matrixelementsRMTSA}

From \cref{eq:HaarAvME} we see that computing the Haar averages of matrix elements amounts to computing the unitary twirl
\begin{equation}
\label{eq:twirl}
    \mathcal T_k(X)=\int d\mu(U) (U^\dag)^{\otimes k} X U^{\otimes k} \ , \qquad X \in {\rm End}((\mathbb C^d)^{\otimes k}) \ .
\end{equation}
Schur-Weyl duality \cite{collins_integration_2006} implies
\begin{equation}
\label{eq:twirlExp}
    \mathcal T_k(X) = \sum_{\sigma \in S_k}c_\sigma(X) P_\sigma \ ,
\end{equation}
where $S_k$ denotes the symmetric group of degree $k$, $c_\sigma(X)$ are coefficients and $P_\sigma$ denotes the global permutation
\begin{equation}
\label{eq:howpermacts}
    P_\sigma(\ket{v_1} \otimes \ldots \otimes \ket{v_{k}}) = \ket{v_{\sigma^{-1}(1)}} \otimes \ldots \otimes \ket{v_{\sigma^{-1}(k)}} \ .
\end{equation}
This definition guarantees $P_{\sigma'}P_\sigma = P_{\sigma'\sigma}$, with $(\sigma' \sigma)(j)=\sigma'(\sigma(j))$, and $P_\sigma^{\dag} = P_{\sigma^{-1}}=P_{\sigma}^{-1}$. A crucial point is that $\mathcal T_k$ is the orthogonal projection onto ${\rm span}\{P_\sigma \,  : \, \sigma \in S_k\}$ with respect to the Hilbert-Schmidt product, and therefore $\Tr[P_\sigma^\dag \mathcal T_k(X)]=\Tr[P_\sigma^\dag X]$. Combining this with \cref{eq:twirlExp} leads to
\begin{equation}
\label{eq:GramMatrix}
    \sum_{\sigma \in S_k}G_{\pi \sigma}c_\sigma(X) = \Tr[P_\pi^\dag X] \ , \qquad G_{\pi \sigma} = \Tr[P_\pi^\dag P_\sigma] = d^{{\rm cycles}(\pi^{-1}\sigma)} \ ,
\end{equation}
where ${\rm cycles}(\tau)$ counts the total number of cycles in the permutation $\tau$.
This $k! \times k!$ system enables us to determine $c_\pi(X)$ in \eqref{eq:twirlExp}. The inversion of the matrix $G$ is what determines the unitary Weingarten coefficients \cite{collins_integration_2006}. For details on all the previous points, see Appendix \ref{sec:appRMTtools}. 

As briefly reviewed in Appendix \ref{sec:appRMTsingle}, \cref{eq:GramMatrix} is enough to immediately obtain the Haar average of matrix elements of fixed 1-replica operators $O$, relevant for the standard ETH. One gets
\begin{align}
\label{eq:diagHaarO}
    &\mathbb E_{\rm Haar}[\braket{m|O|n}] = \delta_{nm} \, \overline O \ , \qquad \overline O=\frac{\Tr[O]}{d}  \ , \\
    \label{eq:HaarETHoff}
    &\mathbb E_{\rm Haar}[|\braket{m|O|n}|^2]-\mathbb E_{\rm Haar}[\braket{m|O|n}]^2  = \frac{\overline{O^2} - \overline O^2}{d} + \mathcal{O}(d^{-2}) \ , \qquad \overline{O^2}=\frac{\Tr[O^2]}{d} \ ,
\end{align}
where we are assuming $\overline O$ and $\overline{O^2}$ are of order one. This yields exactly the skeleton of the standard ETH ansatz from \cref{eq:FloquetStETH} (and its analogue for static Hamiltonians). One can also easily obtain (see Appendix \ref{sec:appRMTsingle}) the Haar average of the spectral term entering the Lehmann representation \eqref{eq:dyn1replicaObs} for $\braket{\psi(t)|O|\psi(t)}$
\begin{equation}
\label{eq:HaarAVspectralquench}
    \mathbb E_{\rm Haar}[\braket{m|O|n}\braket{n|\rho(0)|m}] = \delta_{mn}\left(\frac{1}{d}\overline O + \mathcal O(d^{-3})\right)+\frac{1}{d^2}\Big(\Tr\big[\rho(0) O\big] - \overline O\Big) + \mathcal O (d^{-4}) \ ,
\end{equation}
where we assume $\Tr[\rho(0)O]$ to be of order one. This provides the RMT skeleton of \cref{eq:outETHsinglerep}. In fact, the $\mathcal{O}(d^{-2})$ scale characterizing the off-diagonal term in \eqref{eq:HaarAVspectralquench} reveals the existence of strong phase cancellations: each $\braket{n|\rho(0)|m}$ is typically expected to be $\propto d^{-1}$ (see Appendix \ref{sec:appRMTsingle}), and from \cref{eq:HaarETHoff} we know that $\braket{m|O|n}$ is typically $\propto d^{-1/2}$, which implies an overall scaling for each $\braket{m|O|n}\braket{n|\rho(0)|m}$ of $d^{-3/2}$. The extra $d^{-1/2}$ suppression in \eqref{eq:HaarAVspectralquench} is due to the strongly cancelling phases associated with each of these terms in the Haar average. This Haar perspective supports the claims about the ubiquity of strong cancellations effects in quench dynamics \cite{foini_outofequilibrium_2025, senese2026nonequilibrium},  and it represents the 1-replica Haar analogue of the cancellations discussed in \cref{eq:crossCorrRRtilde}.

\subsubsection{Average diagonal elements of subsystem swap}
To obtain the skeleton of the diagonal ansatz in \cref{eq:ETHansatz1} for the subsystem swap operator $S_A$ we can start from \cref{eq:twirl}, with $k = 2$ and $X = S_A$. For sake of generality we can consider both the symmetrized eigenstates from \cref{eq:symmetricEigen} ($\ket{\bm m_+}\equiv \ket{\bm m}$) and the antisymmetric ones from \cref{eq:antisymmEigen} ($\ket{\bm m_-}\equiv\ket{\widetilde{\bm m}}$)
\begin{equation}
\label{eq:defmsigma}
   \ket{\bm m_\sigma}=\frac{\ket{m,m'}+\sigma \ket{m',m}}{\sqrt{2(1+\delta_{m,m'})}} \ , \qquad   m\le m'\, , \ \ \sigma = \pm 1 \ .
\end{equation}
This is compatible with \cref{eq:antisymmEigen} because $\ket{\bm m_-}=0$ if $m=m'$. 
With this notation $S_{\mathcal H}\ket{\bm m_\sigma} =\sigma \ket{\bm m_\sigma}$, where $S_{\mathcal H}=P_{(12)}$ is the global swap operator (here $\pi = (12)$ denotes cycle notation). Importantly, $\braket{\bm m_\sigma|S_A|\bm n_{\sigma'}} = 0$  $\forall \ \bm m$, $\bm n$ if $\sigma \neq \sigma'$ because $[S_A,P_{(12)}]=0$. Therefore we need to consider only matrix elements within the same $\sigma$ sector. As explained in Appendix \ref{sec:from24to3}, focusing on (anti)symmetrized eigenstates leads to further nice simplifications in the derivation of the statistics of matrix elements.  It is convenient to define the projector onto the $\sigma$ sector
\begin{equation}
\label{eq:Psfnu}
    \mathsf P_\sigma = \frac{1}{2}(\Id^{\otimes 2} + \sigma P_{(12)}) \ ,
\end{equation}
such that
\begin{equation}
\label{eq:agldkgndr}
    \mathsf P_\sigma \ket{m,m'} = \alpha_{\bm m} \ket{\bm m_\sigma}  \ , \qquad  \alpha_{\bm m} = \sqrt{\frac{1+ \delta_{m,m'}}{2}} \ .
\end{equation}
Then the Haar average relevant for matrix elements is simply
\begin{equation}
\label{eq:twirlfordiagSA}
    \alpha_{\bm m} \alpha_{\bm n} \, \mathbb E_{\rm Haar}[\braket{\bm m_\sigma|S_A|\bm n_\sigma}] = \big(\mathsf P_\sigma \, \mathcal T_2(S_A) \, \mathsf P_\sigma\big)_{mm',nn'} \ .
\end{equation}
Calling again $\ket{e_m}$ the $m$-th element in the 1-replica computational basis, the previous equality follows immediately from the definitions \eqref{eq:HaarAvME} and \eqref{eq:twirl}, together wit $[U^{\otimes 2},\mathsf P_\sigma]=0$ and  $U^{\otimes 2}\ket{e_m, e_{m'}}=\ket{m,m'}$, from which
\begin{equation}
\label{eq:U2Paction}
   ( U^{\otimes 2}\mathsf P_\sigma)\ket{e_m, e_{m'}} = \mathsf P_\sigma\ket{m,m'} = \alpha_{\bm m}\ket{\bm m_\sigma} \ .
\end{equation}
Starting from \cref{eq:twirlfordiagSA} and defining $d_\sigma = d(d+\sigma)/2=\dim{\mathcal H_\sigma^{\otimes 2}}$, in Appendix \ref{sec:appRMTmean} we prove that (assuming $m\neq m'$ if $\sigma = -1$)
\begin{equation}
\label{eq:diagHaarSA}
    \mathbb E_{\rm Haar}[\braket{\bm m_\sigma|S_A|\bm m_\sigma}] = \frac{\Tr_{\mathcal H_\sigma^{\otimes 2}}[S_A]}{d_\sigma} = \frac{\sigma d_A+d_B}{d+\sigma} = \left(\frac{1}{d_A}+\sigma\frac{1}{d_B}\right)\left(1 + \mathcal O(d^{-1})\right) \ ,
\end{equation}
while the average vanishes for $\ket{\bm m_\sigma}\neq\ket{\bm n_\sigma}$, as expected (\emph{cf}.~\cref{sec:GUEandCUE}). We note that the first equality in \cref{eq:diagHaarSA} exactly generalizes  \cref{eq:diagHaarO} to the 2-replica case (within each $\sigma$ sector). The average Haar result \eqref{eq:diagHaarSA} (and its Haar fluctuations described in the following section) fixes entirely the form of the diagonal ETH ansatz in \cref{eq:ETHansatz1} (and its antisymmetric counterpart) because, as we discuss in \cref{sec:fullansatzsection}, in the diagonal case there is no need at leading order of a dressing with a continuous slowly varying function.

\subsubsection{Fluctuations of matrix elements of subsystem swap}
\label{sec:fluctuationsSwap}

The Haar fluctuations of the matrix elements, encoded in the Haar average of $|\braket{\bm m_\sigma |S_A|\bm n_\sigma}|^2$ can be accessed from the (anti)symmetrized $k=4$ twirl $\mathcal T_4(X)$ with $X = S_A^{\otimes 2}$
\begin{equation}
\label{eq:offdiagHaarSA}
   \alpha_{\bm m}^2 \alpha_{\bm n}^2\,\mathbb E_{\rm Haar}\left[  |\braket{\bm m_\sigma |S_A|\bm n_\sigma}|^2 \right]= \left( \mathsf P_\sigma^{\otimes 2} \mathcal T_4(S_A^{\otimes 2})\mathsf P_\sigma^{\otimes 2} \right)_{mm'nn',nn'mm'} \ ,
\end{equation}
where $\mathsf P_\sigma^{\otimes 2} = (\Id^{\otimes 4} + \sigma P_{(12)} + \sigma P_{(34)} + P_{(12)(34)})/4$ and the equality follows again from \cref{eq:U2Paction}. In Appendix \ref{sec:appRMTvariance} we compute the Haar average \eqref{eq:offdiagHaarSA} exactly at all orders in $d$. Combining this with \cref{eq:diagHaarSA}, at leading order in $d^{-1}$  and with no assumptions on $d_A$ and $d_B$ individually, the Haar fluctuations read (again we assume $m \neq m'$ and $n \neq n'$ for $\sigma = -1$)
\begin{equation}
\label{eq:HaarFluctOff}
    \mathbb E_{\rm Haar}\left[  |\braket{\bm m_\sigma |S_A|\bm n_\sigma}|^2 \right] - \mathbb E_{\rm Haar}\left[ \braket{\bm m_\sigma |S_A|\bm n_\sigma} \right]^2 = \frac{2}{d^2}\left(1-\frac{1}{d_A^2} - \frac{1}{d_B^2}\right) + \mathcal O(d^{-3}) \ .
\end{equation}
This results proves that for diagonal elements the fluctuations are subleading with respect to the mean in \cref{eq:diagHaarSA} (with the only exception, for $\sigma = -1$, of cases like $d_A = d_B$ for which the mean vanishes or $d_A \approx d_B$ where it is suppressed), while it provides the skeleton for the off-diagonal ETH ansatz in \cref{eq:ETHansatz2}. Indeed, the ansatz \eqref{eq:ETHansatz2} states that a local-in-the-spectrum average over neighbouring 2-replica eigenstates has leading suppression of of $2/d^2$, with the order-one factor $(1-d_A^{-2}-d_B^{-2})$ absorbed in the slowly varying order-one function $f_A^2(\Omega, \omega_1, \omega_2)$. This is perfectly consistent with the sum rule \eqref{eq:sumrulfA}. We note that \cref{eq:HaarFluctOff} is at leading order independent of the choice of the sector $\sigma = \pm 1$. 

\subsection{Haar average of quench spectral weight}

For the quench dynamics, the relevant out-of-equilibrium-ETH ansatz concerns the matrix elements
\begin{equation}
    \braket{\bm m_+|S_A|\bm n_+}\braket{\bm n_+ |\rho(0)^{\otimes 2}|\bm m_+} \ .
\end{equation}
The restriction to the symmetric sector $\sigma = +1$ derives from
\begin{equation}
    \rho(0)^{\otimes 2} = \ket{\psi(0)}^{\otimes2}\bra{\psi(0)}^{\otimes 2} \ , \qquad \braket{\bm m_-|\psi(0)}^{\otimes 2} = 0 \ .
 \end{equation}
The relevant Haar averages can be obtained very similarly to those of \cref{sec:fluctuationsSwap}, using the $k=4$ twirl $\mathcal T_4(X)$ with $X = S_A \otimes \rho(0)^{\otimes 2}$. In particular
\begin{align}
    \alpha_{\bm m}^2 \alpha_{\bm n}^2\,\mathbb E_{\rm Haar}\left[  \braket{\bm m_+|S_A|\bm n_+}\braket{\bm n_+ |\rho(0)^{\otimes 2}|\bm m_+} \right] = \left( \mathsf P_+^{\otimes 2} \mathcal T_4(S_A \otimes \rho(0)^{\otimes 2})\mathsf P_+^{\otimes 2} \right)_{mm'nn',nn'mm'} \ .
\end{align}
By means of the same techniques used for the results in \cref{sec:fluctuationsSwap}, in Appendix \ref{sec:appRMTquench} we obtain
\begin{equation}
\label{eq:nalgp4ggd11}
    \mathbb E_{\rm Haar}\left[  \braket{\bm m_+|S_A|\bm m_+}\braket{\bm m_+ |\rho(0)^{\otimes 2}|\bm m_+} \right] = \frac{2}{d^2}\left(\frac{1}{d_A} + \frac{1}{d_B}\right)\left( 1 + \mathcal O(d^{-1}) \right) \ ,
\end{equation}
\begin{align}
\label{eq:offdiagETHoutHaar}
    \mathbb E_{\rm Haar}\left[  \braket{\bm m_+|S_A|\bm n_+}\braket{\bm n_+ |\rho(0)^{\otimes 2}|\bm m_+} \right]  = \frac{4}{d^4}\left( \mathbb P_A(0) - \frac{1}{d_A} - \frac{1}{d_B}  \right) + \mathcal O(d^{-5}) \ , \qquad \ket{\bm m_+} \neq \ket{\bm n_+} \ . 
\end{align}
For lowly-entangled initial states in the large-subsystems regime, where $\mathbb P_A(0) \gg d_A^{-1} + d_B^{-1}$, the $\mathcal{O}(d^{-5})$ remainder is always subleading. 
\cref{eq:nalgp4ggd11,eq:offdiagETHoutHaar} provide the scaling skeleton of the diagonal and off-diagonal ansätze in \cref{eq:outEqETH1,eq:outEqETH2}. In \cref{eq:outEqETH2} the factor $(\mathbb P_A(0)-d_A^{-1}-d_B^{-1})$ from \eqref{eq:offdiagETHoutHaar} is absorbed into the slowly varying function $g_A(\Omega,\omega_1, \omega_2)$, in perfect agreement with the sum rule in \cref{eq:sumrulgA}. In the off-diagonal part, the emergence of a leading prefactor of $d^{-4}$, instead of $d^{-3}$ that one would predict by the typical Haar scalings $|\braket{\bm m_+|S_A|\bm n_+}| \propto d^{-1}$ and $|\braket{\bm n_+ |\rho(0)^{\otimes 2}|\bm m_+}| \propto d^{-2}$, is due to the Haar counterpart of the strongly cancelling phases leading to \cref{eq:crossCorrRRtilde}. In the diagonal part such strong cancellation effects are absent, indeed the leading scaling emerges directly from the typical Haar values $\braket{\bm m_+|S_A|\bm m_+} = (d_A^{-1} + d_B^{-1})$ (see \cref{eq:diagHaarSA}) and $\braket{\bm m_+ |\rho(0)^{\otimes 2}|\bm m_+} \propto d^{-2}$. This matches \cref{eq:jagkjgV1,eq:jagkjgV2}.

\section{Full ansatz by dressing with slowly varying functions}
\label{sec:fullansatzsection}

In this section we argue for the need of a dressing of the RMT results by continuous slowly varying functions. In doing so, we connect these functions to those entering the 2- and 4-ETH ansätze \eqref{eq:nETHFloquet} for single-replica operators \cite{foini_eigenstate_2019}, and derive the useful representations in \cref{eq:CoffFourier,eq:PAFourier}. 

The crucial difference between the RMT calculations and physical ETH ansätze is that the former involve \emph{global ensemble average}s (for fixed eigenstate indices, e.g.~$m, m',n,n'$), while in the latter the scaling with $d, d_A, d_B$ emerges from \emph{local (in the spectrum) averages} of different matrix elements over neighbouring eigenstates, \emph{cf.}~\cref{eq:Rmn,eq:outEqETH1,eq:outEqETH2,eq:crossCorrRRtilde}. The perfect scaling agreement between these two quite different prescriptions is therefore, a priori, a non-trivial statement. It can be heuristically understood from typicality arguments coupled with Wigner's original intuition on the structural similarity between RMT global statistics and the local-spectrum statistics of real-world systems. 
Other useful perspectives are the one proposed by Deutsch \cite{deutsch_quantum_1991, reimann_eigenstate_2015}, based on the expectation that very weak random perturbations (which can be modelled by more structured ensembles of random matrices) can strongly mix eigenstates while keeping unchanged ETH spectral functions and leading suppressions (\emph{cf}.~also, e.g., Refs.~\cite{foini_eigenstate_2019, vallini_refinements_2026}); or the one based on Berry's conjecture \cite{berry_regular_1977} that motivated Srednicki's work \cite{srednicki_chaos_1994}.

\subsection{Connections with 2- and 4-ETH}
\label{sec:conn2&4ETH}

As a prescription, the dressing of the RMT results by continuous slowly varying functions works extremely well in the case of the established ETH ansätze in \cref{eq:FloquetStETH,eq:nETHFloquet,eq:outETHsinglerep} (and their static-Hamiltonian generalizations). Furthermore, the full-ETH terms in \eqref{eq:nETHFloquet} can be trivially interpreted as matrix elements on multireplica spaces (much like for the subsystem swap operator under consideration). Thanks to this we now show, assuming the validity of the full-ETH ansatz \eqref{eq:nETHFloquet} for standard $1$-replica operators, that a dressing must be present also in the case of the subsystem swap statistics. The key identity is \cref{eq:SAexpanded}
\begin{equation}
\label{eq:SAexpanded2}
    S_A = \sum_{\mu=1}^{d_A^2}O_\mu \otimes O_\mu \ , \qquad \Tr_A[O_\mu O_\nu] = \delta_{\mu, \nu} \ , \ \ O_\mu = O_\mu^\dag \ ,
\end{equation}
where $\{O_\mu\}$ is a orthonormal (with respect to Hilbert-Schmidt product restricted to $A$) Hermitian basis of 1-replica operators with support in $A$. The proof is straightforward: considering any such orthonormal basis, we have
\begin{align}
\label{eq:klhrir49gddc}
    S_A = \sum_{\mu,\nu=1}^{d_A^2}\Tr_{A^{\otimes 2}}[(O_\mu \otimes O_\nu)S_A]O_\mu \otimes O_\nu = \sum_{\mu,\nu=1}^{d_A^2}\Tr_A[O_\mu O_\nu]O_\mu \otimes O_\nu = \sum_{\mu=1}^{d_A^2} O_\mu \otimes O_\mu \ ,
\end{align}
where we used \cref{eq:intertwofS} together with the orthonormality in \eqref{eq:SAexpanded2}.  
We can then use \cref{eq:SAexpanded2} in the definition of the subsystem swap matrix elements
\begin{align}
\label{eq:gbfa3kgerp}
    |\braket{\bm m_\sigma|S_A|\bm n_\sigma}|^2 = \braket{\bm m_\sigma |S_A |\bm n_\sigma}\braket{\bm n_\sigma |S_A |\bm m_\sigma} = \sum_{\mu,\nu=1}^{d_A^2} \braket{\bm m_\sigma |O_\mu \otimes O_\mu |\bm n_\sigma}\braket{\bm n_\sigma |O_\nu \otimes O_\nu |\bm m_\sigma} \ .
\end{align}
Each of the terms in the previous expression, using \cref{eq:agldkgndr}, correspond to
\begin{align}
    &\alpha_{\bm m} \alpha_{ \bm n}\braket{\bm m_\sigma |O_\mu \otimes O_\mu |\bm n_\sigma} = \braket{m,m'|\mathsf P_\sigma (O_\mu \otimes O_\mu)\mathsf P_\sigma|n,n'} \\ \nonumber
    &\qquad = \frac{1}{2}(O_\mu)_{mn}(O_\mu)_{m'n'}+\frac{\sigma}{2}(O_\mu)_{mn'}(O_\mu)_{m'n} \ ,
\end{align}
where we have used the definition \eqref{eq:Psfnu} of $\mathsf P_\sigma$, together with $[(O_\mu \otimes O_\mu),\mathsf P_\sigma] = 0$ and $\mathsf P_\sigma^2 = \mathsf P_\sigma$. Combining this with \eqref{eq:gbfa3kgerp}, we get
\begin{align}
\label{eq:SAmatrixelFromFullETH}
   4\, \alpha_{\bm m}^2 \alpha_{\bm n}^2|\braket{\bm m_\sigma|S_A|\bm n_\sigma}|^2 = \sum_{\mu, \nu=1}^{d_A^2} \bigg\{ \Big[(O_\mu)_{13}(O_\nu)_{31} \cdot (O_\mu)_{24} (O_\nu)_{42}  + \ \sigma \, (O_\mu)_{13}(O_\nu)_{32}(O_\mu)_{24}(O_\nu)_{41} \Big] + \big[{3 \leftrightarrow 4} \big]\bigg\} \ ,
\end{align}
where to ease the identification of the index pattern we have substituted $m\to1$, $m' \to 2$, $n \to 3$, $n' \to 4$. This expression makes apparent that in the case of $m,m',n,n'$ all distinct, i.e. class VII from \cref{tab:classesOfEigenstates}, the matrix elements of $S_A$ \emph{averaged} over neighbouring eigenstates are given by a sum of $d_A^4$ terms each composed of two products of 2-ETH contributions, e.g.
\begin{equation}
\label{eq:connectingFullETH1}
    \overline{(O_\mu)_{13}(O_\nu)_{31}} \cdot \overline{(O_\mu)_{24} (O_\nu)_{42}} \sim \frac{1}{d^2}h^{(2)}_{\mu,\nu}(\omega_{13})h^{(2)}_{\mu,\nu}(\omega_{24})  \ ,
\end{equation}
and two 4-ETH contributions, e.g.
\begin{equation}
\label{eq:connectingFullETH2}
   \overline{(O_\mu)_{13}(O_\nu)_{32}(O_\mu)_{24}(O_\nu)_{41}} \sim \frac{1}{d^3}h^{(4)}_{\mu,\nu}(\omega_{13},\omega_{32},\omega_{24}) \ ,
\end{equation}
where as before “$\sim$” denotes asymptotic equivalence and $\omega_{ab}=\varphi_a - \varphi_b$.
To derive these results we have used the full-ETH ansatz \eqref{eq:nETHFloquet} together with the leading-order factorization \eqref{eq:factorizationFullETH} of disconnected loops. Focusing on the limit of finite but possibly large $d_A$ while $d \to \infty$, where the validity of full-ETH is best established (local operators), we see that even after performing the $\mu,\nu$ sum over $d_A^4$ terms in \eqref{eq:SAmatrixelFromFullETH} the 4-ETH contributions must remain subleading in $d^{-1}$. This implies that a continuous slowly varying function, namely $f_A^2$ in our notation of \cref{eq:ETHansatz2}, must arise at least for finite $d_A$ as a consequence of the presence of the 2-ETH functions $h^{(2)}_{\mu,\nu}(\omega_{13})h^{(2)}_{\mu,\nu}(\omega_{24})$ and $h^{(2)}_{\mu,\nu}(\omega_{14})h^{(2)}_{\mu,\nu}(\omega_{23})$ in \eqref{eq:SAmatrixelFromFullETH}. This perspective also captures the correct $\propto d^{-2}$ leading decay predicted by RMT for off-diagonal terms (although the derivation is easy because limited to finite subsystems $A$).

These considerations at the level of matrix elements translate directly to dynamical correlators. To see this in full generality, we extend the definition of $C_A^{(+)}(t)$ also to the $\sigma = -1$ sector, by replacing $d_+ \to d_-$ and $\mathcal H_+^{\otimes 2}\to\mathcal H_-^{\otimes 2}$. This leads to
\begin{equation}
    C_A^{(\sigma)}(t) = \frac{1}{d_\sigma}\sum_{\mu,\nu=1}^{d_A^2}\Tr_{\mathcal H_\sigma^{\otimes 2}}\left[ O_\mu(t)O_\nu  \otimes O_\mu(t)O_\nu  \right] \ ,
\end{equation}
where $O_\mu(t)=U^\dag(t)O_\mu U(t)$ denotes 1-replica Heisenberg evolution. Each of the terms in the $\mu, \nu$ sum can be expanded as
\begin{align}
\label{eq:gldmmvrp5f}
    &\Tr_{\mathcal H_\sigma^{\otimes 2}}\left[ O_\mu(t)O_\nu  \otimes O_\mu(t)O_\nu  \right] = \Tr_{\mathcal H^{\otimes 2}}\big[\mathsf P_\sigma \big(O_\mu(t)O_\nu  \otimes O_\mu(t)O_\nu\big) \mathsf P_\sigma\big] \\ \nonumber
   &\qquad \qquad =\frac{1}{2}\Tr_{\mathcal H}[O_\mu(t)O_\nu]^2 + \frac{\sigma}{2}\Tr_{\mathcal H}[O_\mu(t)O_\nu O_\mu(t)O_\nu] \ ,
\end{align}
where we have used the property in \cref{eq:intertwofS} of the global swap operator $P_{(12)}$ appearing in the definition \eqref{eq:Psfnu} of $\mathsf P_\sigma$, together with $\mathsf P_\sigma^2 = \mathsf P_\sigma$ and cyclicity of the trace. Therefore $C_A^{(\sigma)}(t)$ amounts to a sum over $d_A^4$ terms that include squares of 2-point functions (coinciding with the 2-ETH part in the matrix elements) as well as 4-point functions (OTOCs, coinciding with the 4-ETH part in the matrix elements for the class VII). For $d_A$ finite while $d \to \infty$ and remembering $d_\sigma \sim d^2/2$, we clearly see, using the full-ETH scalings in \cref{eq:connectingFullETH1,eq:connectingFullETH2} for the Lehmann representation of \cref{eq:gldmmvrp5f} restricted to the all-distinct class VII, that the OTOC part can contribute at most at $\mathcal O(d^{-1})$, while the 2-point functions part can contribute at $\mathcal O(1)$. The fact that in \cref{eq:SAmatrixelFromFullETH,eq:gldmmvrp5f} the contribution multiplied by $\sigma = \pm 1$ is subleading yields a different route to understand why the RMT result in \cref{eq:HaarFluctOff} for off-diagonal elements is independent of the sector $\sigma = \pm 1$. 

At this point we need to remember that the RMT results from \cref{sec:matrixelementsRMTSA} predict a leading scaling of $\mathcal O(d^{-2})$ for off-diagonal terms $|\braket{\bm m_\sigma|S_A|\bm n_\sigma}|^2$ also in all the other off-diagonal classes from \cref{tab:classesOfEigenstates}, i.e. classes II, III, V and VI. This raises a natural question: how do we reconcile such scaling with the fact that for these classes also terms like, e.g., $(O_\mu)_{mm}(O_\nu)_{mm}$ appear, which would normally lead to a different scaling compared to \cref{eq:connectingFullETH1} (\emph{cf.}~\cref{eq:factorizationFullETH})? This apparent paradox is resolved in Appendix \ref{app:coincidentindices} by a simple convenient choice of the basis $\{O_\mu\}$. There we also show how the same basis enables the verification of the diagonal ansatz \eqref{eq:ETHansatz1} in the limit of $d_A$ finite, which provides analytic evidence for the absence of dressing in the diagonal elements. Note also that in the opposite limit where $d_B$ is finite and $d_A$ diverges (as a consequence of $d \to \infty$), identical results aside for $d_A \leftrightarrow d_B$ can be obtained using \cref{eq:excSAandSB} (and its generalizations to antisymmetric states $\ket{\bm m_-}$) and replacing \cref{eq:SAexpanded2} by an expansion of $S_B$. 
  
We stress that, although this connection with $2$-ETH and $4$-ETH is helpful to establish the necessity of a dressing of off-diagonal RMT results by continuous functions (and the absence of dressing for the diagonal terms), per se it does not provide an easy route to the determination of $f_A(\Omega, \omega_1,\omega_2)$ starting from single-replica ETH functions. The reason is that \cref{eq:SAmatrixelFromFullETH} still involves a sum over $d_A^4$ terms, which becomes very large in the (most interesting) limit of $|A|\gg 1$.

\subsection{Connections with 1-replica ansätze for quench dynamics}
\label{sec:conn1repforQuench}

The identity \eqref{eq:SAexpanded2} can be employed also to give a different perspective on the dynamics of the purity when there is no average over initial states (see analogous remarks in Ref.~\cite{ho_entanglement_2017}). From \cref{eq:mainidentity} we get
\begin{equation}
    \mathbb P_A(t)=\braket{\psi(t)|^{\otimes 2}S_A|\psi(t)}^{\otimes 2} = \sum_{\mu=1}^{d_A^2} \braket{\psi(t)|O_\mu|\psi(t)} \braket{\psi(t)|O_\mu|\psi(t)}  \ .
\end{equation}
Using the single-replica Lehmann expansion from \cref{eq:dyn1replicaObs} on the right-hand side of the previous equation one obtains
\begin{align}
\label{eq:bfdvlda59bw}
    \mathbb P_A(t) = \sum_{m,m',n,n'}e^{i(\varphi_m + \varphi_{m'} - \varphi_n-\varphi_{n'})t}\bigg[ &\sum_{\mu=1}^{d_A^2}\braket{m|O_\mu|n}\braket{n|\rho(0)|m}   \braket{m'|O_\mu|n'}\braket{n'|\rho(0)|m'} \bigg] \ .
\end{align}
Note that in this sum we are not requiring $m \le m'$ and $n \le n'$ because, unlike in \cref{eq:mainLehmanSum}, we are \emph{not} resolving eigenstates according to their symmetric component. If we focus on the all-distinct off-diagonal class VII $\to \{(a,b),(c,d)\}$, the statistics of matrix elements relevant in \eqref{eq:bfdvlda59bw} is (we set $\Psi_{nm}\equiv \braket{n|\rho(0)|m}$)
\begin{align}
\label{eq:dgkal4gglv}
  & \overline{(O_\mu)_{mn} \Psi_{nm} (O_\mu)_{m'n'} \Psi_{n'm'}} \\ \nonumber
   &\quad \sim \overline{(O_\mu)_{mn} \Psi_{nm}} \cdot \overline{(O_\mu)_{m'n'} \Psi_{n'm'}} \sim \left(\frac{1}{d^2}g_{O_\mu,\psi(0)}(\omega_{mn})  \right) \left(\frac{1}{d^2} g_{O_\mu,\psi(0)}(\omega_{m'n'})  \right) \ ,
\end{align}
where in the last asymptotic equality we have used \cref{eq:outETHsinglerep}. We note that the first asymptotic equality in \cref{eq:dgkal4gglv} is based on a leading factorization rule that, to our knowledge, has not yet been analyzed in the literature but that it is strongly expected. Summing \cref{eq:dgkal4gglv} over $\mu$, again under the assumption of $d_A$ finite, one recovers the form
\begin{equation}
    \overline{\braket{\bm m_+|S_A|\bm n_+}\braket{\bm n_+ |\rho(0)^{\otimes 2}|\bm m_+}}  \sim \frac{4}{d^4} g_A
\end{equation}
of our ansatz \eqref{eq:outEqETH2} and connect $g_A$ directly to the functions $g_{O_\mu,\psi(0)}(\omega)$ appearing in \cref{eq:dgkal4gglv}. These steps provide a different way of deriving the correct suppression $\propto d^{-4}$ and of arguing for the necessity of dressing the RMT result \eqref{eq:offdiagETHoutHaar} by slowly varying functions (in the limit $d_A$ finite, $d \to \infty$). We expect that similarly to what done in \cref{sec:conn2&4ETH} one can use the convenient choice for $\{O_\mu\}$ of Appendix \ref{app:coincidentindices}, together with factorization rules on repeated indices, to re-derive from such 1-replica perspective the ansätze \eqref{eq:outEqETH1} and \eqref{eq:outEqETH2} for \emph{all} the classes from \cref{tab:classesOfEigenstates} (in the limit $d_A$ finite, $d\to\infty$). A simple example is provided in Appendix \ref{app:coincidentindices}.

\subsection{Dressing of RMT results and their consequences}

\subsubsection{Swap matrix elements and $C_A^{(+)}(t)$}
\label{sec:swapandC+}

Using full- \cite{foini_eigenstate_2019} and out-of-equilibrium-ETH \cite{foini_outofequilibrium_2025} for single-replica operators we just argued in the limit of $d_A$ finite and $d \to \infty$ for the necessity of dressing by slowly varying functions the off-diagonal RMT results, and for a lack of dressing at leading order in the diagonal ones. It seems natural, in the present setting, to expect that the full ansatz in \cref{eq:ETHansatz1,eq:ETHansatz2} holds in the full domain $d_{\rm loc}\le d_A\le d/d_{\rm loc}$, where $d_{\rm loc}$ is the local Hilbert space dimension on a single site (qudits). Indeed, remembering the invariance under $S_A \leftrightarrow S_B$ in \cref{eq:excSAandSB}, we know that if the ansatz is correct at a given finite $d_A=q$, it must also be identical and correct for $d_B=q$, i.e.~at $d_A = d/d_B = d/q$. Furthermore, all ETH ansätze are expected to remain correct whenever the subsystem $A$ of interest and its complement satisfy $|A|/|B| \to 0$, irrespective of whether $|A|$ is finite or diverging. A leap of faith regards the intermediate regime in which both subsystems diverge in size and $|A|/|B| \to x > 0$. This is the regime in which one departs from the “few-body operator” requirement, when comparing with the complement $B$, that usually accompanies ETH statements \cite{essler_quench_2016} (in this case, the operator is the subsystem swap $S_A$ or the single-replica operators it can be decomposed into via \cref{eq:SAexpanded2}). However, we have remarked already that the diagonal ansatz in \cref{eq:ETHansatz1} ensures, for \emph{arbitrary} $d_A$ and $d_B$, that the average purity $\mathbb E_{\ket \psi}[\mathbb P_A(t)]$ relaxes at long times to the expected Haar value \eqref{eq:PurProdRelax} (it also ensures, as expected, that individual Floquet eigenstates have purity matching the Haar value for arbitrary $d_A$ and $d_B$, \emph{cf.}~Refs.~\cite{deutsch_thermodynamic_2010, deutsch_microscopic_2013, lu_renyi_2019, joshi_probing_2022}). We also find that the ED results from \cref{sec:numver} are perfectly consistent with the correctness of our diagonal and off-diagonal ansätze for arbitrary ratios $d_A/d$, although the small system sizes achievable with ED fail to probe a proper “many-body operator” limit. We find that the same is true also for the out-of-equilibrium-ETH ansätze in \cref{eq:outEqETH1,eq:outEqETH2}. In the literature, a few works already support the conjecture that some ETH statements might survive for some operators even when the relevant subsystem lives in such intermediate-size regimes, see e.g.~related discussions and numerical evidence in Refs.~\cite{garrison_does_2018, dymarsky_subsystem_2018, khaymovich_eigenstate_2019, bandyopadhyay_latetime_2023}.\\

The full-ETH arguments from \cref{sec:conn2&4ETH} support the fact that the continuous function associated with off-diagonal terms $|\braket{\bm m_\sigma|S_A|\bm n_\sigma}|^2$, for example in the most relevant all-distinct class VII $\to \{(a,b),(c,d)\}$, should have the functional form $h_1(\omega_{mn},\omega_{m'n'})+h_2(\omega_{mn'},\omega_{m'n})$, where these $h$ functions arise from products of $2$-ETH terms. However, those arguments are restricted to the regime $d_A$ finite, $d \to \infty$ (and do not account for possible simplifications induced by the large sum over $d_A^4$ components, \emph{cf.}~\cref{sec:simplification}). For arbitrary ratios $d_A/d$, we might need to account for a more generic functional dependence, which we encode in the spectral function $f_A$ from \cref{eq:ETHansatz2}. Given the presence of only 4 phases, $\varphi_m$, $\varphi_{m'}$, $\varphi_n$, $\varphi_{n'}$, 3 independent frequencies $\omega$ are sufficient to reconstruct all possible $\omega_{ab} = \varphi_a - \varphi_b$. Given the appearance of 
\begin{equation}
\label{eq:defOmegafre}
    \Omega = \varphi_m + \varphi_{m'} - \varphi_n - \varphi_{n'} \ 
\end{equation}
in the dephasing term of all the Lehmann representations, see \cref{eq:mainLehmanSum,eq:Cplus}, we retain it as the first of our 3 fundamental frequencies. To choose the 2 complementary frequencies $\omega_1$ and $\omega_2$ we require that:
\begin{enumerate}
    \item Physically distinct sets $\{\varphi_m,\varphi_{m'},\varphi_n,\varphi_{n'}\}$ (i.e.~up to $2\pi$ periodicity of each $\varphi_j$ or a common real shift of all $\varphi_j$) map to physically distinct frequency sets $\{\Omega, \omega_1, \omega_2\}$.
    \item Under $(m,m') \leftrightarrow(n,n')$ we have $\{\Omega, \omega_1, \omega_2\} \to \{\widetilde\Omega, \widetilde\omega_1, \widetilde\omega_2\}$, where the new frequencies are related in a \emph{simple way} to the original ones. This is motivated by $|\braket{\bm m_\sigma|S_A|\bm n_\sigma}|^2=|\braket{\bm n_\sigma|S_A|\bm m_\sigma}|^2$, which implies $f_A(\Omega,\omega_1,\omega_2)=f_A(\widetilde\Omega, \widetilde \omega_1, \widetilde \omega_2)$ (remember that $f_A$ is real and positive by construction).
    \item $\omega_1$ and $\omega_2$ do not include the frequencies $\omega_{mm'}$ and $\omega_{nn'}$, which are expected to be relevant only at subleading orders.
\end{enumerate}
A solution to the previous points is to choose
\begin{equation}
\label{eq:defomegasfre}
    \omega_1 = \varphi_m - \varphi_n \ , \qquad \omega_2 = \varphi_m - \varphi_n' \ .
\end{equation}
To show that it is injective (point 1), we start from a generic set $\widetilde {\bm \varphi} =\{\tilde \varphi_m,\tilde \varphi_{m'},\tilde \varphi_n,\tilde\varphi_{n'}\}$ and use the freedom of a common shift to set $\varphi_{n'}=0$ (where $\tilde \varphi_a \to \varphi_a$ after the shift). This fixes $\omega_2 = \varphi_m$, which in turn fixes $\omega_1 = \omega_2 - \varphi_n$, that finally fixes $\Omega = \omega_1 + \varphi_{m'}$. Therefore, two physically distinct sets $\bm \varphi_1$ and $\bm \varphi_2$ map to distinct sets of frequencies $\Omega, \omega_1, \omega_2$. Point 2 above is satisfied because under $(m,m') \leftrightarrow(n,n')$ we have
\begin{equation}
\label{eq:ggldgè45}
    \Omega \to - \Omega \, , \quad \omega_1 \to - \omega_1 \, , \quad \omega_2 \to \omega_2 - \Omega \ ,
\end{equation}
and applying this transformation twice gives the identity. This yields the symmetry $f_A(\Omega,\omega_1, \omega_2) = f_A(-\Omega, -\omega_1, \omega_2 - \Omega)$. 

These variables also allow us to transform, in the limit of $d \to \infty$ and time $t$ arbitrary but finite, sums over eigenstates into convenient integrals. Let's restrict our attention to the all-distinct off-diagonal class VII, which in the thermodynamic limit is the only relevant one for the dynamical part $C_A^{(\rm off)}(t)$ of $C_A^{(+)}(t)$. Setting $M_{mm',nn'} \equiv |\braket{\bm m_+|S_A| \bm n_+}|^2$, we get from \cref{eq:Cplus}
\begin{equation}
\label{eq:ldnglk40g}
    C_A^{(\rm off)}(t) \sim \frac{1}{d_+}\sum_{\substack{m < m', \, n < n'\\(\rm pairwise \ distinct)}}e^{i\Omega t} M_{mm', nn'} = \frac{1}{4\,d_+}\sum_{\substack{m, \, m',\, n,\, n'\\(\rm pairwise \ distinct)}}e^{i\Omega t} M_{mm', nn'} \ .
\end{equation}
In the limit of $d \to \infty$, the Floquet eigenphases are to leading order uniformly distributed in $[0,2\pi)$. Therefore
\begin{equation}
   \Big \lfloor \frac{1}{2\pi}\Delta \varphi_j \Big \rfloor \sim \frac{\# \text{ of phases in } [\varphi_j, \varphi_j+\Delta\varphi_j]}{d}  \ ,
\end{equation}
where $j = m, m',n,n'$. Consider now the index $j$ restricted to a window $\Delta \varphi_j$, which can be chosen vanishingly small in the thermodynamic limit. Denoting the sum where $j\in \Delta \varphi_j \ \forall \ j=m,m',n,n'$ and pairwise coincidences are excluded as
\begin{equation}
\label{eq:agabkvv}
    \sum_{\Delta \varphi} \ \ \equiv \sum_{\substack{j \in \Delta \varphi_j, \, j = m,m',n,n'\\(\rm pairwise \ distinct)}} \ ,
\end{equation}
we can write
\begin{equation}
\label{eq:ngadgk4pog}
     \sum_{\Delta \varphi} e^{i\Omega t} M_{mm', nn'} \sim e^{i\Omega^{(\Delta \varphi)} t}\,d^4 \, \braket{M_{\Delta \varphi}}\prod_{j = m,m',n,n'}\frac{\Delta \varphi_j}{2\pi}  \ ,
\end{equation}
where
\begin{equation}
\label{eq:hhadgk33potr}
    \braket{M_{\Delta \varphi}} = \frac{\sum_{\Delta \varphi} M_{mm', nn'}}{\sum_{\Delta \varphi}} \sim \frac{2}{d^2}f_A^2\big(\Omega^{(\Delta \varphi)}, \omega_1^{(\Delta \varphi)}, \omega_2^{(\Delta \varphi)}\big) \ .
\end{equation}
In the last equation we have used the off-diagonal ETH ansatz in \cref{eq:ETHansatz2} together with \cref{eq:Rmn}. Both in \cref{eq:ngadgk4pog} and \cref{eq:hhadgk33potr} we have used the fact that within vanishingly small windows $\Delta\varphi_i$ the frequencies $\Omega, \omega_1, \omega_2$ remain constant at leading order and we denote as $\Omega^{(\Delta \varphi)}, \omega_1^{(\Delta \varphi)}, \omega_2^{(\Delta \varphi)}$ their average value in the windows. Remembering that $d_+ \sim d^2/2$, from \cref{eq:ldnglk40g} and the previous asymptotic equalities we obtain
\begin{equation}
    C_A^{(\rm off)}(t) \sim\prod_{j=m,m',n,n'}\int_0^{2 \pi}\frac{d\varphi_j}{2\pi}\left[ e^{i \Omega t} f_A^2(\Omega, \omega_1, \omega_2) \right] \ ,
\end{equation}
where we set $\Omega^{(\Delta \phi)} \to \Omega = \Omega(\varphi_m,\varphi_m',\varphi_n, \varphi_n')$ and similarly for $\omega_1$, $\omega_2$. We note that we can remove the $\varphi_{n'}$ integral by employing the freedom of a common real shift of all the $\varphi_j$ eigenphases. The remaining change of variables $\varphi_m, \varphi_{m'}, \varphi_n \to \Omega, \omega_1, \omega_2$ has Jacobian determinant equal to 1. Therefore, we arrive to
\begin{equation}
\label{eq:dpgdagòrlè}
C_A^{(\rm off)}(t) \sim \int_0^{2\pi}\frac{d\Omega}{2\pi} e^{i\Omega t}  \underbrace{\left[\int_0^{2\pi} \frac{d\omega_1\omega_2}{(2\pi)^2}  f_A^2(\Omega,\omega_1,\omega_2)\right]}_{F_A(\Omega)} \ .
\end{equation}
This proves \cref{eq:CoffFourier}. From \cref{eq:ggldgè45}, the associated symmetry of $f_A$, and the $2\pi$ periodicity of all the arguments we obtain
\begin{equation}
    F_A(\Omega) = F_A(-\Omega) \ ,
\end{equation}
which guarantees that $C_A^{(\rm off)}(t) \in \mathbb R$ and $C_A^{(\rm off)}(t)=C_A^{(\rm off)}(-t)$ (from the definition \eqref{eq:Cplus} of $C_A^{(+)}(t)$ in terms of a symmetric-sector trace we know that, despite the lack of time-reversal invariance, $C_A^{(+)}(t) = C_A^{(+)}(-t)$). \\

To provide a few concrete examples on the expected structure of the ETH spectral functions, let us restrict our attention to $D=1$ systems with a finite number of entanglement cuts, such that the boundary $\partial$ between $A$ and $B$ stays finite as $V_{\rm min}\to \infty$. This implies that, at any fixed time $t$, $\mathbb P_A(t)$ is finite too (see \cref{eq:purityboundarea}) and allows us to focus in the simplest possible way on the limit $V_{\rm min }\to \infty$ where universal features of entanglement growth emerge. In this limit \cref{eq:Eprod} yields for $D=1$ systems
\begin{equation}
\label{eq:agrbèdpd}
    \lim_{V_{\rm min} \to \infty} \mathbb E_{\ket{\psi}}[\mathbb P_A(t)] = \lim_{V_{\rm min} \to \infty} C_A^{(+)}(t) = \lim_{V_{\rm min} \to \infty}  C_A^{(\rm off)}(t) \ .
\end{equation}
Given that $\mathbb P_A(t) \ge 0 \ \forall \ t$ then $\lim_{V_{\rm min} \to \infty}C_A^{(\rm off)}(t) \ge 0$. From these properties we deduce a few facts on the marginal $F(\Omega) \equiv \lim_{V_{\rm min} \to \infty}F_A(\Omega)$.
\begin{itemize}
    \item We note that the purity $\mathbb P_A(t)$ is expected to decay with $t$ faster than any power law for a measure 1 of all the initial states in the ensemble \eqref{eq:prodEnsemble}. Indeed, a power law decay would imply a logarithmic growth of Rényi entanglement, which is highly non-generic aside for many-body localized phases (see Ref.~\cite{pain_entanglement_2024} for a spectral treatment similar to the one we employ here) or Hilbert space fragmention scenarios (see, e.g., the entanglement-growth discussion in Ref.~\cite{classen-howes_universal_2025}). Given \eqref{eq:agrbèdpd} a decay faster than any power law applies also to $C_A^{(\rm off)}(t)$. Using
    \begin{equation}
    \label{eq:vprvmpv3}
       F(\Omega) = \sum_{t \in \mathbb Z} e^{-i\Omega t}\left[\lim_{V_{\rm min} \to \infty}C_A^{(\rm off)}(t)\right] \ ,
    \end{equation}
    this implies that $F(\Omega)$ is infinitely differentiable on the real line, i.e.~$F(\Omega) \in C^{\infty}$. Therefore $F(\Omega)$ is a smooth function, in addition to being slowly varying.
    \item In a Floquet model lacking symmetries, the leading decay with $t$ for the purity $P_A(t)$ is expected to be \emph{exponential} for a measure 1 of all the initial states in the ensemble \eqref{eq:prodEnsemble} (typical ballistic growth of entanglement), and therefore it is expected to be exponential also for $C_A^{(\rm off)}(t)$ (see the concentration-of-measure results of Appendix \ref{app:concentrationHaar}). Exceptions to such behaviour in the absence of symmetries can arise in the presence of quenched randomness \cite{nahum_dynamics_2018}. Assuming the exponential decay, we conclude that:
    \begin{quote}
        There exists a complex-analytic continuation $F(z)$ ($z \in \mathbb C$) in a horizontal strip of the complex plane around the real axis. $F(z)$ is analytic in the whole strip.
    \end{quote}
    In particular, if $C_A^{(\rm off)}(t) \le c \, e^{-a|t|}$ for sufficiently large $t$ and for some $a > 0, c >0$, then the analytic strip includes all the points for which $|{\rm Im}(z)|< a$. Importantly, analyticity in the complex strip automatically implies that $F(\Omega)$ is real analytic, and that the Taylor series around any $\Omega_0 \in \mathbb R$ converges at least in a radius $|\Omega-\Omega_0|<a$. From the fact that $\lim_{V_{\rm min} \to \infty}C_A^{(\rm off)}(t)\ge 0$ we also get a simpler necessary and sufficient condition for its leading exponential decay \cite{lukacs1952analytic}:
    \begin{quote}
        The limiting autocorrelation $\lim_{V_{\rm min} \to \infty}C_A^{(\rm off)}(t)$ admits an exponential upper bound in $|t|$ IFF $F(\Omega)$ is real analytic in a open neighbourhood of $\Omega = 0 \ ({\rm mod} \ 2\pi)$.
    \end{quote}
    This result comes from interpreting $\lim_{V_{\rm min} \to \infty}C_A^{(\rm off)}(t)\ge 0$ as an (unnormalized) probability distribution and employing a theorem of Lukacs and Szász \cite{lukacs1952analytic}.
\end{itemize}
The simplest example of such properties is captured by the following, fairly realistic, assumption
\begin{equation}
\label{eq:lkvnpkrv}
    \lim_{V_{\rm min} \to \infty}C_A^{(\rm off)}(t) = C_{\rm exp}(t) + \delta C(t) \ , \qquad C_{\rm exp}(t) = v\,e^{- \gamma |t|} \, , \ \delta C(0)=1-v \, ,
\end{equation}
where $v$ is a constant and $\delta C(t)$ takes into account early-time corrections (therefore $|\delta C(t)|/C_{\rm exp}(t)\ll 1$ past an initial transient $t < t^*$, with $t^*$ finite). Denoting as $\widetilde f$ the Fourier transform of $f$, we get
\begin{align}
\label{eq:pkpviriiii}
    \widetilde C_{\rm exp}(\Omega) &= \sum_{t \in \mathbb Z} e^{-i\Omega t} C_{\rm exp}(t) = \frac{v \, \sinh(\gamma)}{\cosh(\gamma)-\cos(\Omega)} \ , \\
    \label{eq:deltaCtilde}
     \delta \widetilde C(\Omega) &= \sum_{t \in \mathbb Z} e^{-i\Omega t} \delta C(t) = \delta C(0)+  2 \sum_{n=1}^{\infty}\delta C(n) \cos(n \Omega)  \ .
\end{align}
The vary fast decay of $\delta C(n)$ as $n$ grows implies that, in practice, $\delta \widetilde C(\Omega)$ is a simple periodic function composed primarily of a few harmonics, while $\widetilde C_{\rm exp}(\Omega)$ is a simple periodic function whose analytic continuation in $\Omega \to z \in \mathbb C$ features simple poles at $z = \pm i\gamma \ ({\rm mod} \ 2\pi)$, which define the analytic strip $|{\rm Im}(z)|< \gamma$. Given that $F(\Omega)=\widetilde C_{\rm exp}(\Omega)+\delta \widetilde C(\Omega)$, this gives us an idea of what the marginal $F(\Omega)$ from \cref{eq:vprvmpv3} could like. Spectral functions of the same form of \eqref{eq:pkpviriiii} have been shown to arise, in the standard single-replica ETH framework, in connection to dynamical 2-point functions of single-replica operators that decay exponentially with time in dual-unitary circuits \cite{fritzsch_eigenstate_2021}. We note that considerations directly related to entanglement growth similar to the ones above have previously appeared in Ref.~\cite{jindal_generalized_2024}. Similar consideration can be developed also for $D>1$ systems, where however the boundary $|\partial|$ diverges for $V_{\rm min}\to \infty$ and hence all the relevant statements must be developed for the Rényi-2 entropy per boundary area in \cref{eq:entropyperarea}. In such cases, equations like \eqref{eq:agrbèdpd} hold by replacing the limit $V_{\rm min}\to \infty$ with a correction $\mathcal O(e^{-c V_{\rm min}})$ on the right-hand side, \emph{cf.}~\cref{sec:simplification}.

The description of purity dynamics in terms of \cref{eq:dpgdagòrlè} remains useful and exact also for $d_A$ finite and $d \to \infty$ in arbitrary dimensions $D$, although now the offset and proportionality coefficients from \cref{eq:Eprod}, as well as the saturation value due to a nonvanishing $C_A^{(\rm diag)}$, must be accounted for. For $d$ large but finite (finite global system) $C_A^{(\rm off)}(t)$ is still expected to decay exponentially before recurrences' time scales, however it might feature a two-stage relaxation in which the leading-decay exponent $\gamma$ changes around a crossover time which in 1D is proportional to the system size, i.e.~$t_{\rm cross} \propto L$ \cite{znidaric_twostep_2023, jonay_physical_2024} (this is a finite-size effect on time scales significantly shorter than those of recurrences). In any case, any spectral method capable to estimating the underlying function $f_A(\Omega,\omega_1,\omega_2)$ emerging from the matrix element statistics in \cref{eq:ETHansatz2} for $|A|$ and $|B|$ finite but large would automatically gain access to the coarse-grained purity decay via \cref{eq:dpgdagòrlè}. In \cref{sec:numver} we verify this explicitly from ED and exact-time-evolution data. Spectral filter methods \cite{luo_probing_2024} are a promising route to reconstruct $f_A$ for significantly larger (sub)system sizes.

\subsubsection{Quench spectral weight and $\mathbb P_A(t)$}

As anticipated in the previous section, we expect that also the out-of-equilibrium-ETH ansatz in \cref{eq:outEqETH1,eq:outEqETH2}, relevant for the dynamics of the un-averaged purity $\mathbb P_A(t)$ starting from a fixed initial state $\ket{\psi(0)}$, remains correct in the intermediate regime in which $|A|/|B|$ stays finite as $|A|$ and $|B|$ diverge. Indeed, for example, also in this case the diagonal ansatz in \cref{eq:outEqETH1} ensures the expected relaxation of the purity to the Haar value in \cref{eq:PurProdRelax} for \emph{arbitrary} $d_A$ and $d_B$. In \cref{sec:numver} we provide evidence via ED of the correctness of the ansatz for arbitrary ratios $d_A/d$.

Despite in the limit of $d_A$ finite, $d \to \infty$, the frequency structure of the underlying function $g_A$ is related to the frequency structure in \cref{eq:dgkal4gglv}, for arbitrary ratios $d_A/d_B$ we retain the full frequency dependence discussed above in terms of $\Omega$, $\omega_1$ and $\omega_2$ defined in \cref{eq:defOmegafre,eq:defomegasfre}. Also here the simple transformation \eqref{eq:ggldgè45} related to $(m,m') \leftrightarrow (n,n')$ is relevant because the function $g_A$ from \eqref{eq:outEqETH2} satisfies by definition
\begin{equation}
\label{eq:ngnorr}
    g_A(-\Omega,-\omega_1,\omega_2 -\Omega) = g^*_A(\Omega, \omega_1, \omega_2) \ .
\end{equation}
Similarly to before, in the limit of $d\to\infty$ and times $t$ arbitrary but finite, we can transform sums over eigenstates into integrals. We focus again on the leading off-diagonal class VII, the only one relevant for $d \to \infty$ for the dynamical part $ \mathbb P^{\rm (off)}_A(t)$ of the purity. Setting $N_{mm',nn'}\equiv \braket{\bm m|S_A|\bm n}\braket{\bm n |\rho(0)^{\otimes 2}|\bm m}$ we get from \cref{eq:mainLehmanSum}
\begin{equation}
\label{eq:ldnglk40gV2}
    \mathbb P_A^{(\rm off)}(t) \sim  \frac{1}{4}\sum_{\substack{m, \, m',\, n,\, n'\\(\rm pairwise \ distinct)}}e^{i\Omega t} N_{mm', nn'} \ ,
\end{equation}
where we have already removed the constraint $m<m'$ and $n<n'$ from the sum by introducing the factor $1/4$. Using also here the partitioning of the spectrum in vanishingly small windows $\Delta \varphi_j$ and the definition in \cref{eq:agabkvv} for the sum $\sum_{\Delta \varphi}$, we get using \cref{eq:outEqETH2}
\begin{align}
\label{eq:ngadgk4pogV2}
    \sum_{\Delta \varphi} e^{i\Omega t} N_{mm', nn'} &\sim e^{i\Omega^{(\Delta \varphi)} t}\,d^4 \, \braket{N_{\Delta \varphi}}\prod_{j = m,m',n,n'}\frac{\Delta \varphi_j}{2\pi}  \ , \\
\label{eq:hhadgk33potrV2}
    \braket{N_{\Delta \varphi}} = \frac{\sum_{\Delta \varphi} N_{mm', nn'}}{\sum_{\Delta \varphi}} &\sim \frac{4}{d^4}g_A\big(\Omega^{(\Delta \varphi)}, \omega_1^{(\Delta \varphi)}, \omega_2^{(\Delta \varphi)}\big) \ .
\end{align}
Putting together \cref{eq:ldnglk40gV2,eq:ngadgk4pogV2,eq:hhadgk33potrV2} we obtain
\begin{equation}
    \mathbb P_A^{(\rm off)}(t) \sim\prod_{j=m,m',n,n'}\int_0^{2 \pi}\frac{d\varphi_j}{2\pi}\left[ e^{i \Omega t} g_A(\Omega, \omega_1, \omega_2) \right] \ ,
\end{equation}
where as before $\Omega^{(\Delta \phi)} \to \Omega = \Omega(\varphi_m,\varphi_m',\varphi_n, \varphi_n')$ and similarly for $\omega_1$, $\omega_2$. By the same change of variables with unit Jacobian we arrive to
\begin{equation}
\label{eq:dpgdagòrlèV2}
 \mathbb P_A^{(\rm off)}(t) \sim \int_0^{2\pi}\frac{d\Omega}{2\pi} e^{i\Omega t}  \underbrace{\left[\int_0^{2\pi} \frac{d\omega_1\omega_2}{(2\pi)^2}  g_A(\Omega,\omega_1,\omega_2)\right]}_{G_A(\Omega)} \ .
\end{equation}
This proves \cref{eq:PAFourier}. From \cref{eq:ngnorr} and the $2\pi$ periodicity of all the arguments we obtain
\begin{equation}
    G_A(\Omega) = G_A^*(-\Omega) \ ,
\end{equation}
which guarantees $\mathbb P_A^{(\rm off)}(t) \in \mathbb R$. \\

Considerations on the properties of $G_A(\Omega)$ identical to those of \cref{sec:swapandC+} apply also here. In particular: (i) in 1D systems where $|\partial |$ does not grow as $V_{\rm min }\to \infty$, the expected decay of $\lim_{V_{\rm min} \to \infty}\mathbb P_A^{(\rm off)}(t)= \lim_{V_{\rm min} \to \infty} \mathbb P_A(t)$ faster than any power law implies $G(\Omega)=\lim_{V_{\rm min} \to \infty}G_A(\Omega) \in C^{\infty}$, i.e.~$G(\Omega)$ is a smooth function; (ii) assuming, given the absence of symmetries in our setup (and assuming absence of quenched randomness \cite{nahum_dynamics_2018}), ballistic growth of Rényi-2 entropy we obtain that in 1D $G(\Omega)$ is analytic on the real axis and it admits an analytic continuation in a horizontal strip of the complex plane (furthermore, because $\mathbb P_A(t)\ge 0$, also here analyticity around $\Omega =0 \ ({\rm mod} \ 2\pi)$ is actually necessary and sufficient for analyticity in the strip and hence leading exponential decay \cite{lukacs1952analytic}); (iii) for arbitrary $D$, the description of entanglement growth via \cref{eq:dpgdagòrlèV2} remains exact for $d_A$ finite, $d \to \infty$, where it must be combined with the saturation value induced by $\mathbb P_A^{(\rm diag)}$; (iv) for $d_A$, $d_B$ large but finite \cref{eq:dpgdagòrlèV2} remains very useful as it correctly describes the dynamics of the purity on time scales shorter than those characterizing recurrences, as we verify numerically in \cref{sec:numver}. The last point implies that any method, e.g.~spectral filters \cite{luo_probing_2024}, capable of approximately reconstructing $g_A$ in a large but finite (sub)system would automatically determine Rényi-2 dynamics. As before, dealing with the scaling limit $V_{\rm min}\to \infty$ in $D>1$ is possible but requires the use of intensive probes like the entropy per boundary area in \cref{eq:entropyperarea}, or the explicit control of $\mathcal O(e^{-c V_{\rm min}})$ corrections (see \cref{sec:simplification}).

Due to the lack of time-reversal symmetry for $U_F$, we expect $\mathbb P_A(t)\neq \mathbb P_A(-t)$. Therefore the simplest example of a 1D spectral function $G(\Omega)$, paralleling what done in \cref{eq:lkvnpkrv,eq:pkpviriiii,eq:deltaCtilde}, is given by the following fairly realistic assumption
\begin{align}
\label{eq:lkvnpkrvV2}
    &\lim_{V_{\rm min} \to \infty}\mathbb P_A^{(\rm off)}(t) = \mathbb P_{\rm exp}(t) + \delta \mathbb P(t) \ , \\
    &\mathbb P_{\rm exp}(t) =
    \begin{cases}
    v_+e^{- \gamma_+ t} \, , & t>0   \\
    v_+ + v_- \, , & t=0 \\
    v_- e^{\gamma_- t} \, , & t<0 
    \end{cases} \, ,
\end{align}
where $v_{\pm},\gamma_{\pm}>0$ and $\delta \mathbb P(t)$ takes care of early-time deviations from the asymptotic exponential decays. We then get
\begin{align}
\label{eq:pkpviriiiiV2}
    \widetilde{\mathbb P}_{\rm exp}(\Omega) = \frac{v_+}{1-e^{-\gamma_+}e^{-i\Omega}}+\frac{v_-}{1-e^{-\gamma_-}e^{i\Omega}} \ , \qquad \quad 
     \delta \widetilde{\mathbb P}(\Omega) = \sum_{t \in \mathbb Z} e^{-i\Omega t} \delta \mathbb P(t)   \ .
\end{align}
Similarly to $\delta \widetilde C(t)$ from \cref{sec:swapandC+}, also $\delta \widetilde{\mathbb P}(\Omega)$ is a simple function composed of very few harmonics. This demonstrates that $G(\Omega)=\widetilde{\mathbb P}_{\rm exp}(\Omega) + \delta \widetilde{\mathbb P}(\Omega)$ is a simple $2\pi$-periodic function that can be analytically continued in the strip $-\gamma_- < {\rm Im}(z) < \gamma_+$ and that features simple poles at $z = \pm i\gamma_{\pm} \ ({\rm mod} \ 2\pi)$.

\subsection{A simplification in the limit of large subsystems}
\label{sec:simplification}

Focusing on the limit of large subsystems $V_{\rm min}\gg 1$, we now demonstrate
\begin{equation}
\label{eq:simptoprove}
    |r_A^{(f)}(\Omega,\omega_1,\omega_2)| \le \exp(-\alpha' V_{\rm min}) \ ,    \qquad \alpha' > 0  \ ,
\end{equation}
in relation to \cref{eq:adjkdafjh1,eq:adjkdafjh1bis}. The argument uses locality of the evolution together with Floquet ETH \eqref{eq:FloquetStETH} and standard assumptions on bulk spectral dephasing. Using locality together with established results on return amplitudes during global quantum dynamics we also establish the analogous result for the individual-quench function $g_A$ (see \cref{eq:adjkdafjh2,eq:adjkdafjh2bis})
\begin{equation}
\label{eq:simptoproveG}
    |r_A^{(g)}(\Omega,\omega_1,\omega_2)| \le \exp(-\beta' V_{\rm min}) \ ,    \qquad \beta' > 0  \ .
\end{equation}
Both arguments rely on the assumption that the boundary $\partial$ between $A$ and $B$ is subextensive, i.e.~$|\partial|=o(V_{\rm min})$. The derivation of \eqref{eq:simptoproveG} is developed for an initial product state $\ket{\psi(0)}$, and then generalized with further assumptions to initial states with subextensive entanglement with respect to the bipartition into $A$ and $B$. An important technical remark is that our derivation establishes the decompositions in \cref{eq:adjkdafjh1,eq:adjkdafjh2} and the bounds \eqref{eq:simptoprove}-\eqref{eq:simptoproveG} at any fixed finite resolution in frequency space, implemented by a finite Fourier cutoff. Arbitrarily fine resolution is obtained by taking $V_{\rm min}\to\infty$ before removing the cutoff.

We first discuss the swap matrix elements, and then turn to the quench spectral weight.

\subsubsection{Swap matrix elements}
\label{sec:simplificationSwap}

To test whether the dependence on $\omega_1$ and $\omega_2$ disappears for $V_{\rm min} \gg 1$, it is convenient to Fourier transform with respect to all three frequencies from the outset. We temporarily use unsymmetrized eigenstates and denote their swap matrix elements by $X_{mm',nn'}=\braket{m,m'|S_A|n,n'}$. Consider 
\begin{align} 
\label{eq:SIspectral} 
\mathcal Z_A(t,s,r) &=\frac{1}{d^2}\sum_{m,m',n,n'} e^{i(\Omega t+\omega_1s+\omega_2r)}|X_{mm',nn'}|^2\\ 
\label{eq:SIkernel} &=\frac{1}{d^2}\Tr_{\mathcal H^{\otimes2}}\left[ (U_F^{-t-s-r}\otimes U_F^{-t})S_A (U_F^{t+s}\otimes U_F^{t+r})S_A\right]\ , 
\end{align} 
where $t,s,r\in\mathbb Z$ count Floquet periods and are conjugate to $\Omega,\omega_1,\omega_2$, respectively. 
At $s=r=0$, both replicas evolve for the same time $t$, and $\mathcal Z_A(t,0,0)$ is the swap autocorrelation before projection onto a replica-symmetry sector, which would turn it into $C_A^{(\sigma)}(t)$ for $\sigma = \pm 1$. The reason for introducing $\mathcal Z_A(t,s,r)$ is that a spectral function that is constant in $\omega_1$ ($\omega_2$) has Fourier coefficients with $s\neq0$ ($r \neq 0$) equal to zero. We will show that, as the subsystems $A$ and $B$ grow at arbitrary but fixed $t,s$ and $r$, one in fact obtains
\begin{equation} 
\label{eq:SItarget} 
\mathcal Z_A(t,s,r)= \delta_{s,0}\delta_{r,0} \, \mathcal Z_A(t,0,0) + \mathcal O(e^{-\alpha' V_{\rm min}}) \ . 
\end{equation} 
This will automatically imply \cref{eq:simptoprove} (more precisely, this implies \eqref{eq:simptoprove} at an arbitrary but fixed resolution in $\Omega$, $\omega_1$, $\omega_2$, determined by a constraint like $r,s,t\le T_{V_{\rm min}}$; however, $T_{V_{\rm min}}$ can be chosen arbitrarily large in the limit of $V_{\rm \min}\to \infty$, which leads to our conclusions at arbitrarily fine-grained scales in frequency space).
As in the preceding sections, the connection to the smooth ETH functions is made by replacing spectral sums by integrals in the limit of $d \to \infty$ and finite (but otherwise arbitrary) times, see \cref{eq:dpgdagòrlè}. Note that coincident-indices contributions must be treated separately. To understand \cref{eq:SItarget}, it is useful to employ an “interaction picture” that separates the dynamics entirely within $A$ or $B$ from that due to terms crossing the boundary $\partial$ between the two. For example, a Floquet unitary $U_F$ induced by a periodic time-dependent Hamiltonian $H(y)$ with period $\tau$ and boundary component $H_\partial(y)$ reads
\begin{equation}
    H(y) = H_A(y) + H_B(y) + H_\partial(y) \ , \qquad U_F = \mathcal T \exp\left[ -i\int_{0}^{\tau}dx H(x) \right] \ .
\end{equation}
By standard manipulations we can introduce an interaction picture unitary $U_I$
\begin{align}
    U_F &=  U_0 \, U_I  \ , \qquad U_0 = u_0(\tau) \, , \ U_I = u_I(\tau) \ , \\
    u_0(y) &= u_A(y) \otimes u_B(y)  \ , \qquad  u_{A/B}(y) = \mathcal T \exp\left[ -i\int_{0}^{y}dx H_{A/B}(x) \right] \ , \\
    \label{eq:gpanvkr4r}
    u_I(y) &= \mathcal T \exp\left[-i \int_{0}^{y} dx H_\partial^{(I)}(x)\right] \ , \qquad H_\partial^{(I)}(x) = u_0^\dag(x) H_\partial(x) u_0(x) \ ,
\end{align}
such that
\begin{equation}
\label{eq:p4lkvnvakdè}
    U_F^n = U_0^n \, \Big( \mathcal U_I(n-1) \mathcal U_I(n-2)\ldots \mathcal U_I(1) \mathcal U_I(0) \Big) \ , \qquad \mathcal U_I(n) = U_0^{-n} U_I U_0^{n} \ .
\end{equation}
The crucial point of this terms is that, under the assumption that $H(y)$ is local (e.g.~sum of finite-range operators), for any finite $n$ each term $\mathcal U_I(n)$ is localized (aside for exponentially small tails) around the boundary $\partial$ between $A$ and $B$, as a consequence of the Lieb-Robinson bound \cite{lieb_finite_1972, bravyi_liebrobinson_2006, nachtergaele_quasilocality_2019}. Indeed, by the latter bound \cite{bravyi_liebrobinson_2006} we know that $H_\partial^{(I)}(x)$ from \eqref{eq:gpanvkr4r} has subextensive-range support around $\partial$, up to exponentially small corrections, for any finite continuous time $x$. This in turn implies that $U_I$ has subextensive range due to its definition, and therefore each $\mathcal U_I(n)$ from \eqref{eq:p4lkvnvakdè} has subextensive range around $\partial$ for any finite discrete time $n$, again by Lieb-Robinson. The exponentially small spatial tails do not change the following physical argument: we expect the standard Floquet ETH and bulk-dephasing estimates used below, including their exponential suppression in subsystem volume, to remain valid for the quasi-local boundary operators generated by finite-time evolution. These operators are localized within a finite-width neighbourhood of the boundary $\partial$, up to exponentially decaying spatial tails \cite{bravyi_liebrobinson_2006}. Also in cases where the Floquet unitary $U_F$ is a quantum circuit built from stacking of finite-range gates one can obtain an expression similar to \cref{eq:p4lkvnvakdè}, which separates the decoupled part of evolution within $A$ and within $B$ from the one involving gates that straddle the boundary $\partial$. In such case the Lieb-Robinson bound is not needed because the circuit features an exact light cone due to its ultralocal nature, and therefore there are no correction tails. 

We can first consider the effect of $U_0^n$ from \eqref{eq:p4lkvnvakdè}, for the time being neglecting the interaction-picture terms $\mathcal U_I$. Setting $U_0 = U_A \otimes U_B = u_A(\tau) \otimes u_B(\tau)$, we evaluate \cref{eq:SIkernel} with this uncoupled evolution in place of $U_F$. 
The subsystem swap exchanges the two $U_A^{t+s}$ and $U_A^{t+r}$ factors between the two replicas, while leaving the analogous $B$ factors in place. Consequently, in the first copy of $A$ the powers combine as $(-t-s-r)+(t+r)=-s$ and in the second as $-t+(t+s)=s$. In $B$, the corresponding powers are instead $-r$ and $r$. Taking the four traces gives 
\begin{equation} 
\label{eq:SIdecoupled} 
\mathcal Z_{A,0}(t,s,r) =\frac{\big|\Tr_A[U_A^s]\big|^2}{d_A^2} \frac{\big|\Tr_B[U_B^r]\big|^2}{d_B^2}\ , 
\end{equation} 
where the subscript $0$ in $\mathcal Z$ denotes the uncoupled evolution. We note that the time $t$ has cancelled everywhere. Let us now assume that the eigenphase density $D(\varphi)$ of each isolated subsystem $A$ and $B$, i.e.~the one associated individually with $U_A$ and $U_B$, approaches the uniform density on the circle $D(\varphi)\sim d/2\pi$ for large $d$, as expected in the chaotic Floquet regime. This is generally expected to imply that 
\begin{equation}
\label{eq:aldl48bcnd4}
    \frac{\Tr_A[U_A^s]}{d_A} = \mathcal O(e^{-c |A|}) \ , \quad   s \neq 0 \ ,
\end{equation}
where $c>0$. This scaling is justified by the strong cancelling effects between the eigenphases uniformly distributed around the circle (note that $\Tr_A[U_A^s]=\sum_{m_A=1}^{d_A}e^{-is \varphi_{m_A}}$) together with $d_A = \exp(- |A| \ln d_{\rm loc})$. Clearly the same holds for $B$. Therefore from \eqref{eq:SIdecoupled} we obtain
\begin{equation}
\label{eq:dpgrivnvr}
    \mathcal Z_{A,0}(t,s,r) = \delta_{s,0}\delta_{r,0} + \mathcal O(e^{-c V_{\rm min}}) \ , \quad c > 0 \ ,
\end{equation}
which has precisely the selection of times in \cref{eq:SItarget}. Note that the constants $c$ appearing both in \eqref{eq:aldl48bcnd4} and \eqref{eq:dpgrivnvr} are not in general the same, but for simplicity of notation we always use the letter $c$ to denote the prefactor of an exponential scaling.

We now restore the coupling across the cut $\partial$. This is responsible for the nontrivial time dependence of the swap autocorrelation so its effect must be retained. The locality of the coupling terms in \cref{eq:p4lkvnvakdè} allows us to do so without changing the residual bulk powers identified in \cref{eq:SIdecoupled} . We rewrite \cref{eq:p4lkvnvakdè} compactly as
\begin{equation} 
\label{eq:SIboundaryfactor} 
U_F^n=(U_A\otimes U_B)^n \mathcal V_{\partial}(n)\ ,
\end{equation} 
remembering that, up to exponentially small corrections, $\mathcal V_{\partial}(n)$ acts in a region around the boundary whose width grows with $|n|$ but \emph{not} with the subsystem sizes. Therefore $\mathcal V_{\partial}(n)$ might have an order-one effect but it cannot reach the asymptotically large interiors of $A$ and $B$ at fixed times $n$. Substituting \cref{eq:SIboundaryfactor} in \cref{eq:SIkernel}, we can collect the uncoupled evolution factors exactly as before. Commuting them past a boundary operator only evolves that operator for a finite time, so it remains near the cut. More explicitly, using also $S_A ^2 = \Id$
\begin{align}
\label{eq:SIvldfffkvndèf}
    \mathcal Z_A(t,s,r)&=\frac{1}{d^2}\Tr_{\mathcal H^{\otimes 2}} [T_1 V_1 T_2 V_2] \ ,\\
    \label{eq:SI1oOo}
    T_1 & \equiv (U_A \otimes U_B)^{-t-s-r} \otimes (U_A \otimes U_B)^{-t} \ , \\
    \label{eq:SI2oOo}
    V_1 & \equiv \big[\mathcal V_{\partial}(-t-s-r) \otimes \mathcal V_\partial(-t) \big] \ , \\
    \label{eq:SI3oOo}
    T_2 & \equiv S_A \big[(U_A \otimes U_B)^{t+s} \otimes (U_A \otimes U_B)^{t+r}\big] S_A \ , \\
    \label{eq:SI4oOo}
    V_2 & \equiv S_A\big[\mathcal V_{\partial}(t+s) \otimes \mathcal V_\partial(t+r) \big] S_A \ .
\end{align}
At this point we can use that $\widetilde V_1\equiv T_2^{-1}V_1 T_2$ is also, by Lieb-Robinson, a boundary term. Writing $\mathcal W_\partial(t,s,r)\equiv \widetilde V_1 V_2$, we obtain
\begin{align}
\label{eq:SIdressedtrace} 
    \mathcal Z_A(t,s,r)&=\frac{1}{d^2}\Tr_{\mathcal H^{\otimes 2}} [(T_1 T_2) \mathcal W_\partial(t,s,r)] = \\ \nonumber
    & = \frac{1}{d^2}\Tr_{\mathcal H^{\otimes 2}}\left\{ \left[(U_A^{-s} \otimes U_B^{-r}) \otimes (U_A^{s} \otimes U_B^{r})\right] \mathcal W_\partial(t,s,r)\right\} \ .
\end{align}
The first term on the last line of this equation is exactly the one yielding the uncoupled result \eqref{eq:SIdecoupled} if we replace $\mathcal W_\partial \to \Id$, while $\mathcal W_\partial$ has collected all the boundary factors including their finite-time evolution. Therefore the bulk part of this expression is the same as in \cref{eq:SIdecoupled} and all effects of the coupling are contained in $\mathcal W_\partial$. It remains to check that this boundary operator cannot prevent a scaling like \eqref{eq:dpgrivnvr} when $s$ or $r$ is nonzero. We expand $\mathcal W_\partial$ in products of operators acting on the four boundary regions, one in each copy of $A$ and $B$. Each term in \cref{eq:SIdressedtrace} is then of the form 
\begin{equation} 
\label{eq:SIfourtraces} 
\frac{\Tr_A[U_A^{-s}O_{A_1}]\Tr_A[U_A^sO_{A_2}]}{d_A^2} \frac{\Tr_B[U_B^{-r}O_{B_1}]\Tr_B[U_B^rO_{B_2}]}{d_B^2} \ , 
\end{equation} 
where all four operators $O$ have subextensive support (up to exponentially small tails \cite{bravyi_liebrobinson_2006}, which do not change the following conclusions) and bounded norm. For any finite values of $t,s$ and $r$ the boundary expansion in each of the 4 subsystem-replicas $Y=A_1,A_2,B_1,B_2$ contains an order $\simeq \exp(c |\partial_Y|)$ terms (each with bounded coefficient), where $|\partial_Y|$ is subextensive with respect to $|Y|$. Therefore if each one of the expansion terms vanishes as $\mathcal O[\exp(-c|Y|)]$ unless $r=s=0$, that is sufficient to establish \eqref{eq:SItarget}. Making use of diagonal Floquet ETH \eqref{eq:FloquetStETH} within each subsystem $Y$, we obtain
\begin{equation} 
\label{eq:SIlocaltrace} 
\frac{\Tr_Y[U_Y^s O_Y]}{d_Y} \sim \frac{\Tr_Y[O_Y]}{d_Y} \frac{\Tr_Y[U_Y^s]}{d_Y} = \mathcal O(e^{-c |Y|})\ ,\qquad s \neq 0\ , 
\end{equation}
where we used again \eqref{eq:aldl48bcnd4} (and its analogue for $B$) together with the fact that $\Tr_Y[O_Y]/d_Y$ is finite. Note that in \eqref{eq:SIlocaltrace} we have neglected the diagonal ETH fluctuations at leading order, but even if they were the primary contribution (e.g.~if $\Tr_Y[O]=0$) the claim on $\Tr_Y[U_Y^s O_Y]/d_Y$ would still hold. 
\cref{eq:SIlocaltrace} demonstrates that if $s\neq0$ ($r \neq 0$), \cref{eq:SIdressedtrace} vanishes at least exponentially with $V_{\rm min}$ as a consequence of the subextensive nature of the boundary $\partial$. This establishes \cref{eq:SItarget}. 
When instead $s=r=0$, there is no residual bulk evolution and the physical swap autocorrelation $\mathcal Z_A(t,0,0)$ merely decays exponentially with $|\partial|$ (see \cref{eq:adjkdafjh1bis} and the discussion around \cref{eq:purityboundarea,eq:entropyperarea}). This is why the locality, ETH and bulk dephasing arguments above remove the dependence on $\omega_1,\omega_2$ while retaining the nontrivial one on $\Omega$. 

We can now return to (anti)symmetrized eigenstates. For the all-distinct class VII, the matrix element is \begin{equation} 
\label{eq:SIsymmetry} 
\braket{\bm m_\sigma|S_A|\bm n_\sigma} =X_{mm',nn'}+\sigma X_{mm',n'n}\ . 
\end{equation} 
This yields
\begin{equation}
    |\braket{\bm m_\sigma|S_A|\bm n_\sigma}|^2 = |X_{mm',nn'}|^2+ |X_{mm',n'n}|^2 + 2 \sigma {\rm Re}[X_{mm',nn'}X^*_{mm',n'n}] \ ,
\end{equation}
which is the matrix element containing the spectral function $f_A$.  
The first term on the right-hand side is exactly the one appearing in the definition \eqref{eq:SIspectral} of $\mathcal Z_A(t,s,r)$, which we just demonstrated to yield a spectral function that only depends on $\Omega$. The second term is obtained from the first by $n \leftrightarrow n'$, which leaves $\Omega$ invariant but exchanges $\omega_1$ and $\omega_2$. Therefore the arguments above apply identically and one obtains again a spectral function that only depends on $\Omega$ in the limit of large subsystem sizes. The last interference term proportional to $\sigma$ is subleading (in $d$) and therefore we can neglect it. This can be understood from locality arguments similar to the ones above, starting from a modified correlator $\widetilde{\mathcal Z}_A(t,r,s)$ defined as in \eqref{eq:SIspectral} but via the interference term. We do not report such argument because we already have strong evidence that this term is subleading from: (i) the off-diagonal RMT result in \cref{eq:HaarFluctOff}, which at leading order does not contain any dependence on $\sigma$; (ii) the connection with full-ETH from \cref{sec:conn2&4ETH}, where we showed that the terms proportional to $\sigma$ are subleading because related to 4-ETH contributions instead of products of $2$-ETH ones (although that was only explicitly established for $d_A$ arbitrarily large but finite). The results above therefore demonstrate the validity of \cref{eq:simptoprove} for the class VII from \cref{tab:classesOfEigenstates}. Similar results can be derived for all the other off-diagonal classes whose frequency dependence involves more than one frequency. In \cref{sec:numver} and Appendix \ref{sec:furthernumerics} we show numerical evidence from ED supporting our claim for all the off-diagonal classes.

\subsubsection{Quench spectral weight}

For the quench function $g_A$, we can repeat the same steps using the initial state $\ket{\psi(0)}$ instead of a normalized trace. The appropriate three-time quantity is 
\begin{align} 
\label{eq:SIquench} 
\mathcal Q_A(t,s,r) =\braket{\psi(t+s+r),\psi(t)|S_A|\psi(t+s),\psi(t+r)}=\sum_{m,m',n,n'}e^{i(\Omega t+\omega_1s+\omega_2r)} X_{mm',nn'}\Psi_{nm}\Psi_{n'm'}\ , 
\end{align} 
where $\Psi_{nm}=\braket{n|\rho(0)|m}$ and $\ket{\psi(t)}=U_F^t\ket{\psi(0)}$. At $s=r=0$ this is the purity $\mathbb P_A(t)$. For an initial \emph{product state} $\ket{\psi(0)}=\ket{\psi_A(0)}\otimes\ket{\psi_B(0)}$, the uncoupled calculation via $U_0 = U_A \otimes U_B$ now gives 
\begin{equation} 
\label{eq:SIecho} \mathcal Q_{A,0}(t,s,r) =|\braket{\psi_A(0)|U_A^s|\psi_A(0)}|^2 |\braket{\psi_B(0)|U_B^r|\psi_B(0)}|^2\ . 
\end{equation} 
The residual powers are the same as before, but their normalized traces have been replaced by return probabilities of the initial subsystem states. For a generic global quench from a short-range-correlated state, one expects these return probabilities to vanish exponentially in the subsystem sizes $|A|,|B|$ at any fixed nonzero time \cite{heyl_dynamical_2018}. Therefore we find the analogue of \cref{eq:dpgrivnvr} in the quench context, namely $\mathcal Q_{A,0}(t,s,r)=\delta_{s,0}\delta_{r,0} + \mathcal O(e^{-c V_{\rm min}})$. Restoring the coupling between $A$ and $B$ thanks to \cref{eq:SIboundaryfactor}, we obtain
\begin{equation}
\label{eq:dlropwvmasw}
    \mathcal Q_A(t,s,r)= \bra{\psi(0)}^{\otimes 2} T_1 V_1 T_2 V_2 \ket{\psi(0)}^{\otimes 2} \ ,
\end{equation}
where $T_j$, $V_j$ are defined in \cref{eq:SI1oOo,eq:SI2oOo,eq:SI3oOo,eq:SI4oOo} and we have used $S_A\ket{\psi(0)}^{\otimes 2}=\ket{\psi(0)}^{\otimes 2}$ for an initial product state (together with $S_A^2 = \Id$). As in \cref{eq:SIdressedtrace} , this leads us to
\begin{equation}
\label{eq:òddèadvjvddd}
    \mathcal Q_A(t,s,r)= \bra{\psi(0)}^{\otimes 2} \left[(U_A^{-s} \otimes U_B^{-r}) \otimes (U_A^{s} \otimes U_B^{r})\right] \mathcal W_\partial(t,s,r) \ket{\psi(0)}^{\otimes 2} \ ,
\end{equation}
As before, we can expand $\mathcal W_\partial$ into products of operators acting on the four boundary regions (two per replica). Each of these contributions to $\mathcal Q_A$, whose total number scales exponentially with $|\partial|$, reads
\begin{align}
    &\braket{\psi_A(0) |U_A^{-s}O_{A_1}|\psi_A(0)}\braket{\psi_A(0) |U_A^{s}O_{A_2}|\psi_A(0)} \\
    &\quad\times \braket{\psi_B(0) |U_B^{-r}O_{B_1}|\psi_B(0)}\braket{\psi_B(0) |U_B^{r}O_{B_2}|\psi_B(0)}  \ ,
\end{align}
where up to exponentially small corrections each $O_Y$ has subextensive support in a finite width around the boundary. Due to this, we expect that the return amplitude to $\ket{\psi_Y(0)}$ of $U_Y^{s}(O_Y \ket{\psi_Y(0)})$ is still exponentially suppressed in the subsystem volume $|Y|$ for any $s \neq 0$, exactly as for $O_Y \to \Id_Y$ discussed in the uncoupled case. This is because the evolution induced by $U_Y^s$ changes the state also in the (asymptotically large) bulk of $Y$, where $O_Y$ has no effect. This implies
\begin{equation} 
\label{eq:SIresultg} 
\mathcal Q_A(t,s,r)= \delta_{s,0}\delta_{r,0}\mathbb P_A(t) + \mathcal O(e^{-c V_{\rm min}}) \ . 
\end{equation} 
The symmetric quench weight combines the two unsymmetrized weights related by $n\leftrightarrow n'$, so it inherits their dependence on $\Omega$. This establishes the validity of \eqref{eq:simptoproveG} for an initial product state.

The arguments above generalize easily to sub-volume-law entangled initial states: the main difference is that in \cref{eq:òddèadvjvddd} we need to expand both $\mathcal W_\partial$ into subsystem operators and the initial state $\ket{\psi(0)}$ into a sum of product states with respect to the bipartition into $A$ and $B$. Furthermore, the identity $S_A\ket{\psi(0)}^{\otimes 2}=\ket{\psi(0)}^{\otimes 2}$ does not hold here, hence there is an $S_A$ left acting on the rightmost replicated initial state in \eqref{eq:òddèadvjvddd}. Assume that the initial state has subextensive entanglement for arbitrary fixed Rényi index $\alpha \in \mathbb R^+$, i.e.~$\mathbb S^{(\alpha)}_A(0)\propto |\partial_E|=o(V_{\rm min})$ (actually, it is sufficient to assume this for $\alpha = 1/2$). Call $\lambda_i$ the squared Schmidt coefficients of the Schmidt expansion of $\ket{\psi(0)}$ into product states with respect to the the $A$, $B$ bipartition. The \emph{overall} Schmidt weight (i.e. neglecting the extra terms in the expression) is
\begin{equation}
    \sum_{i,j,k,\ell}\sqrt{\lambda_i \lambda_j\lambda_k\lambda_\ell} = \big( \sum_i \sqrt{\lambda_i} \big)^4 = e^{2\mathbb S_A^{(1/2)}(0)} \propto e^{c |\partial_E|} \ ,
\end{equation}
which is \emph{merely} exponentially large in $|\partial_E|=o(V_{\rm min})$. It seems natural in the present context to assume that the return amplitude to $\bra{\psi'_Y(0)}$ of $U_Y^{s}(O_Y \ket{\psi_Y(0)})$, where $\ket{\psi_Y(0)}$ is a subsystem-$Y$ component appearing in the Schmidt expansion of $S_A\ket{\psi(0)}^{\otimes 2}$ while $\bra{\psi'_Y(0)}$ appears in the Schmidt expansion of $\bra{\psi(0)}^{\otimes 2}$, is exponentially suppressed in the subsystem volume $|Y|$ for any $s \neq 0$. Given that an overall Schmidt-weight factor $\propto \exp(c|\partial_E|)$ cannot counterbalance the exponential suppression in the \emph{bulk subsystem volume} $|Y|\ge V_{\rm min}$ of each of the terms with $s \neq 0$ or $r \neq 0$, \cref{eq:simptoproveG} is expected to hold also for sub-volume-law entangled initial states.

By translating sums over eigenstates into integrals in the limit of $d \to \infty$ for the class VII that dominates, as done when deriving \cref{eq:dpgdagòrlèV2}, the results above demonstrate the validity of \cref{eq:simptoproveG} under the stated well-established assumptions. The other off-diagonal classes could be treated in a similar fashion. In \cref{sec:numver} we show numerical evidence from ED supporting \cref{eq:simptoproveG}.

\section{Numerical verification}
\label{sec:numver}

\subsection{The model}
\label{sec:themodel}
\begin{figure}[t]
    \centering
    \includegraphics[width=0.4\linewidth]{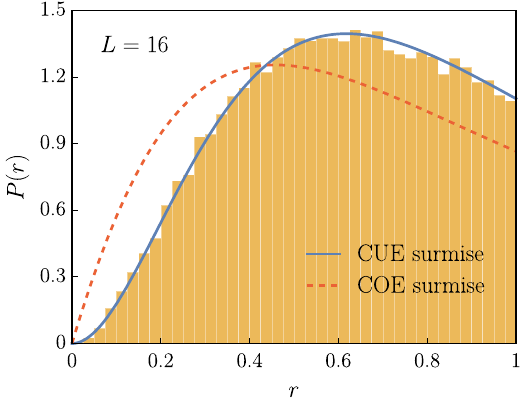}
    \caption{Histogram of the gap ratios \eqref{eq:gapratiodef} from the eigenphases of $U_F$ for a chain with $L = 16$. The continuous and dashed lines are, respectively, the CUE and COE surmises from \cite{atas_distribution_2013}.}
    \label{fig:gap_ratio_L16}
\end{figure}

In this section we numerically verify the results above for a deterministic chaotic spin-$1/2$ Floquet chain using ED and exact time evolutions. We consider a $U_F$ that lacks any symmetry, including time reversal invariance, which places it into the CUE class used in our RMT calculations. It corresponds to a ternary drive in a each period $\tau$
\begin{align}
\label{eq:UFdefinition}
    U_F &= U_z \, U_x \, U_h \ , \\ 
\label{eq:UFdefinition1}
    U_z &= \exp\Big[ -i\frac{\tau}{3}\Big(J_z\sum_{j=1}^{L-1}Z_j Z_{j+1}\Big)\Big]  \ , \\
\label{eq:UFdefinition2}
    U_x &= \exp\Big[ -i\frac{\tau}{3}\Big(J_x\sum_{j=1}^{L-1}X_j X_{j+1}\Big)\Big] \ , \\
\label{eq:UFdefinition3}
    U_h & = \exp\Big[ -i\frac{\tau}{3}\sum_{j=1}^{L}\Big(h_x X_j + (h_z + \varepsilon \delta_{j1}) Z_j \Big)\Big] \ .
\end{align}
Here $L$ is the length of the chain with open boundary conditions (OPB), $X_j$ and $Z_j$ are Pauli matrices on site $j$, and $\varepsilon$ sets the strength of a magnetic impurity on site $j=1$. A similar ternary drive was introduced in \cite{yates_almost_2019} and run on a quantum simulator in \cite{harle_observing_2023}. Absence of ordinary symmetries is enforced as follows: (i) OPB and $\varepsilon \neq 0$ break translation symmetry in space; (ii) $\varepsilon \neq 0$ breaks reflection symmetry around the center of the chain; (iii) the presence of nonzero $h_x$ and $h_z$ breaks spin-flip symmetries; (iv) $J_z \neq J_x$ and $h_z \neq h_x$ break possible symmetries exchanging $X \leftrightarrow Z$. Furthermore, despite individually $U_z$, $U_x$, $U_h$ are time-reversal invariant (symmetric matrices in the computational basis), there is no obvious basis in which the global matrix $U_F$ is symmetric (and hence time-reversal invariant), not even if we allow a shift in the defining period $\tau$, e.g.~$U_F'=U_h^{1/2}U_z U_x U_h^{1/2}$. Therefore, despite each component of the ternary drive belongs to the COE class, the full Floquet operator is expected to belong to CUE \cite{regnault_floquet_2016}. We choose the following non-fine-tuned set of parameters, which also comply with the constraints above: $J_z = 1$, $J_x = 0.75$, $h_z = 1.1$, $h_x =0.85$, $\varepsilon = 0.4$, $\tau = 2$. To verify that $U_F$ belongs to the CUE class we can use the gap ratio statistics \cite{oganesyan_localization_2007, atas_distribution_2013, nishigaki_distributions_2024, nishigaki_distributions_2026} for
\begin{equation}
\label{eq:gapratiodef}
    r_j = \frac{\min(\Delta \varphi_j,\Delta \varphi_{j-1})}{\max(\Delta \varphi_j,\Delta \varphi_{j-1})} \in [0,1] \ , \qquad \Delta \varphi_j = \varphi_{j+1} - \varphi_j \ ,
\end{equation}
where $\varphi_j$ are the eigenphases of the Floquet operator $U_F \ket{j} = e^{-i\varphi_j}\ket j$ which we order as $0 \le \varphi_1 \le \ldots \le \varphi_d < 2\pi$. 
\begin{table}[t!]
    \centering
    \renewcommand{\arraystretch}{1.3}
    \setlength{\tabcolsep}{18pt}
    \begin{tabular}{ccc}
        \toprule
        & CUE (exact) & ED $(L=16)$ \\
        \midrule
        $\braket{r}$   & $0.5997...$ & $0.6000...$\\
        $\braket{r^2}$ & $0.4132...$ & $0.4136...$ \\
        $\braket{r^3}$ & $0.3100...$ & $0.3105...$ \\
        $\braket{r^4}$ & $0.2460...$ & $0.2466...$ \\
        \bottomrule
    \end{tabular}
    \caption{Comparison of the exact CUE moments $\braket{r^k}$ for the gap ratio \eqref{eq:gapratiodef} from Refs.~\cite{nishigaki_distributions_2024, nishigaki_distributions_2026} and the average moments obtained with $L=16$ ED for $U_F$.}
    \label{tab:CUEratio}
\end{table}
A few values for the CUE moments $\braket{r^k}$ have been determined exactly \cite{nishigaki_distributions_2024, nishigaki_distributions_2026} and are reported in \cref{tab:CUEratio}, together with the average values obtained from ED of $U_F$ at $L = 16$. They match very well, confirming that our $U_F$ belongs to the CUE class. In \cref{fig:gap_ratio_L16} we also plot the histogram of the ED ratios $r$ against the approximate (but accurate) GUE and GOE surmises from \cite{atas_distribution_2013} (note that in the limit $d \to \infty$ the probability distribution $P(r)$ for GUE (GOE) Hamiltonians in the bulk and CUE (COE) unitaries is the same \cite{nishigaki_distributions_2026, mehta_random_matrices}). The plot makes apparent that the ED distribution is far from COE while it matches very well CUE. We also verified that the density of eigenphases $\varphi_j$ tends towards the uniform limit $D(\varphi)=d/(2\pi)$ in the limit of large $d$.  

Aside for providing a simple example of a chaotic CUE-class Floquet operator, $U_F$ from  \eqref{eq:UFdefinition} is also particularly simple to apply to a state. Working in the $Z$ computational basis $\ket{e_n}$, we indeed have
\begin{align}
    &U_z \ket{e_n} = e^{-i\phi(e_n)}\ket{e_n} \ ,\\
    &U_x = \prod_{j=1}^{L-1}\Big( \cos(c_x) - i \sin(c_x)X_j X_{j+1} \Big) \ , \\
    &U_h = \prod_{j=1}^L \bigg( \cos(c_{h,j}) - i \frac{\tau \sin(c_{h,j})}{3\,c_{h,j}}(h_x X_j + h_{z,j}Z_j) \bigg) \ ,
 \end{align}
where $\phi(e_n)$, $c_x$, $c_{h,j}$ and $h_{z,j}$ are all real numbers obtained from the definitions in \cref{eq:UFdefinition1,eq:UFdefinition2,eq:UFdefinition3}. This allows us to construct $\ket{\psi(t)}=U_F^t\ket{\psi(0)}$, up to very large $t \in \mathbb Z$, without the need to diagonalize $U_F$. For the dynamics we are therefore able to reach system sizes up to $L = 30$. On the other end, whenever ED is required (e.g.~for ETH-related quantities) we are constrained to $L \le 16$.

The numerical construction of the matrix elements $\braket{\bm m_{\sigma}|S_A|\bm n_{\sigma}}$ is achieved efficiently as follows. Given an eigenstate $\ket m$ obtain by ED, it is straightforward to numerically decompose it as
\begin{equation}
\label{eq:SchmidtIdentity1}
    \ket m = \sum_{a=1}^{d_A}\sum_{b=1}^{d_B} \Psi_{ab}^{(m)}\ket{a}_A\ket{b}_B \ ,
\end{equation}
where $\Psi^{(m)}$ is a $d_A \times d_B$ matrix and $\ket{a}_A,\ket{b}_B$ denote computational-basis states in $A$ and $B$. From this we get the following identity for the entries of the reduced transition matrices $\rho_A^{(nm)}$ from \cref{eq:redtranmatrix2}
\begin{equation}
\label{eq:SchmidtIdentity2}
    \big(\rho_A^{(nm)}\big)_{aa'} = \big( \Psi^{(n)} (\Psi^{(m)})^\dag \big)_{aa'} \ ,
\end{equation}
where $\Psi^{(n)}(\Psi^{(m)})^\dag$ is a $d_A \times d_A$ matrix. Using \cref{eq:redtranmatrix1} we then get for the unsymmetrized matrix elements
\begin{equation}
\label{eq:SchmidtIdentity3}
    X_{mm',nn'}=\braket{m,m'|S_A|n,n'} = \sum_{a=1}^{d_A}\Big(\rho_A^{(nm)}\rho_A^{(n'm')}\Big)_{aa} \ .
\end{equation}
The (anti)symmetrized matrix elements are obtained simply as
\begin{equation}
\label{eq:fld4t39dl50d}
    \braket{\bm m_\sigma |S_A|\bm n_\sigma} = \frac{X_{mm',nn'}+\sigma X_{mm',n'n}}{\sqrt{(1+\delta_{mm'})(1+\delta_{nn'})}} \ .
\end{equation}
These simple manipulations allow us to obtain the matrix elements of interest only by elementary products and traces of $d_A \times d_A$ matrices, hence avoiding the need to work with any $d \times d$ or larger matrices. Similar steps also allow us to efficiently obtain the purity $\mathbb P_A(t)=\Tr_A[\rho_A(t)^2]$ from the decomposition $\ket{\psi(t)}=\sum_{a=1}^{d_A}\sum_{b=1}^{d_B}\chi_{ab}(t)\ket{a}_A\ket{b}_B$ of the time-evolving state
\begin{equation}
    \mathbb P_A(t) = \sum_{a=1}^{d_A}\Big( \chi(t)\chi^\dag(t)\chi(t)\chi^\dag(t)\Big)_{aa}
\end{equation}

\begin{figure}[t]    
\centering
\begin{tikzpicture}[
  x=0.8cm,y=1cm,
  site/.style={circle,draw=black,fill=white,
               minimum size=5mm,inner sep=0pt}
]
  \fill[blue!10,rounded corners=4pt]
    (0.55,-0.48) rectangle (7.45,0.48);
  \fill[orange!20,rounded corners=4pt]
    (7.55,-0.48) rectangle (12.45,0.48);

  \node at (4,0.78) {$B$};
  \node at (10,0.78) {$A$};

  \draw[thick] (1,0) -- (12,0);
  \foreach \i in {1,...,12} {
    \node[site] at (\i,0) {};
  }

  \fill[red!75!black] (1,0) circle (2pt);
  \node[above,text=red!75!black]
    at (1,0.27) {$\varepsilon$};
\end{tikzpicture}
\caption{Example for $L = 12$ and $|A|=5$ of the bipartition used for the ED and time-evolution numerics. Subsystem $A$ is placed at the right end of the chain, on the opposite side with respect to the inversion-symmetry-breaking impurity $\varepsilon$. None of the main conclusions from this section change when $A$ is placed differently, e.g., at the center of the chain.}
    \label{fig:subsystemplacem}
\end{figure}
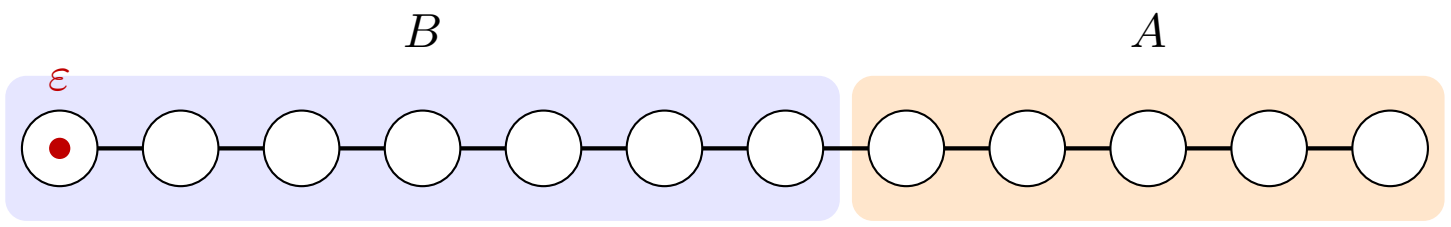

In the following, we place subsystem $A$ at right end of the chain, as depicted in \cref{fig:subsystemplacem}. This produces a single entanglement cut at the boundary $\partial$ between $A$ and $B$. We have verified that none of our main conclusions change when $A$ is placed differently, e.g.~at the centre of the chain, featuring 2 entanglement cuts.

\subsection{Swap matrix elements and $C_A^{(+)}(t)$}
\label{sec:swapmatrixED}

\begin{figure}[b]
    \centering
    \includegraphics[width=0.45\linewidth]{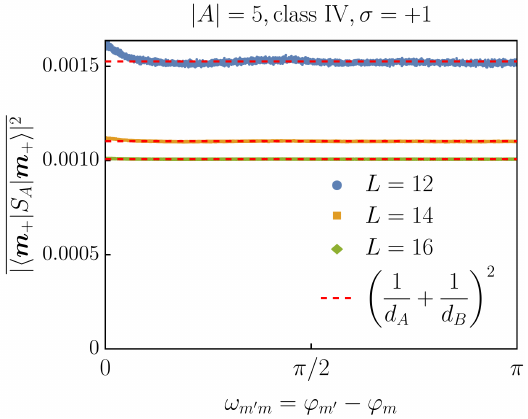}
    \caption{Local-spectrum-averaged diagonal elements $\overline{|\braket{\bm m_+ |S_A|\bm m_+}|^2}$ from class IV (\cref{tab:classesOfEigenstates}), as functions of $\omega_{m'm}$ for $|A| = 5$ and $L=12,14,16$. The local-spectrum radii are, respectively, $\mathcal R = 8, 8,16$ ($\mathcal M = 2$ for class IV). The red dashed line is the leading diagonal prediction from \cref{eq:ETHansatz1}. We restrict the $x$-axis to $[0,\pi]$ because the outputs are symmetric around $\pi$, as expected. For each system size the plot shows around $5\times 10^3$ sampled points (quartets $\bm q$).}
    \label{fig:class_IV_eth}
\end{figure}
Due to the expected presence of the erratically varying numbers $R^{(A)}_{\bm m,\bm n}$ with order-one variance in the off-diagonal ETH ansatz \eqref{eq:ETHansatz2}, it is convenient to plot the numerical data already averaged over neighbouring eigenstates. We refer to these arithmetic means as \emph{local-spectrum averages}. The procedure we employ is as follows. 
\begin{enumerate}
    \item We randomly sample quartets $\bm q=\{m, m',n,n'\}$ (with the usual convention $m \le m'$, $n\le n'$) according to the specifics of the class from \cref{tab:classesOfEigenstates} under consideration (i.e.~accounting for coincidences, for example $m=m'$ implies that we effectively sample triplets). 
    \item For each $\bm q=\{m, m',n,n'\}$ sampled, we consider for each index $\ell = m,m',n,n'$ the ($2\mathcal R+1$) neighbouring indices $j_\ell=\ell-\mathcal R,\ldots,\ell,\ldots,\ell+\mathcal R$. This produces $(2\mathcal R+1)^{\mathcal M}$ overall quartets $\bm j =\{j_m,j_{m'},j_n, j_{n'}\}$, where $\mathcal M$ denotes the number of distinct indices (from $1$ to $4$) in the class under consideration. If one of these quartets contains a pair for which $j_{\ell_1} = j_{\ell_2}$ although $\ell_1 \neq \ell_2$, we discard it in order to preserve the structure of the class chosen (this happens if two neighbourhoods of radius $\mathcal R$ centred around $\ell_1$ and $\ell_2$ overlap). We call $\mathcal Q_{\bm q,\mathcal R}$ the set of valid quartets $\bm j$ obtained for a given $\bm q=\{m, m',n,n'\}$ and $\mathcal W_{\bm q,\mathcal R} = (2\mathcal R+1)^{\mathcal M} - (\#\text{ of discarded $\bm j$})$ their total number. 
    \item Each data point in this section's plots corresponds to (we set $M^{(\sigma)}_{\bm q} \equiv |\braket{\bm m_\sigma|S_A|\bm n_\sigma}|^2$)
    \begin{equation}
        \overline{M_{\bm q}^{(\sigma)}} = \frac{1}{\mathcal W_{\bm q,\mathcal R}}\sum_{\bm j \in \mathcal Q_{\bm q,\mathcal R}}\ M^{(\sigma)}_{\bm j} \ .
    \end{equation}
    This makes precise the meaning of the local-spectrum average symbol. We also save the values of the 2nd and 4th cumulants
    \begin{align}
    \label{eq:kappa2ED}
        \kappa_2^{(\sigma)}(\bm q) = \overline{\left(M_{\bm q}^{(\sigma)} - \overline{M_{\bm q}^{(\sigma)}}\right)^2} \ , \qquad \quad \kappa_4^{(\sigma)}(\bm q) = \overline{\left(M_{\bm q}^{(\sigma)} - \overline{M_{\bm q}^{(\sigma)}}\right)^4} - 3 \kappa_2(\bm q)^2 \ ,
    \end{align}
    which enable us to check whether the moments of $R^{(A)}_{\bm m, \bm n}$ are compatible with a complex Gaussian distribution.
\end{enumerate}

\subsubsection{Diagonal elements}
We start from the dominant diagonal class IV $\to \{(a,b),(a,b)\}$ from \cref{tab:classesOfEigenstates}, which admits matrix elements in both sectors $\sigma = \pm 1$. A priori, the only possible frequency dependence is on $\omega_{m'm}=\varphi_{m'}-\varphi_m$. However, from the previous sections we expect that such dependence can only arise at subleading orders. Indeed, at leading order the diagonal ETH ansatz from \cref{eq:ETHansatz1} predicts no frequency dependence at all: $\braket{\bm m_\sigma |S_A|\bm m_\sigma} \sim (d_A^{-1} + \sigma d_B^{-1})$. This is verified in \cref{fig:class_IV_eth} for $\sigma = 1$, where we fix $|A|=5$ and vary $L = 12, 14, 16$. As $L$ (and hence $d$) increases, the weak dependence on $\omega_{m'm}$ disappears and all the points $\overline{|\braket{\bm m_+ |S_A|\bm m_+}|^2}$ converge on the line $(d_A^{-1} + d_B^{-1})^2$. The local-spectrum standard deviation (square root of the variance \eqref{eq:kappa2ED}) is negligible for each data point, confirming the absence of erratically varying fluctuations $R_{\bm m, \bm m}^{(A)}$ at leading order. These results are consistent with those of Ref.~\cite{kaneko_characterizing_2020} which investigated diagonal elements. \cref{fig:class_IV_eth_sigmaMinus} in Appendix \ref{app:addFigures} reports an analogous plot for $\sigma = -1$. We have verified that the decay of the subleading fluctuations $|\braket{\bm m_\sigma |S_A|\bm m_\sigma}-(d_B+\sigma d_A)/(d+\sigma)|^2$ is compatible with a scaling $\sim d^{-2}$, as predicted from \cref{eq:ETHansatz2}. We have also verified the correctness of the diagonal ETH ansatz \eqref{eq:ETHansatz1} for the subdominant class I $\to \{(a,a),(a,a)\}$. In this case the match was highly expected from the fact that the matrix elements $\braket{m,m|S_A|m,m}$ coincide with the subsystem purity of the eigenstate $\ket m$ thanks to \cref{eq:redtranmatrix1}. All the previous conclusions remain the same for arbitrary values of $|A|$ accessible from our ED numerics.

\subsubsection{Off-diagonal elements}
\begin{figure*}[b]
    \centering
    \includegraphics[width=0.9\linewidth]{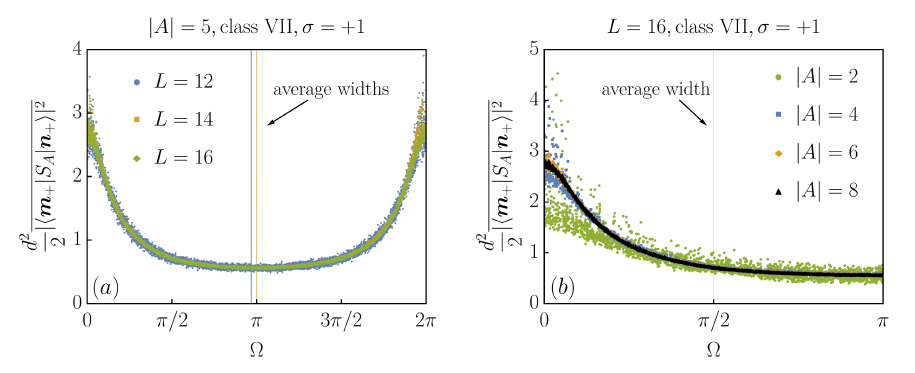}
    \caption{Plots for the rescaled local-spectrum-averaged off-diagonal elements $d^2/2\overline{|\braket{\bm m_+|S_A|\bm n_+}|^2}$ from class VII (\cref{tab:classesOfEigenstates}), as functions of $\Omega = \varphi_m + \varphi_{m'}-\varphi_n - \varphi_{n'} \ ({\rm mod} \ 2 \pi)$. For each (sub)system size the plot shows around $5$--$10\times 10^3$ sampled quartets $\bm q$. The vertical strips around the centre of the plots have widths equal to $2\pi/d \cdot 8\mathcal R$, i.e.~the size of the small $\Omega$-windows over which we perform the local-spectrum averages. \textbf{(a)} For fixed $|A|=5$ we vary $L = 12, 14, 16$ ($d=4096,16384,65536$). The local-spectrum radii are, respectively, $\mathcal R = 2, 5,5$, yielding $\approx (2\mathcal R+1)^{\mathcal M}=625,14641,14641$ terms per average ($\mathcal M=4$ for class VII). \textbf{(b)} For fixed $L=16$ we vary $|A|=2,4,6,8$. The local-spectrum radius is $\mathcal R=5$ ($14641$ terms per average). We restrict the $x$-axis to $[0,\pi]$ because of symmetry around $\pi$ (see subplot (a)).}
    \label{fig:class_VII_eth}
\end{figure*}
We turn to the leading off-diagonal class VII $\to \{(a,b),(c,d)\}$, which admits both sectors $\sigma = \pm 1$. Here the expectations from our off-diagonal ETH ansatz \eqref{eq:ETHansatz2} (and its $\sigma = -1$ generalization from the previous sections) are: (i) multiplying $\overline{M^{(\sigma)}_{\bm q}} =\overline{|\braket{\bm m_\sigma|S_A|\bm n_\sigma}|^2}$ by $d^2/2$ should yield order-one outputs that reveal the function $f_A^2$; (ii) in the limit of large subsystems $A,B$ the function $f_A$ should depend only on the main frequency $\Omega = \varphi_m + \varphi_{m'}-\varphi_n -\varphi_{n'} \ ({\rm mod} \ 2 \pi)$. In \cref{fig:class_VII_eth} we verify both points for $\sigma = +1$ by plotting $d^2/2\, \overline{M^{(+)}_{\bm q}}$ against $\Omega$. \cref{fig:class_VII_eth}(a) showcases the convergence of $d^2/2\, \overline{M^{(+)}_{\bm q}}$ to order-one data as we increase $L =12,14,16$ at fixed $|A|=5$ (note that $d^2/2 \approx 8.4 \times 10^6$ for $L = 12$ and $\approx 2.15 \times 10^9$ for $L=16$). We also observe that already at $|A|=5$ the data collapse quite well onto a single line $(f_A^2)$ that only depends on $\Omega$. Absence of collapse is more pronounced around $\Omega \approx 0 \ ({\rm mod} \ 2 \pi)$, which is the frequency-region governing the long-time decay of $\mathbb E_{\ket{\psi}}[\mathbb P_A(t)]$. This implies that in these small-frequency regions $f_A$ retains relevant dependence on its other two arguments $\omega_1,\omega_2$ from \cref{eq:defomegasfre}. We also notice that the sampled data are symmetric around $\pi$: this is expected from $|\braket{\bm m_\sigma|S_A|\bm n_\sigma}|^2=|\braket{\bm n_\sigma|S_A|\bm m_\sigma}|^2$ and the $2\pi$-periodicity of all phases $\varphi_j$. 
Importantly, the local-spectrum-average widths depicted as narrow vertical stripes around the centre of \cref{fig:class_VII_eth}(a) showcase that the averages of the erratically varying numbers $R_{\bm m , \bm n}^{(A)}$ yielding $\overline{\big|R_{\bm m, \bm n}^{(A)}\big|^2}=1$ (see \cref{eq:ETHansatz2}) happen within $\Omega$-windows that are \emph{extremely} small compared to the typical scale of variation of $f_A$. The surviving small vertical width of the data also away from $\Omega\approx0 \ ({\rm mod}\ 2 \pi)$ is due to unresolved dependence on $\omega_1,\omega_2$ (but also $R_{\bm m, \bm n}^{(A)}$ not having perfectly converged yet to its unit variance can contribute).
In \cref{fig:class_VII_eth}(b) we plot the same quantities but this time we consider $L=16$ fixed and vary $|A|=2,4,6,8$ (note that we restrict the $x$-axis to $[0,\pi]$ thanks to the symmetry around $\pi$). This plot showcases that, as expected from \cref{sec:simplificationSwap}, at small $|A|$ there is \emph{no} collapse of the data onto a single function of $\Omega$ (for example, the vertical spread of the $|A|=2$ data at each $\Omega$ is substantial), while the collapse onto a single $F(\Omega)$ gets better and better as we increase $|A|$. \cref{fig:class_VII_eth}(b) also makes apparent that the bulk of the $[0,2\pi]$ $\Omega$-dependence of $f_A$ is almost invariant under changes in $|A|$: indeed, this is the frequency region that predominantly influences the short-time decay of $\mathbb E_{\ket{\psi}}[\mathbb P_A(t)]$, which is approximately independent of $|A|$. On the contrary, in the small frequency region $\Omega \approx 0 \ ({\rm mod} \ 2\pi)$, in addition to having a lack of collapse for small $|A|$, the data converge to quite distinct values and shapes for different $|A|$. This pronounced size dependence near zero frequency is consistent with changes in the later-time relaxation of $\mathbb E_{\ket{\psi}}[\mathbb P_A(t)]$ as $|A|$ is varied. We have verified that plots analogous to \cref{fig:class_VII_eth} but for $\sigma = -1$ appear identical on the scales shown, as expected from the prediction in the previous section that the off-diagonal ansatz \cref{eq:ETHansatz2} applies identically to $\sigma = \pm 1$ at leading order in $d$. 
\begin{figure*}[t]
    \centering
    \includegraphics[width=0.9\linewidth]{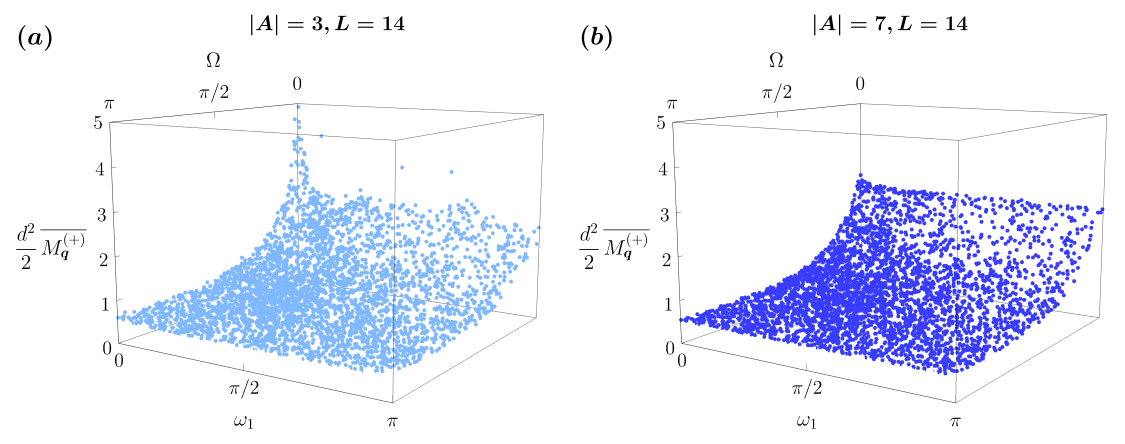}
    \caption{3D plots for rescaled off-diagonal elements $d^2/2\, \overline{M_{\bm q}^{(+)}}$ from class VII (\cref{tab:classesOfEigenstates}), as functions of $\Omega$ and $\omega_1 = \varphi_m - \varphi_n \ ({\rm mod} \ 2 \pi)$. Here $L = 14$, with local-spectrum radius $\mathcal R = 7$ ($\approx 50625$ elements per average). Thanks to the invariance under \eqref{eq:ggldgè45} we folded $\omega_1$ to $ [0,\pi]$ thanks to $(\Omega, \omega_1)\to(-\Omega \ {\rm mod}\ 2\pi, 2\pi - \omega_1)$ for $\omega_1 > \pi$. \textbf{(a)} $|A|=3$. \textbf{(b)} $|A|=7$.}
    \label{fig:classVII_eth_3D}
\end{figure*}

To better visualize the neglected dependence on $\omega_1,\omega_2$ at $\Omega \approx 0 \ ({\rm mod} \ 2\pi)$, in \cref{fig:classVII_eth_3D} we 3D plot the rescaled $d^2/2\,\overline{M^{(+)}_{\bm q}}$ as functions of $\Omega$ and $\omega_1 = \varphi_m - \varphi_n \ ({\rm mod} \ 2\pi)$, at $L = 14$. We see that while in \cref{fig:classVII_eth_3D}(b) for fairly large $|A|=7$ the emerging 2D surface has no dependence on $\omega_1$, in \cref{fig:classVII_eth_3D}(a) for $|A|=3$ an explicit dependence on $\omega_1$ appears. In particular, there is an enhancement of the rescaled matrix elements when $\Omega \approx 0 \ ({\rm mod} \ 2\pi)$ \emph{and} $\omega_1 \approx 0 \ ({\rm mod} \ 2\pi)$. This is at the origin of the strong vertical fluctuations we observe in \cref{fig:class_VII_eth}(b) for $|A|\le4$, where the $\omega_1$-dependence is not resolved. For such small $|A|$ the smooth curve $f_A^2$ can hence be reconstructed in its entirety only as a function of all the three frequencies $\Omega,\, \omega_1$ and $\omega_2$.

\begin{figure}[t]
    \centering
    \includegraphics[width=0.4\linewidth]{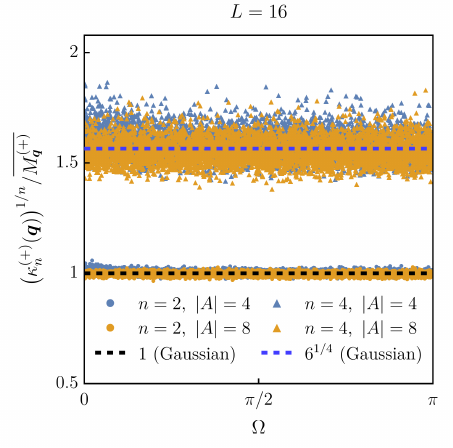}
    \caption{Rescaled 2nd and 4th cumulants from \cref{eq:rescaledcums} as functions of $\Omega$, for class VII (\cref{tab:classesOfEigenstates}). The dashed lines with values $1$ (for $\kappa_2$) and $6^{1/4}$ (for $\kappa_4$) are the predictions for $R_{\bm m, \bm n}^{(A)}$ modelled as a complex Gaussian random variable with unit variance. The data are for $L = 16$ and $|A|=4,8$. The local-spectrum radius is $\mathcal R = 5$ ($\approx 14641$ terms per average).}
    \label{fig:classVII_eth_R}
\end{figure}

From the ED output it is also possible to reconstruct the cumulants of the erratically varying numbers $Z\equiv \big|R^{(A)}_{\bm m, \bm n}\big|^2$ from \cref{eq:ETHansatz2}. Indeed, according to our ansatz, we have (note that $\kappa_4^{(\sigma)}(\bm q)>0$ in all our off-diagonal outputs)
\begin{equation}
    \frac{d^2}{2}\overline{M_{\bm q}^{(\sigma)}} \sim f_A^2 \, \overline{Z}  \ , \qquad \frac{d^2}{2}\big(\kappa_n^{(\sigma)}\big)^{1/n}(\bm q) \sim f_A^2 \, \big( \kappa_n(Z)\big)^{1/n} \ {\rm for} \ n = 2, 4\ ,
\end{equation}
where $\kappa_n(Z)$ are the cumulants of the random variable $Z$ and our definition of $f_A$ is based on $\overline Z=1$ (by construction). We can then estimate directly from the ED data the $f_A$-independent values
\begin{equation}
\label{eq:rescaledcums}
    \big(\kappa_n^{(\sigma)}(\bm q)\big)^{1/n}/\overline{M_{\bm q}^{(\sigma)}}  \sim \big(\kappa_n(Z)\big)^{1/n}/\overline Z \ , \qquad n = 2, 4 \ .
 \end{equation}
For $R_{\bm m, \bm n}^{(A)}$ compatible (at the level of its moments) with a complex Gaussian distribution of unit variance, i.e.~$R = X + i Y$ where $X$ and $Y$ are Gaussian distributed with zero mean and $1/2$ variance, we would get 
\begin{equation}
\label{eq:GaussPredZ}
    \overline Z = 1 \ ,  \quad \big(\kappa_2(Z)\big)^{1/2}/\overline Z = 1 \ , \quad \big(\kappa_4(Z)\big)^{1/4}/\overline Z= 6^{1/4} \ .
\end{equation}
In \cref{fig:classVII_eth_R} we verify the match of \eqref{eq:rescaledcums} with the Gaussian prediction \eqref{eq:GaussPredZ} for $L = 16$ and $|A|=4,8$. This supports approximate independence on $A$ of the tested normalized moments of $R_{\bm m , \bm n}^{(A)}$ for sufficiently large subsystem sizes $|A|$, as anticipated in \cref{sec:ETHforsubswap}. However, for smaller $|A|\lesssim 4$ we obtain clear deviations from the Gaussian predictions, and the moments of $R$ retain a dependence on $|A|$. This is shown in \cref{fig:classVII_eth_R_small_subsystem} in Appendix \ref{app:addFigures}. Related deviations from Gaussian statistics at small subsystem sizes were observed in Ref.~\cite{hahn_eigenstate_2024} in the probability distribution of their fundamental quantity $M_A(nmn'm')$,  which coincides with the unsymmetrized matrix element $\braket{m,m'|S_A|n,n'}$. The approximate frequency independence of the normalized cumulants in \cref{fig:classVII_eth_R}, despite the frequency dependence of the rescaled matrix elements in \cref{fig:class_VII_eth,fig:classVII_eth_3D}, is consistent with the cancellation of $f_A^2$ in \cref{eq:rescaledcums}. Together with the agreement with the Gaussian predictions, this supports the expectation that $f_A$ is effectively constant within the local-spectrum averaging windows.

We have explicitly verified via ED that \emph{all} the subleading off-diagonal classes from \cref{tab:classesOfEigenstates} respect our off-diagonal ansatz \eqref{eq:ETHansatz2} too. We also obtained evidence that $f_A$ is continuous with respect to the limit $\omega_{jk} \approx 0\  \to \ \omega_{jk} =0$ induced by coincidences among the indices $m,m',n,n'$ in these subleading classes. We discuss these cases and provide supporting plots in Appendix \ref{app:swapforsubleadingclasses}. As we explain there, the subleading off-diagonal classes are an ideal venue to investigate the underlying dependence of $f_A$ on $\omega_1,\omega_2$ (e.g.~its enhancement at small frequencies).

\subsubsection{Time-dependence of $C_A^{(+)}$ from spectral functions}
\label{sec:FTofCA+}
\begin{figure*}
    \centering
    \includegraphics[width=0.9\linewidth]{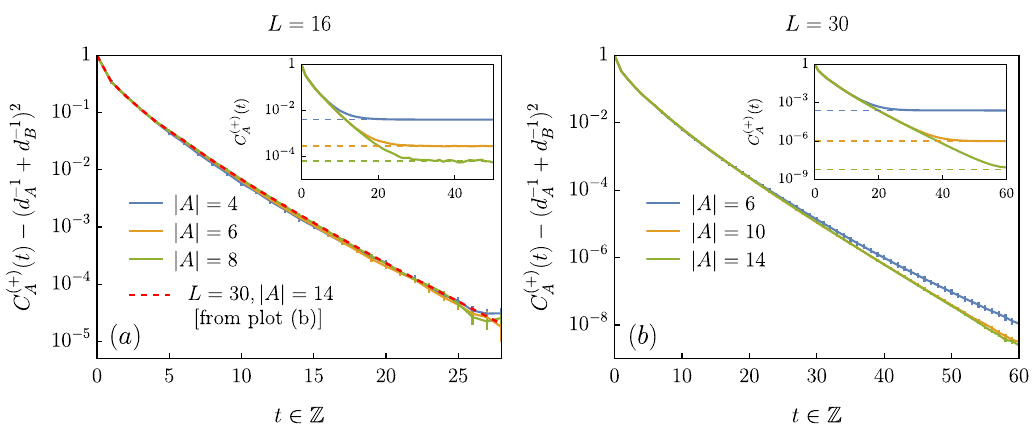}
    \caption{Log-scale plots of the evolution of $C_A^{(+)}(t)$, obtained numerically via its relation to $\mathbb E_{\ket \psi}[\mathbb P_A(t)]$ (see main text). In the main plots, we subtract from $C_A^{(+)}(t)$ its expected long-time limit $(d_A^{-1}+d_B^{-1})^2\sim C_A^{(\rm diag)}$. The error bars denote the standard error on the mean with respect to the average $\mathbb E_{\ket \psi}[\ldots]$, where we used $20$ sampled initial states $\ket{\psi}$. The insets (with no error bars) show the full $C_A^{(+)}(t)$ with no subtraction, and the horizontal dashed lines coincide with $C_A^{(\rm diag)}$. \textbf{(a)} $L = 16$ and subsystems $|A|=4,6,8$. The red dashed line plots data for $L=30,|A|=14$ from plot (b). \textbf{(b)} $L = 30$ and subsystems $|A|=6,10,14$.}
    \label{fig:C_plus_dynamics}
\end{figure*}
Here we present numerical results for the evolution of $C_A^{(+)}(t)$, reconstruct its Fourier transform, and compare it with the spectral function $f_A^2$ emerging in \cref{fig:class_VII_eth}.

For the computation of $C_A^{(+)}(t)$ we can use the fundamental identity \eqref{eq:Eprod} that relates it to the average purity $\mathbb E_{\ket \psi}[\mathbb P_A(t)]$, where the ensemble of initial product states $\ket \psi$ is defined in \cref{eq:prodEnsemble}. The strategy is therefore to sample a few $\ket \psi$, compute their exact time evolution $\ket{\psi(t)}$ without the need of ED (see \cref{sec:themodel}), from this extract $\mathbb P_A(t)$, and then average all these purities to get an estimate of $\mathbb E_{\ket \psi}[\mathbb P_A(t)]$ (and its related standard error on the mean). This program requires generating Haar random states $\ket s_{\mathcal S}$ independently in $\mathcal S=A,B$. This can be achieved by drawing complex vectors $\bm z = (z_1, \ldots, z_{\dim(\mathcal S)})$ from any distribution such that $P(\bm z)=\widetilde P(\|\bm z\|)$, and constructing the normalized $\ket{s}_{\mathcal S}=\sum_{j=1}^{\dim(\mathcal S)}z_j \ket{e_j}/\|\bm z\|$, where $\ket{e_j}$ is a computational basis in $\mathcal S$. Indeed, by construction such $\ket{s}_{\mathcal S}$ is uniformly distributed on the $(2\dim(\mathcal S)-1)$-dimensional unit sphere. An efficient method is drawing the real and imaginary parts of each $z_j$ from a real Gaussian distribution of unit variance. We then construct the initial states $\ket \psi$ as product states of such Haar vectors: $\ket{\psi}=\ket{a}_A\ket{b}_B$. 

In \cref{fig:C_plus_dynamics} we report $C_A^{(+)}(t)$ estimated via the average purity from 20 independent initial states $\ket \psi$. Unsurprisingly given the concentration-of-measure result from Appendix \ref{app:concentrationHaar}, we find that despite the fairly small number of samples and subsystem sizes the standard uncertainty on the mean is very low. Indeed, by plotting together all the 20 purities $\mathbb P_A(t)$ one can verify that they all approximately collapse on the same curve on the scale of our logarithmic or linear plots.
In \cref{fig:C_plus_dynamics} we plot both the full $C_A^{(+)}(t)$ (insets), which as expected relaxes to the asymptotic value $C_A^{(\rm diag)}$, and the time-dependent part $C_A^{(\rm off)}(t)=C_A^{(+)}(t)-C_A^{(\rm diag)}$ alone (main plots), which keeps decaying beyond the saturation time scale. As already observed before, the relaxation of $C_A^{(+)}(t)$ to $C_A^{(\rm diag)}$ implies the expected relaxation of $\mathbb E_{\ket \psi}[\mathbb P_A(t)]$ to the subsystem purity of \emph{global} Haar random states (at leading order in $d$). After a transient period, we observe both in \cref{fig:C_plus_dynamics}(a) for $L = 16$ and in \cref{fig:C_plus_dynamics}(b) for $L = 30$ the onset of exponential decay with time (before the saturation for $C_A^{(+)}$, and even beyond it for $C_A^{(\rm off)}$). This is consistent (see, e.g., \cref{eq:agrbèdpd}) with the expectation that for our Floquet unitary lacking symmetries the decay of the purity $\mathbb P_A(t)$ is exponential (ballistic growth of entanglement) with identical rate for typical initial states $\ket \psi$ due to concentration (Appendix \ref{app:concentrationHaar}). We remark that beyond the saturation time scale for the purity, where $P_A^{(\rm off)}(t)$ (see \cref{eq:sumrulgA}) keeps decaying, there could be finite-size effects (in $L$) that hide the true asymptotic decay one would observe for $|L|\to \infty, |A|\to\infty$, \emph{cf.}~the two-step relaxation phenomenon from Refs.~\cite{znidaric_twostep_2023, jonay_physical_2024}. We rule out similar effects for $C_A^{\rm (off)}(t)$ at least on the time scales of \cref{fig:C_plus_dynamics}(a), where we check the convergence of the curves for $L = 16$ and $L = 30$.  

\begin{figure}[t]
    \centering
    \includegraphics[width=0.4\linewidth]{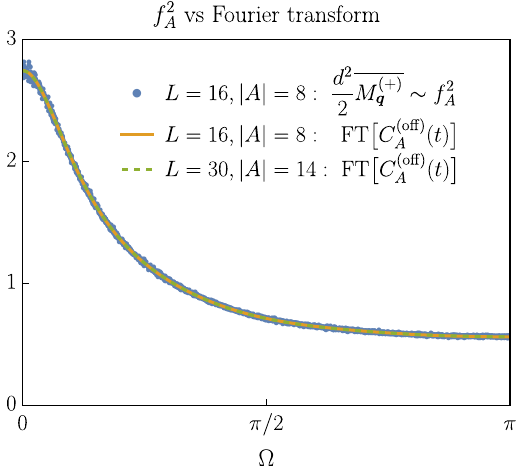}
    \caption{Comparison of (i) the function $f_A^2\approx f_A^2(\Omega)$ emerging from the data of \cref{fig:class_VII_eth}(b) for class VII, $L = 16$, $|A| = 8$, with (ii) the FT of $C_A^{(\rm off)}(t)$ (see \cref{eq:FTofCA}) for $L = 16,|A|=8$ and $L = 30,|A| = 14$. We restrict the $\Omega$-range from $[0,2\pi]$ to $[0,\pi]$ thanks to the symmetry around $\pi$.}
    \label{fig:fA_vs_Fourier}
\end{figure}

Having access for both $L = 16$ and $L = 30$ to the dynamics of $C_A^{(\rm off)}(t)$ we can verify the conclusions of \cref{sec:swapandC+}, in particular \cref{eq:CoffFourier} which relates the Fourier transform (FT) of $C_A^{(\rm off)}(t)$ to the marginal of $f_A^2(\Omega,\omega_1,\omega_2)$ over $\omega_1, \omega_2$. Remembering that $C_A^{(\rm off)}(t)$ is symmetric under time-reversal symmetry $t \to -t$, we compute its FT as
\begin{equation}
\label{eq:FTofCA}
    {\rm FT}\big[C_A^{(\rm off)}\big](\Omega) = C_A^{(\rm off)}(0) + 2\sum_{t=1}^{\infty}\cos(\Omega t)C_A^{(\rm off)}(t) \ ,
\end{equation}
where beyond the time-scales to which we have numerical access we simply set $C_A^{(\rm off)}(t) \to 0$ (this has negligible impact on the FT at the scales of our plots considering that at late times $C_A^{(\rm off)}$ is already exponentially small). Clearly, the overall shape of the FT is dominated by the short-time transient where $C_A^{(\rm off)}$ takes values of order one, while the subsequent asymptotic decay at late times only affects it on finer scales (which are however very relevant for the analytic properties of the FT in the thermodynamic limit, \emph{cf}.~\cref{sec:swapandC+}). Given that the emergent function $f_A^2$ from \cref{fig:class_VII_eth}(b) is already almost independent of $\omega_1,\omega_2$ for $|A|=8$, i.e.~$f_A^2\approx f_A^2(\Omega)$, we can use those data points to compare directly with ${\rm FT}\big[C_A^{(\rm off)}\big](\Omega)$. The comparison is shown in \cref{fig:fA_vs_Fourier} and the agreement is excellent on the scale of the plot. The agreement of the FTs from the distinct data sets $L=16,|A|=8$ and $L=30,|A| = 14$ was expected from the very good collapse already shown in \cref{fig:C_plus_dynamics}(a) (see red dashed line there).

\subsection{Quench spectral weights and $\mathbb P_A(t)$}

\subsubsection{Time-dependence of $\mathbb P_A(t)$}
\begin{figure}[t]
    \centering
    \includegraphics[width=0.85\linewidth]{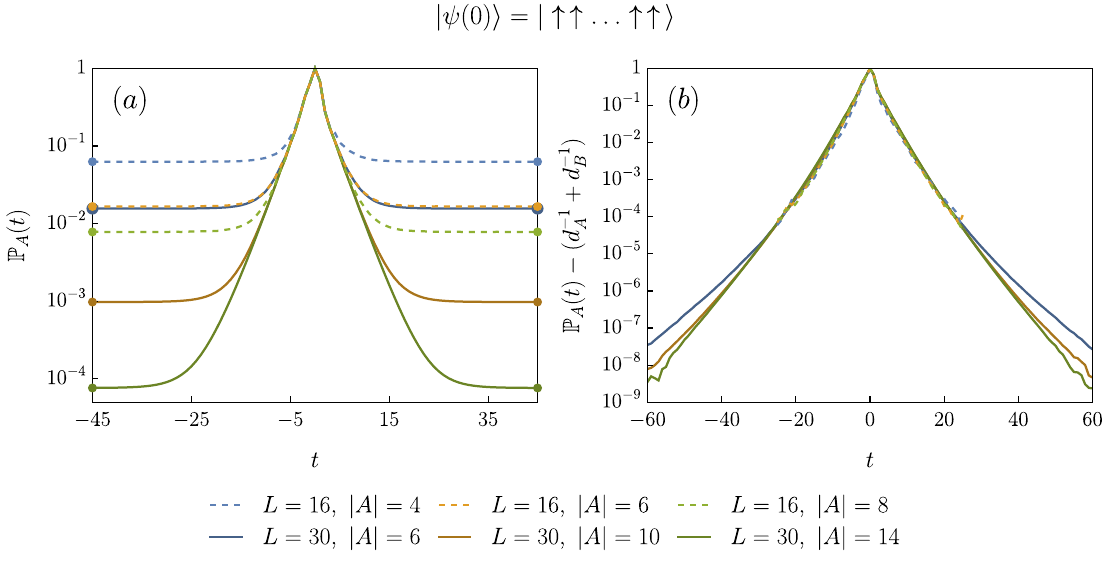}
    \caption{Log-scale plots for the exact time evolution of the purity $\mathbb P_A(t)$ from a fully polarized initial state $\ket{\psi(0)}=\ket{\uparrow \uparrow \ldots \uparrow \uparrow}$. Due to lack of time-reversal invariance for the Floquet $U_F$ in \cref{eq:UFdefinition}, we plot evolution both for $t<0$ and $t>0$. We compare numerical outputs for $L = 16$ and $|A| = 4,6,8$ with those for $L = 30$ and $|A| = 6,10,14$. \textbf{(a)} Evolution of the full $\mathbb P_A(t)$. The dots on the left and right axes correspond to the expected relaxation values $(d_A^{-1}+d_{B}^{-1})\sim \mathbb P^{(\rm diag)}_A$. \textbf{(b)} Evolution of $\mathbb P_A(t)-(d_A^{-1}+d_B^{-1})$. Here the $L=16$ data is plotted up to $|t|$ values for which noisy finite-size effects are not predominant in log-scale.}
    \label{fig:purity_dynamics_all_up}
\end{figure}

We numerically perform time evolution up to sizes $L = 30$ (see \cref{sec:themodel}) to determine the exact purity $\mathbb P_A(t)$ after the quench with $U_F^{t}$ from simple initial states $\ket{\psi(0)}$. In the log-plots of \cref{fig:purity_dynamics_all_up} we show results for $L = 16,30$ starting from a fully polarized initial state $\ket{\psi(0)}=\ket{\uparrow \uparrow \ldots \uparrow \uparrow}$, for various subsystem sizes $|A|$. As expected for our Floquet unitary, after a short initial transient the purity decays exponentially with time (ballistic growth of Rényi-2 entropy) and relaxes at late times to the Haar value $\mathbb P_A^{(\rm diag)}\sim (d_A^{-1}+d_B^{-1})$. In \cref{fig:purity_dynamics_all_up}(b) we plot $\mathbb P_A(t)-(d_A^{-1}+d_B^{-1})\sim \mathbb P_A^{(\rm off)}(t)$, which keeps decaying even beyond the saturation time scales. Here we observe a change in decay rate at late times with respect to the one at earlier times, consistent with the two-step relaxation phenomenon discussed in \cite{znidaric_twostep_2023, jonay_physical_2024} (which is a finite-size effect in $L$). Given that our Floquet unitary \eqref{eq:UFdefinition} is not time-reversal symmetric, we plot evolution both at $t>0$ and $t<0$. In \cref{fig:purity_dynamics_plus_x} of Appendix \ref{app:addFigures} we show a similar plot for a fully polarized initial state along the $x$ axis, i.e.~$\ket{\psi(0)}=\ket{\rightarrow \rightarrow \ldots \rightarrow \rightarrow}$. For the comparisons in the next sections, we have also considered evolutions from other initial states, e.g.~Néel $\ket{\uparrow \downarrow \ldots \uparrow \downarrow}$ or domain-walls $\ket{\uparrow\ldots\uparrow\downarrow\ldots\downarrow}$, although we do not report them here. We find that none of our conclusions depend on the choice of such initial states.

Similarly to what done in \cref{sec:FTofCA+}, also here we can numerically compute the FT of $\mathbb P_A^{(\rm off)}(t)$, to compare it with the spectral function emerging from the ETH analysis performed below. Given the absence of time-reversal symmetry, in this case we have
\begin{equation}
\label{eq:FTofPAoff}
    {\rm FT}\big[\mathbb P^{(\rm off)}_A\big](\Omega) = \sum_{t \in \mathbb Z}e^{-i\Omega t}\mathbb P_A^{(\rm off)}(t) \ , \qquad {\rm FT}\big[\mathbb P_A^{(\rm off)}\big] \in \mathbb C \ ,
\end{equation}
where as before we simply set $\mathbb P_A^{(\rm off)}(t) \to 0$ for times beyond what accessible from our numerics (again, this only affects very fine-grained scales of ${\rm FT}\big[\mathbb P_A^{(\rm off)}\big]$ due to the smallness of $\mathbb P_A^{(\rm off)}(t)$ at late times).

\subsubsection{Diagonal elements}
\begin{figure}[t]
    \centering
    \includegraphics[width=0.45\linewidth]{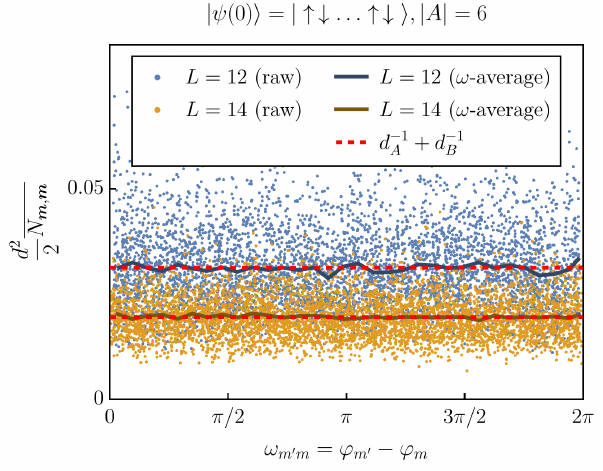}
    \caption{$(d^2/2)$-rescaled local-spectrum averages $\overline{N_{\bm m, \bm m}}=\overline{\braket{\bm m|S_A|\bm m}\braket{\bm m |\rho(0)^{\otimes 2}|\bm m}}$ from class IV for the Néel initial state $\ket{\uparrow \downarrow\ldots\uparrow\downarrow}$, as functions of $\omega_{m'm}$. The subsystem size is $|A|=6$ and we vary $L = 12,14$. The red dashed line represents the expected value $(d_A^{-1}+d_B^{-1})$. The radii are $\mathcal R=16$ and $20$  respectively for the two system sizes ($1089$ and $1681$ averaged terms per data point). Due to the considerable fluctuations surviving such local-spectrum averages, we perform additional averages in windows $\Delta \omega_{m'm}=0.2$ along the $x$-axis (continuous lines).}
    \label{fig:diagonal_outETH}
\end{figure}

For the diagonal elements $N_{\bm m, \bm m}=\braket{\bm m|S_A|\bm m}\braket{\bm m |\rho(0)^{\otimes 2}|\bm m}$, similarly to what done in \cref{sec:swapmatrixED}, we can simply draw at random quartets $\bm q = \{m,m',n,n'\}$ complying with classes I or IV from \cref{tab:classesOfEigenstates} (e.g.~$m = n$ and $m'=n'$ for class IV), and then average $N_{\bm m, \bm m}$ in a neighbourhood of $\bm q$ set by a radius $\mathcal R$. The main difference compared to the statistics of $\braket{\bm m|S_A|\bm m}$ is that $N_{\bm m, \bm m}$ features at leading order also the erratically varying numbers $\widetilde R_{\bm m, \bm m}$ from \eqref{eq:Rtilde} (due to $\braket{\bm m |\rho(0)^{\otimes 2}|\bm m}$), therefore the local-spectrum average is essential. Having already verified with ED $\braket{\bm m|S_A|\bm m}\sim (d_A^{-1}+d_B^{-1})$, then our ansatz $\overline{N_{\bm m, \bm m}}\sim 2/d^2 (d_A^{-1}+d_B^{-1})$ is correct if \cref{eq:jagkjgV1,eq:jagkjgV2} are correct (\emph{cf.}~also Refs.~\cite{haake_quantum_2010, herrmann_characterizing_2023, schmid2026genuine}). The agreement is shown in \cref{fig:diagonal_outETH} for the dominant class IV and the Néel initial state, where we plot $d^2/2\, \overline{N_{\bm m, \bm m}}$ and verify convergence to $(d_A^{-1}+d_B^{-1})$. As expected, no dependence on $\omega_{m'm}$ (the only nonzero frequency) is visible. Note that the radii $\mathcal R$ we could employ at $L = 12,14$ while retaining a fairly small local-spectrum window are not sufficient to obtain direct collapse onto $(d_A^{-1}+d_B^{-1})$, and significant fluctuations remain (this implies that in \cref{eq:jagkjgV1,eq:jagkjgV2} there are strong fluctuations around the average). We perform an additional average over windows along the $x$-axes, which reveals the very good agreement. If one had access to data for larger $L$ this additional step would not be required because $\mathcal R$ could be chosen significantly larger. We obtain similar results for other initial states, subsystem sizes $|A|$ and for class I.

\subsubsection{Off-diagonal elements and their spectral function}
As anticipated in \cref{sec:summaryPurityForSinglePsi}, the statistics of the off-diagonal elements $N_{\bm m, \bm n}=\braket{\bm m|S_A|\bm n}\braket{\bm n |\rho(0)^{\otimes 2}|\bm m}$ presents a difficulty that was absent in previous cases: due to the strongly cancelling phases predicted from the scaling with $d$ in \cref{eq:crossCorrRRtilde}, the local-spectrum averages of $N_{\bm m, \bm n}$ require \emph{at least} an order $\simeq d^2$ terms to resolve the underlying smooth function $g_A$ (previously, instead, we have always worked with large but $d$-independent averages). Indeed, in \cref{eq:crossCorrRRtilde} the terms entering the mean are expected to be of order one, and yet their average decays as $\propto 1/d$. Therefore the uncertainty on the numerically-computed mean must be at least as small as $\simeq 1/d$, which suggests the need for averaging over at least $\simeq d^2$ terms (standard error on the mean under independent-samples assumptions). Luckily, there is not need to perform an \emph{explicit} average over $\simeq d^2$ terms, which would be quite costly at $L\ge 14$ (considering that we need to perform one such average per quartet $\bm q$).

To show this, let us consider the dominant off-diagonal class VII from \cref{tab:classesOfEigenstates}. For each $\bm q=\{m,m',n,n'\}$, we want to average $N_{\bm m, \bm n}$ over quartets $\bm j=\{j_m,j_{m'},j_n,j_{n'}\}$ in a neighbourhood of $\bm q$. As before, we consider for each index $\ell = m,m',n,n'$ the ($2\mathcal R+1$) neighbouring indices in a window $j_\ell=\ell-\mathcal R,\ldots,\ell,\ldots,\ell+\mathcal R$. For class VII, this produces $(2\mathcal R+1)^4$ overall quartets $\bm j$ under the simplifying assumption that \emph{no two such windows overlap} (we can always enforce this by sampling only centres $m,m',n,n'$ that satisfy the no-overlap requirement for a given $\mathcal R$). Now let us denote with $\ket{\phi_\ell}$ the following unnormalized states 
\begin{equation}
\label{eq:glkng49fbaojvnd}
    \ket{\phi_{\ell}}=\sum_{j_\ell} c_{j_\ell}\ket{j_\ell} \, \qquad c_a = \braket{a|\psi(0)} \, , \ \ell = m,m',n,n' \ .
\end{equation}
We have shown in \cref{eq:SchmidtIdentity1,eq:SchmidtIdentity2,eq:SchmidtIdentity3} that it is very easy to compute the unsymmetrized matrix element $X_{mm',nn'}=\braket{m,m'|S_A|n,n'}$ numerically. The same remains true when in place of eigenstates $\ket{m}$ one considers (possibly unnormalized) states $\ket{\phi}$, i.e.~$X_{\phi_a\phi_b,\phi_c\phi_d}=\braket{\phi_a,\phi_b|S_A|\phi_c,\phi_d}$. Given that the states $\ket{\phi_\ell}$ from \eqref{eq:glkng49fbaojvnd} are very easy to construct numerically with ED, we can make use the of the following exact identity to obtain our average very efficiently in class VII
\begin{equation}
\label{eq:maintrickd2av}
    \overline{N_{\bm m, \bm n}} = \frac{2(X_{\phi_m\phi_{m'},\phi_n\phi_{n'}}+X_{\phi_m\phi_{m'},\phi_{n'}\phi_{n}})}{(2\mathcal R+1)^4} \ .
\end{equation}
Here we used \cref{eq:Rtilde} together with \cref{eq:fld4t39dl50d}. This avoids the need of explicit loops over the $(2\mathcal R+1)^4$ terms, making the computation of $\overline{N_{\bm m,\bm n}}$ efficient. The remaining question is how large do we need to take $\mathcal R$? As noted above we need at least $\simeq d^2$ terms in each local-spectrum average, which suggests $\mathcal R \propto d^{1/2}$ or larger. For arbitrarily large $L$, the scaling $d^{1/2}$ is perfectly fine because the spectrum is composed of $d$ eigenstates and therefore each such average involves only eigenstates in vanishingly small windows of the total spectrum $[0,2\pi]$. However, at the sizes $L\le 16$ we have access to this poses strict limitations: a window of $(2\mathcal R+1)$ elements, where $\mathcal R \propto d^{1/2}$, can represent a considerable fraction of the global window $[0,2\pi]$. This can partially spoil the locality in the spectrum needed to resolve correctly the frequency dependence of the underlying smooth function $g_A$. Below we use values of $\mathcal R$ which represent a good trade-off between having fairly local windows and having enough elements in the average to overcome the signal-to-noise-ratio problem induced by the cancelling phases. 
\begin{figure*}[t]
    \centering
    \includegraphics[width=0.85\linewidth]{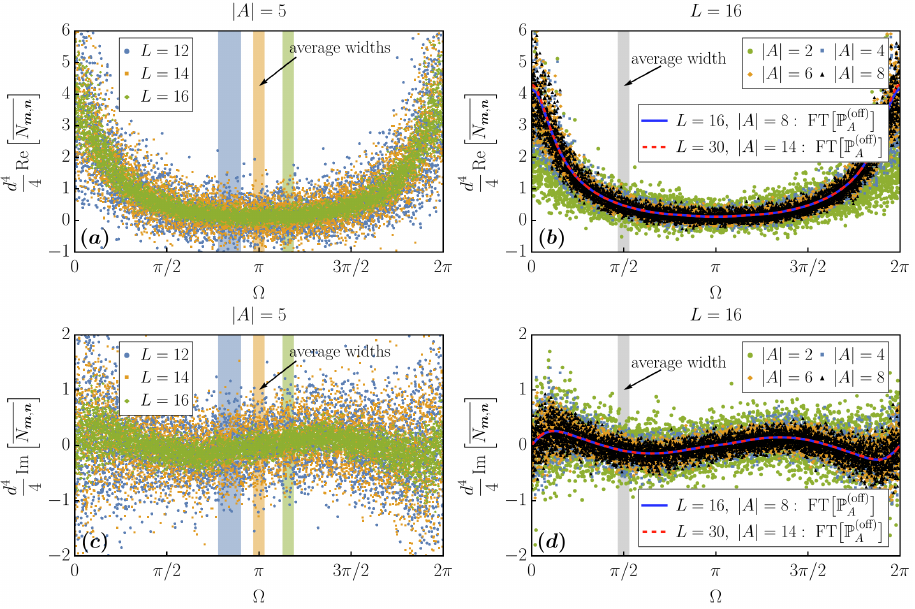}
    \caption{Real [(a) and (b)] and imaginary [(c) and (d)] parts of the $(d^4/4)$-rescaled and local-spectrum averaged quench spectral weights $N_{\bm m, \bm n}$ for class VII (\cref{tab:classesOfEigenstates}) and initial state $\ket{\psi(0)}=\ket{\uparrow \uparrow \ldots \uparrow \uparrow}$, as functions of $\Omega$. For each (sub)system size we show $5$$\times 10^3$ sampled quartets $\bm q=\{m,m',n,n'\}$. The vertical strips have widths equal to $2\pi/d \cdot 8\mathcal R$, i.e.~the size of $\Omega$-windows over which we perform the local-spectrum averages. \textbf{(a)} and \textbf{(c)}: for fixed $|A|=5$ we vary $L = 12, 14, 16$; the local-spectrum radii are, respectively, $\mathcal R = 32, 64, 256$, yielding $(2\mathcal R+1)^{4}\approx 1.8, 27.7,6926 \times 10^7$ terms per average. \textbf{(b)} and \textbf{(d)}: for fixed $L=16$ we vary $|A|=2,4,6,8$ with $\mathcal R = 256$; the continuous and dashed lines are the numerical FT \eqref{eq:FTofPAoff} obtained from the data for $\mathbb P_A^{(\rm off)}(t)$ in \cref{fig:purity_dynamics_all_up}(b).}
    \label{fig:offdiagonal_outETH_all_up}
\end{figure*}

In \cref{fig:offdiagonal_outETH_all_up} we report the rescaled values $d^4/4 \, \overline{N_{\bm m, \bm n}}$ for many sampled quartets $\bm q \in $ class VII as functions of $\Omega$, for $\ket{\psi(0)}=\ket{\uparrow \uparrow \ldots \uparrow \uparrow}$. We note that the data are (anti)symmetric around $\Omega=\pi$, as expected from $N_{\bm m, \bm n}=N^*_{\bm n, \bm m}$ and the $2\pi$-periodicity of all phases. Similarly to our analysis of the matrix elements of the swap operator, the expectations from our off-diagonal out-of-equilibrium-ETH ansatz \eqref{eq:outEqETH2} are: (i) $d^4/4 \, \overline{N_{\bm m, \bm n}}$ should converge to order one data that reveal the function $g_A$; (ii) in the limit of large subsystems $A,B$ the function $g_A$ should depend only on $\Omega$. \cref{fig:offdiagonal_outETH_all_up}(a) and (c) provide strong evidence for (i) by showcasing convergence of $d^4/4 \, \overline{N_{\bm m, \bm n}}$ (both for its real and imaginary parts) as we increase $L=12,14,16$ at fixed $|A| = 5$ (note that $d^4/4 \approx 7.04 \times 10^{13}$ for $L = 12$ and $\approx 4.6 \times 10^{18}$ for $L = 16$). Despite the presence of significantly larger fluctuations compared to the analogous plots for the swap matrix elements, we can clearly distinguish the emerge of the continuous profile of $g_A$ as a function of $\Omega$. The inability at system sizes $L \le 16$ to perform local-spectrum averages that fully overcome the signal-to-noise-ratio issue is at the origin of the stronger fluctuations. Indeed, already with the current radii $\mathcal R$ employed, the local-spectrum widths depicted as vertical stripes in \cref{fig:offdiagonal_outETH_all_up} occupy a significant fraction of the $2\pi$-circle, hence increasing $\mathcal R$ further would undermine our ability to resolve the frequency dependence of $g_A$. Even in absence of the signal-to-noise problem (which one would overcome by employing larger $\mathcal R$ values at larger sizes $L$) we would expect some fluctuations to survive at $|A|=5$ due to the dependence on $\omega_1$, $\omega_2$ that we are neglecting, exactly as we saw for the swap matrix elements. 
In \cref{fig:offdiagonal_outETH_all_up}(b) and (d) we plot the same quantities but now considering $L=16$ fixed and varying $|A|=2,4,6,8$. The data are consistent with the expected property (ii): despite the strong signal-to-noise fluctuations, we can observe that the fluctuations at $|A|=2$ are significantly larger than those at $|A|\ge 4$ although we are working with fixed $L$ and $\mathcal R$. This suggests that the degree of unresolved $\omega_1$, $\omega_2$ dependence at small $|A|$ is more significant than at larger values of $|A|$. The most important test of the meaningfulness of the emerging $g_A$ envelope is given by the comparison, again in \cref{fig:offdiagonal_outETH_all_up}(b) and (d), with the FT \eqref{eq:FTofPAoff} of the exact time-evolution data for $\mathbb P_A^{(\rm off)}(t)$ from \cref{fig:purity_dynamics_all_up}. Indeed, the match between these FT curves and the emerging profile from the ETH data is excellent.

We have verified that similar results and conclusions also hold for other initial states, e.g.~Néel and domain-wall product states. An explicit example is shown for the $x$-polarized state $\ket{\psi(0)}=\ket{\rightarrow \rightarrow\ldots\rightarrow\rightarrow}$ in \cref{fig:offdiagonal_outETH_plus_x} of Appendix \ref{app:addFigures}, where the FT curves derive from \cref{fig:purity_dynamics_plus_x}(b). Given the presence of strong signal-to-noise fluctuations, we find it difficult to reconstruct the dependence on $\omega_1$, $\omega_2$ via multidimensional plots like those in \cref{sec:swapmatrixED}: such analysis would be meaningful only at larger $L$, where bigger $\mathcal R$ values can be employed in a way compatible with spectral locality.

We note that the tricks in \cref{eq:glkng49fbaojvnd,eq:maintrickd2av} that allowed us to reconstruct a partial profile for $g_A$ are not limited to class VII and can be extended to other off-diagonal classes. In the cases of classes V and VI from \cref{tab:classesOfEigenstates}, one applies \eqref{eq:glkng49fbaojvnd} to the two unreplicated free indices, while the sum on the remaining replicated index is performed via an explicit loop. Having multiplicity $\mathcal M =3$ instead of $4$ for such classes, we are forced to require at least $\mathcal R \propto d^{2/3}$. This means that the signal-to-noise issue is worse than in class VII but still perfectly fine in the limit of $L$ arbitrarily large. Examples are shown in \cref{fig:off_diagonal_outETH_class_V} (class V) and \cref{fig:off_diagonal_outETH_class_VI} (class VI) of Appendix \ref{app:addFigures}, where we plot the real part of the rescaled $d^4/4 \, \overline{N_{\bm m, \bm n}}$ as a function of $\Omega$, again for $\ket{\psi(0)} = \ket{\uparrow \uparrow \ldots \uparrow \uparrow}$. The convergence of the data with increasing $L$, and the similar profile for the emergent $g_A$ (in particular in \cref{fig:off_diagonal_outETH_class_V}, while \cref{fig:off_diagonal_outETH_class_VI} presents significantly stronger fluctuations hiding the continuous profile of $g_A$), provide supporting evidence that indeed our off-diagonal ansatz \eqref{eq:outEqETH2} applies equally well also to the subleading off-diagonal classes.

As already discussed in \cref{sec:summaryPurityForSinglePsi}, for classes II and III from \cref{tab:classesOfEigenstates} each local average would need to include an order $\propto d$ terms, representing a finite fraction of the spectrum even for $d \to \infty$. Given that these classes do not contribute at leading order to the dynamics of the purity (and not even at the next leading order), we do not report results for them.

\section{Extension to higher Rényi entropies}
\label{sec:higherRenyi}

In this section we briefly outline how the spectral description of purity and our previous results (are expected to) extend to higher moments of the reduced density matrix. We retain the same physical assumptions on the deterministic chaotic Floquet evolution and focus on arbitrary \emph{integer} Rényi indices $k \ge 2$. The dynamical ETH ansätze below are conjectured extensions of our $k=2$ results. Their verification, together with a more detailed analysis of the $k>2$ generalizations below, is left to future work.

\subsection{Generalized purities and replica eigenstates}

The higher-order generalizations of the purity and Rényi-2 entropy are
\begin{equation}
\label{eq:higherRenyiMoments}
    \mathbb P_A^{(k)}(t)=\Tr_A[\rho_A(t)^k]\ ,\qquad
    \mathbb S_A^{(k)}(t)=-\frac{1}{k-1}\ln\mathbb P_A^{(k)}(t) \ ,
\end{equation}
so that $\mathbb P_A^{(2)}=\mathbb P_A$ and $\mathbb S_A^{(k)}$ represents the Rényi-$k$ entropy. Among all the $k!$ permutations $\sigma$ in the symmetric group of degree $k$, consider the $(k-1)!$ full cyclic permutations $\pi \in \mathcal C_k = \{k\text{-cycles in }S_k\}$, e.g.~$(12\ldots k)$. For example, for $k = 3$ they are $(123)$ and $(132)$. Let $P_{\pi}^{(A)}$, $\pi \in \mathcal C_k$, be the representation that cyclically permutes the $k$ copies of $A$ according to $\pi$, leaving the copies of $B$ untouched (the action of the permutation on the replicated Hilbert space is the same of \cref{eq:howpermacts}). Then the replica identity \eqref{eq:mainidentity} generalizes to the following standard identity \cite{ekert_direct_2002}
\begin{equation}
\label{eq:higherRenyiReplica}
    \mathbb P_A^{(k)}(t) =\Tr_{\mathcal H^{\otimes k}}\big[P_{\pi}^{(A)}\rho(t)^{\otimes k}\big] =\bra{\psi(t)}^{\otimes k}P_{\pi}^{(A)}\ket{\psi(t)}^{\otimes k} \ , \qquad \pi \in \mathcal C_k \ ,
\end{equation}
irrespective of the $\pi \in \mathcal C_k$ chosen. \cref{eq:higherRenyiReplica} shows that $S_A$ from $k=2$ is generalized to $k>2$ by \emph{any} $P^{(A)}_\pi$ with $\pi \in \mathcal C_k$. As before, only the fully symmetric replica sector $\mathcal H_+^{\otimes k}$ has nonzero overlap with $\ket{\psi(0)}^{\otimes k}$ in \eqref{eq:higherRenyiReplica}. Its projector and dimension are \cite{harrow_church_2013}
\begin{equation}
\label{eq:higherRenyiSector}
    \mathsf P_+^{(k)}=\frac{1}{k!}\sum_{\sigma \in S_k}P_\sigma \ ,\qquad d_+^{(k)}=\dim\mathcal H_+^{\otimes k}=\binom{d+k-1}{k}\sim\frac{d^k}{k!} \ ,
\end{equation}
where $P_\sigma=P_\sigma^{(A)}P_\sigma^{(B)}$ permutes \emph{global} $AB$ replicas.
Similarly to before, we denote by $\ket{\bm m} \in \mathcal H_+^{\otimes k}$ the normalized symmetrization of $\ket{m_1,\ldots,m_k}$, with $m_1\leq\cdots\leq m_k$. Its Floquet eigenphase under $U_F^{\otimes k}$  is $\Phi_{\bm m}=\sum_{j=1}^k\varphi_{m_j} \ ({\rm mod} \ 2 \pi)$. 

We note that instead of the individual $P_\pi^{(A)}$ with $\pi \in \mathcal C_k$, we can also define another generalization of $S_A$ to $k>2$, by taking an average over $\mathcal C_k$
\begin{equation}
\label{eq:higherRenyiProjected}
    S_A^{(k)} \equiv \frac{1}{(k-1)!}\sum_{\pi \in \mathcal C_k} P_\pi^{(A)} \ .
\end{equation}
Such $S_A^{(k)}$ can be used in \eqref{eq:higherRenyiReplica} in place of any of the individual $P_\pi^{(A)}$.
Unlike $P_\pi^{(A)}$, $S_A^{(k)}$ features some properties very similar to those of $S_A$ from $k=2$: (i) $[S_A^{(k)},P_\sigma]=0 \ \forall \ \sigma \in S_k$, which implies that $S_A^{(k)}$ does \emph{not} connect $\mathcal H_+^{\otimes k}$ to other replica symmetry sectors ($P_\pi^{(A)}$ generally does); (ii) $S_A^{(k)}$ is Hermitian but not unitary, while each $P_\pi^{(A)}$ is unitary but not Hermitian ($S_A$ for $k=2$ was both). We therefore have many possible choices for an appropriate counterpart of \eqref{eq:mainLehmanSum} (here $\Omega=\Phi_{\bm m}-\Phi_{\bm n}$)
\begin{equation}
\label{eq:higherRenyiLehmann}
    \mathbb P_A^{(k)}(t) =\sum_{\bm m,\bm n}e^{i\Omega t} \braket{\bm m|\mathbb X_A^{(k)}|\bm n} \braket{\bm n|\rho(0)^{\otimes k}|\bm m}\ ,\qquad  \mathbb X_A^{(k)} = P_\pi^{(A)} \ (\pi \in \mathcal C_k), \, \text{or } S_A^{(k)}\ .
\end{equation}
All such choices of $\mathbb X_A^{(k)}$ yield identical matrix elements $\braket{\bm m |\mathbb X_A^{(k)}|\bm n}$.

\subsection{Autocorrelation and averaging over initial states}

The natural counterpart of the swap autocorrelation $C_A^{(+)}(t)$ from $k=2$ is
\begin{align}
\label{eq:higherRenyiCorrelator}
    C_A^{(k)}(t)=\frac{(k-1)!}{d_+^{(k)}}
    \Tr_{\mathcal H_+^{\otimes k}}\big[S_A^{(k)}(t)S_A^{(k)}\big] =\frac{(k-1)!}{d_+^{(k)}}\sum_{\bm m,\bm n}
    e^{i\Omega t}\big|\braket{\bm m|S_A^{(k)}|\bm n}\big|^2
    \ ,
\end{align}
where $S_A^{(k)}(t)$ is the Heisenberg evolution of $S_A^{(k)}$ via $(U_F^t)^{\otimes k}$. For $k = 2$ we recover $C_A^{(2)}(t)=C_A^{(+)}(t)$. Note that the use of $\mathbb X_A^{(k)} \to S_A^{(k)}$ is convenient here because the latter is block diagonal in the replica sectors, and therefore $S_A^{(k)}(t)$ preserves $\mathcal H_+^{\otimes k}$ at all times (the use of any of the other possible $\mathbb X_A^{(k)}$ would require insertion of additional projectors $\mathsf P_+^{(k)}$ in the expressions).
Consider the same initial-state ensemble as in \eqref{eq:Eprod}: $\ket\psi=\ket a_A\ket b_B$, with independent Haar distributions in $A$ and $B$. The following Haar moment identity \cite{harrow_church_2013} generalizes the $k=2$ case reported in Appendix \ref{sec:appRMTproduct}
\begin{equation}
\label{eq:aldhdalad54}
    \mathbb E_{\rm Haar(A)} \left[(\ket a\bra a)^{\otimes k}\right] = \frac{\sum_{\sigma \in S_k}P_\sigma^{(A)}}{(d_A)_k} \ ,
\end{equation}
where $(x)_k=x(x+1)\cdots(x+k-1)$. The analogous identity holds for $B$. This leads to
\begin{equation}
    \mathbb E_{\ket \psi}[\rho(0)^{\otimes k}] = \frac{k!}{(d_A)_k(d_B)_k}\left( \sum_{\sigma \in S_k} P_\sigma^{(A)} \right) \mathsf P_+^{(k)} \ .
\end{equation}
Combining the previous identity with \eqref{eq:higherRenyiReplica} (where we use $S_A^{(k)}$ in place of $P_\pi^{(A)}$), we obtain
\begin{equation}
\label{eq:higherRenyiProductAverage}
    \mathbb E_{\ket\psi}[\mathbb P_A^{(k)}(t)]=\frac{\eta_k}{d_+^{(k)}}\Tr_{\mathcal H_+^{\otimes k}}\left[S_A^{(k)}(t)\sum_{\sigma\in S_k}P_\sigma^{(A)}\right] \ ,\qquad \eta_k=\frac{(d)_k}{(d_A)_k(d_B)_k}\ .
\end{equation}
Using the definition \eqref{eq:higherRenyiCorrelator} together with \eqref{eq:higherRenyiProjected}, we get
\begin{equation}
    \mathbb E_{\ket\psi}[\mathbb P_A^{(k)}(t)] = \eta_k \, C_A^{(k)}(t) + R^{(k)}(t) \ ,
\end{equation}
where $R^{(k)}$ is equal to the right-hand side of the first equality in \eqref{eq:higherRenyiProductAverage} but with $\sigma \in S_k$ replaced by $\sigma \in S_k, \sigma \notin \mathcal C_k$ (i.e.~the sum excludes the $k$-cycles already included in $C_A^{(k)}$). Under the assumption of locality of the evolution $U_F$ (\emph{cf}.~\cref{sec:simplification}), the crucial simplification comes in the limit of large subsystems $V_{\min }\gg 1$: for any fixed $k$ and any fixed time $t$, one finds
\begin{equation}
\label{eq:higherRenyiTypicalBridge}
    \mathbb E_{\ket\psi}[\mathbb P_A^{(k)}(t)] = C_A^{(k)}(t) + \mathcal O(e^{-cV_{\rm \min}}) \ , \qquad c> 0 \,  .
\end{equation}
This generalizes \cref{eq:agrbèdpd} (and its $D>1$ analogues) from $k=2$. The $\mathcal O(e^{-cV_{\rm \min}})$ suppression of $R^{(k)}(t)$ can be understood as follows. At any fixed time $t$, as a consequence of the Lieb-Robinson bound \cite{lieb_finite_1972, bravyi_liebrobinson_2006} or of exact light cones in local circuit models, the vast majority of the support of the operator $S_A^{(k)}(t)$ in the limit $V_{\rm min}\gg 1$ is the same as for $S_A^{(k)}$ (up to controllably small Lieb-Robinson tails), with deviations only arising close to the boundary $\partial$ between $A$ and $B$. Let us denote as $D_A$ and $D_B$ the dimensions of the large subregions of $A$ and $B$ where $S_A^{(k)}(t)$ is still equal to $S_A^{(k)}$ (the crucial point being that $D_A$ and $D_B$ are at least exponentially large in $V_{\rm min}$). Now choose a $k$-cycle $\pi \in \mathcal C_k$ from the expansion of $S_A^{(k)}$, a permutation $\sigma$ from the sum in the square bracket in \eqref{eq:higherRenyiProductAverage} and a global permutation $\sigma'$ from the expansion of $\mathsf P_+^{(k)}$ (which is implicitly present in \eqref{eq:higherRenyiProductAverage} because the trace is already restricted to $\mathcal H_+^{\otimes k}$). The contribution of such term in \eqref{eq:higherRenyiProductAverage} comes from a \emph{bounded} boundary trace (carrying the time dependence) multiplied by
\begin{equation}
    D_A^{\text{cycles}(\pi \sigma \sigma')-k}D_B^{\text{cycles}(\sigma')-k} \ .
\end{equation}
Given that $\text{cycles}(\sigma)\le k \ \forall \ \sigma$ with equality only for the identity permutation $\sigma = e$, the only way to avoid a large suppression is to have $\sigma'=e$ and $\sigma = \pi^{-1} \in \mathcal C_k$. Given that $R^{(k)}(t)$ involves only $\sigma \notin \mathcal C_k$, the previous equality cannot be satisfied and hence every term in $R^{(k)}(t)$ contains at least one factor $D_A^{-1}$ or $D_B^{-1}$.

The concentration-of-measure arguments from Appendix \ref{app:concentrationHaar}, based on Ref.~\cite{hayden_aspects_2006}, apply also to $k>2$. This implies that also here, at any fixed time, deviations of $\mathbb P_A^{(k)}(t)$ from its ensemble average $\mathbb E_{\ket\psi}[\mathbb P_A^{(k)}(t)]$ become exponentially unlikely with $V_{\rm min}$. Together with \eqref{eq:higherRenyiTypicalBridge}, this gives $\mathbb P_A^{(k)}(t)=C_A^{(k)}(t)+\mathcal O(e^{-cV_{\rm min}})$ in probability for $V_{\rm min}\gg 1$ (\emph{cf}. \cref{eq:ncxmbvmxcx}).

\subsection{Expected ETH structure}

Similarly to the case $k = 2$, the diagonal ansatz for the matrix elements of $\mathbb X_A^{(k)}$ should be governed by the generalized purity $\mu_A^{(k)}$ of global Haar-random pure states, i.e.~we expect (\emph{cf}.~also \cite{kaneko_characterizing_2020})
\begin{equation}
\label{eq:higherRenyiDiagonal}
    \overline{\braket{\bm m|\mathbb X_A^{(k)}|\bm m}} \sim \mu_A^{(k)} \ ,\qquad \mu_A^{(k)}= \mathbb E_{\rm Haar}\left[\Tr_A[\rho_A^k]\right] =\frac{\Tr_{\mathcal H_+^{\otimes k}}[\mathbb X_A^{(k)}]}{d_+^{(k)}} \ .
\end{equation}
The exact Haar identity $\mathbb E_{\rm Haar}[\rho^{\otimes k}]=\mathsf P_+^{(k)}/d_+^{(k)}$~\cite{harrow_church_2013} (the same used in \cref{eq:aldhdalad54}) gives the last equality, as well as the compact expression
\begin{equation}
\label{eq:higherRenyiHaarMoment}
    \mu_A^{(k)}=\frac{1}{(d)_k}\sum_{\sigma\in S_k}
    d_A^{{\rm cycles}(\pi\sigma)}d_B^{{\rm cycles}(\sigma)} \ , \qquad \pi \in \mathcal C_k\ ,
\end{equation}
where we have used the substitution $\mathbb X_A^{(k)} \to P_\pi^{(A)}$. For $k=2$, this gives $\mu_A^{(2)}=(d_A+d_B)/(d+1)$, reproducing our diagonal RMT prediction from Appendix \ref{sec:appRMTmean}. Given that for fixed $d_A$ with $d_B\to\infty$ one obtains $\mu_A^{(k)}\to d_A^{1-k}$, \cref{eq:higherRenyiDiagonal} is consistent with \emph{local} relaxation to the infinite temperature ensemble for any $k$.

For the off-diagonal elements, we restrict our attention to the generalization of class VII (\cref{tab:classesOfEigenstates}): all $2k$ indices in $\ket{\bm m}$ and $\ket{\bm n}$ are \emph{distinct}. We use a complete set of independent phase differences $(\Omega,\boldsymbol\omega)$, with $\Omega=\Phi_{\bm m}-\Phi_{\bm n}$ and $\boldsymbol\omega$ containing $2k-2$ auxiliary frequencies. Similarly to what done in \cref{sec:swapandC+}, they are chosen as periodic coordinates with unit absolute Jacobian after removing the common eigenphase, so that the integration measure remains uniform. The analogues of \eqref{eq:ETHansatz2} and \eqref{eq:outEqETH2} are then
\begin{align}
\label{eq:higherRenyiVariance1}
    \overline{\big|\braket{\bm m|\mathbb X_A^{(k)}|\bm n}\big|^2} &\simeq\frac{k}{d^k}\big[f_A^{(k)}(\Omega,\boldsymbol\omega)\big]^2 \ ,\\
\label{eq:higherRenyiVariance2}
    \overline{\braket{\bm m|\mathbb X_A^{(k)}|\bm n}\braket{\bm n|\rho(0)^{\otimes k}|\bm m}}&\simeq\frac{(k!)^2}{d^{2k}}g_A^{(k)}(\Omega,\boldsymbol\omega)\ .
\end{align}
As before, the bar denotes local-spectrum average for a fixed Floquet operator. Unlike the $k=2$ case, for certain choices of $k>2$ and finite $d_A$ it is possible that $f_A^{(k)}$ and $g_A^{(k)}$ exactly vanish (due to $\braket{\bm m |\mathbb X_A^{(k)}|\bm n}=0$ for the present all-distinct class). When these have nonzero limits, the scalings $d^{-k}$ and $d^{-2k}$ in \cref{eq:higherRenyiVariance1,eq:higherRenyiVariance2} allow the present all-distinct class to contribute at order one to both $C_A^{(k)}(t)$ and $\mathbb P_A^{(k)}(t)$. In this regime, for a delocalized initial state, the product in \eqref{eq:higherRenyiVariance2} has typical magnitude $d^{-3k/2}$ but proposed mean $d^{-2k}$. This would generalize the strong phase cancellations found for the purity to any $k > 2$. 

For $k>2$, one cannot in general assume that the scalings in \cref{eq:higherRenyiVariance1,eq:higherRenyiVariance2} extend unchanged to off-diagonal classes with coincident indices. A systematic RMT analysis of these class-dependent scalings is left for future work.
The most important implication is that, unlike for $k=2$, some of the classes with coincident indices can contribute at leading order to the Lehmann expansions. However, it remains possible that in the limit of asymptotically large subsystems $V_{\min} \to \infty$ the all-distinct class represents the leading contribution. This is an important possibility to investigate.

\subsection{Connections with full-ETH}

The operator expansion \eqref{eq:SAexpanded} used for the subsystem swap has a direct generalization. Let $\{O_\mu\}_{\mu=1}^{d_A^2}$ be the same Hermitian basis in $A$, normalized by $\Tr_A[O_\mu O_\nu]=\delta_{\mu\nu}$. Let us choose the simplest $\pi \in \mathcal C_k$, namely $\pi^* = (12\ldots k)$, and set $\mathbb X_A^{(k)} = P_{(\pi^*)^{-1}}^{(A)}$. Then
\begin{equation}
\label{eq:higherRenyiOperatorExpansion}
    \mathbb X_A^{(k)}=\sum_{\mu_1,\ldots,\mu_k=1}^{d_A^2}
    T_{\mu_1\ldots\mu_k}^{(k)}\, O_{\mu_1}\otimes\cdots\otimes O_{\mu_k}\ ,\qquad
    T_{\mu_1\ldots\mu_k}^{(k)} = \Tr_A[O_{\mu_1}\cdots O_{\mu_k}]\ .
\end{equation}
This is elementary to prove similarly to what done in \cref{eq:klhrir49gddc}, but now using the intertwining effect of $P_{(\pi^*)^{-1}}^{(A)}$ on $k$ replicas. A matrix element $\braket{\bm m|\mathbb X_A^{(k)}|\bm n}$ is therefore a sum of products of $k$ single-replica matrix elements. To connect with the full-ETH framework \cite{foini_eigenstate_2019,pappalardi_eigenstate_2022}, as before we keep $d_A$ fixed while $d\to\infty$. Let us restrict again, for simplicity, to the \emph{all-distinct} class of $2k$ indices in $\ket{\bm m}$ and $\ket{\bm n}$.
For the statistics relevant for \eqref{eq:higherRenyiVariance1}, expanding the symmetrized eigenstates into their unsymmetrized components using $\ket{\bm m} \propto \mathsf P_+^{(k)}\ket{m_1, \ldots, m_k}\propto\sum_{\sigma \in S_k}P_\sigma\ket{m_1,\ldots,m_k}$, and using \eqref{eq:higherRenyiOperatorExpansion}, we obtain that $|\braket{\bm m |\mathbb X_A^{(k)}|\bm n}|^2$ equals a sum of terms of the form
\begin{equation}
    \prod_{j=1}^k \big(O_{\mu_{\tau_1(j)}}\big)_{m_jn_{\tau_2(j)}} \big(O_{\nu_{\tau_3(j)}}\big)_{n_jm_{\tau_4(j)}} \ .
\end{equation}
The expansion involves two sums over indices $\{\mu_1, \ldots, \mu_k\}$ and $\{\nu_1,\ldots,\nu_k\}$ from \eqref{eq:higherRenyiOperatorExpansion} and sums over permutations $\tau_1, \ldots, \tau_4 \in S_k$. Similarly to what seen for $k=2$, the crucial point is that once we reintroduce the local-spectrum average in this expansion, the sum is dominated by products of 2-ETH terms (see \cref{eq:nETHFloquet}) like
\begin{equation}
\label{eq:adghlj4ca222}
    \overline{\prod_{j=1}^k \big(O_{\mu_{\tau_1(j)}}\big)_{m_jn_j}\big(O_{\nu_{\tau_3(j)}}\big)_{n_jm_j} } \sim \frac{1}{d^k}
    \prod_{j=1}^k h_{\mu_{\tau_1(j)}\nu_{\tau_3(j)}}^{(2)}(\varphi_{m_j}-\varphi_{n_j}) \ ,
\end{equation}
which came from the choice $\tau_2 = \tau_4 = e$ and where we have used the full-ETH factorization rule \eqref{eq:factorizationFullETH}. The other 2-ETH-only contributions come from all the terms in which $\tau_4 = \tau_2^{-1}$, and take a form identical to \eqref{eq:adghlj4ca222} aside for a relabelling of the eigenstate indices. Other contributions to the expansion coming from terms with $\tau_4 \neq \tau_2^{-1}$ necessary involve at least one $n$-ETH term with $n>2$, which induces an overall scaling of $d^{-k-q}$, $q\ge 1 \in \mathbb Z$. In the present fixed-$d_A$ limit, the sum over $\simeq (k!)^4d_A^{4k}$ such terms cannot contrast the additional $d^{-q}$ suppression, and therefore the $2$-ETH terms \eqref{eq:adghlj4ca222} dominate the expansion. Note that the scaling in \eqref{eq:adghlj4ca222} is consistent with the $d^{-k}$ decay in the generalized ansatz \eqref{eq:higherRenyiVariance1}. In the present limit of $d_A$ finite, $d \to \infty$, \eqref{eq:adghlj4ca222} also allows us to interpret $\big[f_A^{(k)}\big]^2$ as a sum over products $ T_{\mu_1\ldots\mu_k}^{(k)} T_{\nu_1\ldots\nu_k}^{(k)}\prod_{j=1}^k h_{\mu_{\tau_1(j)}\nu_{\tau_3(j)}}^{(2)}(\varphi_{m_j}-\varphi_{n_j})$ involving 2-ETH functions.

For an individual quench, the same expansion gives the exact identity
\begin{equation}
\label{eq:higherRenyiSingleReplicaQuench}
    \mathbb P_A^{(k)}(t)=
    \sum_{\mu_1,\ldots,\mu_k=1}^{d_A^2}
    T_{\mu_1\ldots\mu_k}^{(k)}
    \prod_{j=1}^k\bra{\psi(t)}O_{\mu_j}\ket{\psi(t)}
    \ .
\end{equation}
Each factor has a Lehmann representation involving $(O_\mu)_{mn}\rho(0)_{nm}$, whose local-spectrum average is given by \cref{eq:outETHsinglerep} in terms of a spectral function $g_{O_\mu,\psi(0)}$ multiplied by a prefactor $d^{-2}$. In the present finite-$d_A$ limit, the conjectured factorization for $k$ pairs with all indices distinct (which generalizes \cref{eq:dgkal4gglv}) would give $d^{-2k}$, motivating the scaling in \eqref{eq:higherRenyiVariance2}, while also connecting $g_A^{(k)}$ to products of $g_{O_\mu,\psi(0)}$ functions.

For a complementary connection between spectral representation for arbitrary-$k$ Rényi entropies and the full-ETH framework see Ref.~\cite{jindal_generalized_2024}.

\subsection{Simplifications for large subsystems}

The Fourier coefficients of the spectral functions $f_A^{(k)}(\Omega,\bm \omega)$ and $g_A^{(k)}(\Omega,\bm \omega)$ encode the all-distinct-class contributions to $C_A^{(k)}(t)$ and $\mathbb P_A^{(k)}(t)$, generalizing \cref{eq:CoffFourier,eq:PAFourier} to Rényi indices $k>2$. We conjecture that the frequency-dependence simplification derived for $k=2$ in \cref{sec:simplification} applies to all $k>2$ (under the same assumptions used there). Explicitly
\begin{equation}
\label{eq:higherRenyiCollapse}
\big[f_A^{(k)}(\Omega,\boldsymbol\omega)\big]^2 = \mathsf F_A^{(k)}(\Omega) + \mathcal{O}(e^{-\alpha V_{\rm min}}) \ , \qquad g_A^{(k)}(\Omega, \bm \omega) = \mathsf G_A^{(k)}(\Omega) + \mathcal{O}(e^{-\beta V_{\rm min}}) \ ,
\end{equation}
with $\alpha,\beta > 0$ constants. Although an explicit generalization of the steps from \cref{sec:simplification} are beyond the scope of the present work, it is clear the core of the argument, namely the locality of the time evolution together with bulk spectral dephasing or suppression of bulk return amplitudes, can be applied similarly to before. Also here, for the second equation in \eqref{eq:higherRenyiCollapse}, subextensive entanglement of the initial state $\ket{\psi(0)}$ is a needed assumption.

If, at fixed $k\ge 2$ and fixed time $t$, the all-distinct class represents the leading contribution in the limit of $V_{\rm min} \to \infty$, then \eqref{eq:higherRenyiCollapse} would imply that $\mathsf F_A^{(k)}$ and $\mathsf G_A^{(k)}$ encode alone the dynamics of the Rényi-$k$ entropies per boundary area (i.e.~the $k>2$ generalizations of \cref{eq:entropyperarea}).

\section{Discussion}
\label{sec:discussion}

In this work, we have introduced simple ETH-like ansätze (\cref{eq:ETHansatz1,eq:ETHansatz2,eq:outEqETH1,eq:outEqETH2}) relevant for the description of Rényi-2 dynamics and its late-time saturation in chaotic Floquet systems. These ansätze feature slowly varying smooth functions that emerge from the statistics of matrix elements in the eigenstate basis, and govern the evolution of subsystem purity. Their functional form is very similar to the one of the standard ETH statements in \cref{eq:FloquetStETH,eq:outETHsinglerep}, enabling a description of observable and entanglement evolution within a common spectral perspective. In the limit of large subsystems, we have argued analytically and verified numerically that the spectral functions develop a striking simplification in their frequency dependence (see \cref{eq:adjkdafjh1,eq:adjkdafjh1bis,eq:adjkdafjh2,eq:adjkdafjh2bis}). In \cref{sec:higherRenyi} we have also discussed how all our results for Rényi-2 (are expected to) generalize to arbitrary integer-index Rényi entropies.

Our results warrant several directions for future work. (i) A proper derivation and verification of generalized ETH statements for higher Rényi entropies is desirable. One of the most pressing questions is about the role of coincident-indices classes of multireplica eigenstates (which in the case of purity are subleading, but might cease to be so for higher Rényi index). (ii) Generalizations to static-Hamiltonian dynamics and Floquet systems with local conservation laws would complement our ansätze by incorporating energy dependence and symmetry resolution into the corresponding ETH statements. The latter is particularly interesting, as exact theoretical results for symmetry-resolved dynamics are currently available only for integrable models \cite{parez_quasiparticle_2021, parez_exact_2021, bertini_nonequilibrium_2023}, despite the existence of experimental observations   \cite{lukin_probing_2019,vitale_symmetryresolved_2022, neven_symmetryresolved_2021}. In such physical scenarios, Rényi entropies with index $k>1$ are known to grow sub-ballistically with time after quenches, as a consequence of the underlying hydrodynamic modes \cite{rakovszky_subballistic_2019, huang_dynamics_2020, zhou_diffusive_2020, znidaric_entanglement_2020, rakovszky_entanglement_2021, znidaric_reply_2021, lastres2026geometry}. It would be hence valuable to connect this to specific differences in the structure of their ETH ansätze. (iii) The existence of slowly varying ETH functions that fully encode Rényi entanglement growth sets the stage for future applications of spectral filter methods \cite{luo_probing_2024} to extract them efficiently by means of classical computational resources. Indeed, the crucial requirement for the success of such schemes is that the spectral functions to be reconstructed have very slow variations compared to level-spacing scales, which we have demonstrated. Conceptually this implies that “broadened” eigenstates, which possess drastically lower entanglement compared to individual exact eigenstates \cite{banuls_entanglement_2020}, could be employed to reconstruct the spectral functions from their broadened matrix-element statistics.


\begin{acknowledgments}
We thank John Chalker, Fabian Essler, Laura Foini, Dominik Hahn, Jorge Kurchan and Lev Vidmar for useful discussions. We also thank Angelo Russotto and Filiberto Ares for valuable comments during the early stages of this project. This work was supported by the European Research Council (ERC) through the Advanced Grant MOSE No. 101199196.\\
\end{acknowledgments}

\noindent \textbf{Use of generative AI statement}  Generative AI systems from the GPT-5.6 and GPT-6 model families were used at various stages of this research, including for extensive literature searches and code writing. They also: provided substantial insights and guidance in developing the analytical arguments of \cref{sec:simplification} and Appendix~\ref{app:concentrationHaar}; helped significantly shorten an earlier derivation of the RMT results in Appendix \ref{appendix:RMT}; contributed important observations for the generalization of our results to higher Rényi entropies (\cref{sec:higherRenyi}). All code was verified by the authors, and the manuscript was entirely written by the authors.

\appendix

\section{Concentration of purity for Haar product states}
\label{app:concentrationHaar}

Let us denote $\mu_t \equiv \mathbb E_{\ket{\psi}}[\mathbb P_A(t)]$, where $\mathbb E_{\ket \psi}[\ldots]$ is the average over the ensemble  \eqref{eq:prodEnsemble} of initial product states. Let us denote as $\mathbb P_t(a,b)$ the subsystem purity $\mathbb P_A(t)$ from a specific $\ket \psi = \ket{a}_A \ket{b}_B$ in this class. We now prove
\begin{equation}
\label{eq:targetLevy}
    {\rm Pr}\big(|\mathbb P_t(a,b) - \mu_t|>\varepsilon\big) \le 8 \exp\big(-c d_{\min} \varepsilon^2\big) \ ,
\end{equation}
where $\varepsilon>0$ is an arbitrarily small constant, $c>0$,  $d_{\min} = \min(d_A,d_B)$ and ${\rm Pr}$ denotes the probability induced by the two independent Haar distributions in $A$ and $B$. This bound implies that, at every fixed time, the purity of a randomly sampled initial state is arbitrarily close to its ensemble average with probability approaching one as $d_{\min}\to\infty$. 

We need two facts, one detailed and one proved, in Ref.~\cite{hayden_aspects_2006} (we denote the ordinary vector norm as $\|v\|=\sqrt{\braket{v|v}}$):
\begin{enumerate}
    \item \emph{Lévy's lemma on concentration for Haar vectors}. Let $x\in \mathbb C^d$ be a Haar-random normalized vector. Consider a global real Lipschitz function $f$, i.e.~$|f(x)-f(y)|\le \eta \, \|x-y\| \ \ \forall \ x,y$, where $\eta > 0$ is a constant. Then Lévy's lemma \cite{ledoux_concentration_2005} applied to this case states
    \begin{equation}
    \label{eq:Levyslemma}
        {\rm Pr}\big(|f(x)-\mathbb E[f]|>\alpha\big) \le 4 \exp\left(- c_0 d\frac{\alpha^2}{\eta^2}\right) \ ,
    \end{equation}
    where $c_0>0$ is a universal constant and $\alpha>0$ is an arbitrarily small number. In the limit $d\to\infty$ at any fixed $\alpha$, this implies concentration of typical $f(x)$ values around the average $\mathbb E[f]$.
    \item \emph{Square-root purity is globally Lipschitz}. Consider the subsystem purity $\mathbb P(\psi)$ of a normalized bipartite pure state $\ket \psi$. Then $\sqrt{\mathbb P(\psi)}$ is globally Lipschitz, in particular
    \begin{equation}
        \left|\sqrt{\mathbb P(\psi)}-\sqrt{\mathbb P(\phi)}\right| \le 2 \|\ket\psi-\ket\phi\| \ \ \forall \ \ket{\psi}, \ket \phi \ .
    \end{equation}
    This bound is valid irrespective of the chosen bipartition.
    Given that $0\le \mathbb P(\psi)\le 1$, this automatically implies
    \begin{equation}
    \label{eq:PisLipschitz}
        \big|\mathbb P(\psi) - \mathbb P(\phi)\big| = \left( \sqrt{\mathbb P(\psi)} + \sqrt{\mathbb P(\phi)}\right)\left|\sqrt{\mathbb P(\psi)}-\sqrt{\mathbb P(\phi)}\right| \le 4\|\ket \psi-\ket \phi\| \ .
    \end{equation}
    Therefore also $\mathbb P(\psi)$ is globally Lipschitz.
\end{enumerate}
The crucial fact for our purposes is that unitary evolution $U(t)$ preserves distances. In particular
\begin{equation}
    \|U(t) \ket a\otimes \ket b - U(t)\ket{a}\otimes \ket b'\| = \|\ket a \otimes (\ket b - \ket{b'})\| = \|\ket b - \ket{b'}\| \ .
\end{equation}
Using \cref{eq:PisLipschitz} we then get
\begin{equation}
\label{eq:pdbcmqfjsnepf}
    |\mathbb P_t(a,b) - \mathbb P_t(a,b')|\le 4 \|\ket b - \ket{b'}\| \ \ \forall\ t, \ \forall a, b, b' \ .
\end{equation}
A similar inequality is obtained for $A$ inputs by exchanging the roles of $a$ and $b$. Therefore purity is Lipschitz in each input $a$ and $b$, at every time, and for any $U(t)$. 

We now apply these results keeping in mind that our initial-state ensemble consists of two \emph{independent} Haar distributions in $A$ and $B$. For this purpose, let us define
\begin{equation}
\label{eq:intermstepLevy}
    p_t(b) = \mathbb E_a[\mathbb P_t(a,b)]  \quad \longrightarrow  \quad \mu_t = \mathbb E_b[p_t(b)] \ .
\end{equation}
Crucially, $p_t(b)$ is Lipschitz in $b$. Indeed, the average $\mathbb E_a$ cannot remove the Lipschitz property inherited from $\mathbb P_t(a,b)$, because
\begin{equation}
    |p_t(b)-p_t(b')|\le \mathbb E_a\left[ |\mathbb P_t(a,b) - \mathbb P_t(a,b')|\right]\le 4 \|\ket b - \ket{b'}\| \ ,
\end{equation}
where in the last inequality we used \eqref{eq:pdbcmqfjsnepf}. Looking at \eqref{eq:intermstepLevy}, there are now two possible sources of deviations: $\mathbb P_t(a,b)$ might deviate from $p_t(b)$ for a specific $a$, and $p_t(b)$ might deviate from the total average $\mu_t$ for a specific $b$. Given the Lipschitz properties of both $\mathbb P_t(a,b)$ and $p_t(b)$, we can now apply Levy's lemma \eqref{eq:Levyslemma} (with $\eta = 4$ and $\alpha = \varepsilon/2$) to both such scenarios
\begin{align}
\label{eq:Levybound1}
    {\rm Pr}_a\left( |\mathbb P_t(a,b) - p_t(b)|>\varepsilon/2\right) &\le 4 \exp\left(-cd_A \varepsilon^2\right) \ ,\\
\label{eq:Levybound2}
    {\rm Pr}_b \left(|p_t(b)-\mu_t|>\varepsilon/2\right)&\le 4 \exp\left(-cd_B \varepsilon^2\right) \ ,
\end{align}
where $c=c_0/64$. Importantly, the right-hand side of the first inequality is independent of $b$, so the inequality remains valid under ${\rm Pr}_a \to {\rm Pr}_{a,b}={\rm Pr}$. This allows us to combine the bounds thanks to the triangle inequality
\begin{equation}
    |\mathbb P_t(a,b)-\mu_t|\le |\mathbb P_t(a,b) - p_t(b)| + |p_t(b)-\mu_t| \ .
\end{equation}
For the left-hand side to exceed $\varepsilon$, at least one of the two terms on the right-hand side must exceed $\varepsilon/2$. The probability that either event occurs is at most the sum of their probabilities, yielding
\begin{equation}
    {\rm Pr}\left( |\mathbb P_t(a,b)-\mu_t|>\varepsilon \right) \le 4\left( e^{-cd_A \varepsilon^2} +e^{-cd_B \varepsilon^2} \right) \le 8 \exp\left( -c d_{\min}\varepsilon^2 \right) \ .
\end{equation}
This proves \cref{eq:targetLevy}.

\section{RMT calculations}
\label{appendix:RMT}

In this Appendix we derive the Haar averages used in the main text and reported in \cref{sec:RMTbriefresults}. As discussed in \cref{sec:GUEandCUE}, the eigenvector averages are identical for GUE Hamiltonians and CUE Floquet operators. We can therefore work directly with a Haar-distributed unitary $U\in\mathrm U(d)$, whose columns are the single-replica eigenstates $\ket m=U\ket{e_m}$ in a fixed computational basis $\{\ket{e_m}\}$. We use $d=d_A d_B$ and assume $d_A,d_B\geq2$.

\subsection{The twirl as a projection onto permutation operators}
\label{sec:appRMTtools}

We first collect the ingredients common to all the Haar calculations below. The aim is to replace integration over eigenvectors by a finite set of trace constraints. The single-replica calculations in \cref{sec:appRMTsingle} provide the simplest application and hence a useful warm-up. The same construction will then be used for the subsystem swap.

The unitary twirl introduced in \cref{eq:twirl} is
\begin{equation}
\label{eq:appRMTtwirl}
    \mathcal T_k(X)=\int d\mu(U)\,(U^\dag)^{\otimes k}XU^{\otimes k}\ ,
\end{equation}
where $d\mu(U)$ is the normalized Haar measure. Haar invariance implies
\begin{equation}
    (V^\dag)^{\otimes k}\mathcal T_k(X)V^{\otimes k}
    =\mathcal T_k(X)\qquad\forall\ V\in\mathrm U(d)\ .
\end{equation}
Thus the twirl takes values in the space of operators that commute with every $V^{\otimes k}$. By Schur--Weyl duality, this space is (see, e.g., Ref.~\cite{collins_integration_2006})
\begin{equation}
\label{eq:appRMTcommutant}
    \mathcal C_k
    =\big\{Y:[Y,V^{\otimes k}]=0\ \ \forall\ V\in\mathrm U(d)\big\}
    =\operatorname{span}\{P_\pi:\pi\in S_k\}\ ,
\end{equation}
i.e.~it is spanned by the representations $P_\pi$ of the symmetric group $S_k$ of degree $k$. 
We use the permutation convention of the main text,
\begin{equation}
\label{eq:appRMTperm}
    P_\pi\ket{v_1,\ldots,v_k}
    =\ket{v_{\pi^{-1}(1)},\ldots,v_{\pi^{-1}(k)}}\ ,
    \qquad P_\pi P_\tau=P_{\pi\tau}\ .
\end{equation}
Every $Y\in\mathcal C_k$ is unchanged by the twirl, since $(U^\dag)^{\otimes k}YU^{\otimes k}=Y$ for every $U$.

Knowing the space in which $\mathcal T_k(X)$ lies is only the first step: we still need the expansion coefficients in the basis $\{P_\pi\}$. The useful additional fact \cite{collins_integration_2006} is that the twirl is the \emph{orthogonal projection} onto $\mathcal C_k$, with respect to the Hilbert--Schmidt product $(Y,X)_{\rm HS}=\Tr[Y^\dag X]$ (from now on we denote $\Tr_{\mathcal H^{\otimes n}}$ simply as $\Tr$, with context providing the value of $n$).
This means that $\mathcal T_k(X)$ is the component of $X$ in $\mathcal C_k$, and that the remainder is orthogonal to every operator in that space. To see this directly, let $Y\in\mathcal C_k$. Cyclicity of the trace gives
\begin{align}
\label{eq:appRMTprojection}
    \Tr[Y^\dag\mathcal T_k(X)]
    =\int d\mu(U)\,\Tr\!\left[
        U^{\otimes k}Y^\dag(U^\dag)^{\otimes k}X\right]=\Tr[Y^\dag X]\ .\nonumber
\end{align}
Indeed this implies
\begin{align}
    X&=\underbrace{\mathcal T_k(X)}_{\in \, \mathcal C_k}
       +\underbrace{\big[X-\mathcal T_k(X)\big]}_{\perp \, \mathcal C_k}\ ,
\end{align}
which together with $\mathcal T_k(\mathcal T_k(X)) = \mathcal T_k(X)$ establishes the orthogonal-projection property.

The practical consequence of \cref{eq:appRMTprojection} is that the twirl can be removed from any trace against a permutation. Writing
\begin{equation}
\label{eq:twirlexpApp}
    \mathcal T_k(X)=\sum_{\tau\in S_k}c_\tau(X)P_\tau\ ,
\end{equation}
this leads us directly to the linear system
\begin{equation}
\label{eq:appRMTG}
    \Tr[P_\pi^\dag X]=\sum_{\tau\in S_k}G_{\pi\tau}c_\tau(X)
    \ ,\qquad
    G_{\pi\tau}=\Tr[P_\pi^\dag P_\tau]
    =d^{{\rm cycles}(\pi^{-1}\tau)}\ .
\end{equation}
Here ${\rm cycles}(\pi)$ counts cycles in $\pi$ (including cycles of length one) and the last equality follows from $P_\pi^{\dag} = P_{\pi^{-1}}=P_{\pi}^{-1}$ together with 
\begin{equation}
\label{eq:TrPpiAPPRMT}
    \Tr[P_\pi]=d^{{\rm cycles}(\pi)} \ .
\end{equation}
The last equality is due to the fact that the trace identifies all basis indices within each cycle and leaves one independent sum per cycle. Importantly, $G$ is the same for every operator $X$, therefore the only input to be computed for each $X$ are the traces $\Tr[P_\pi^{\dag}X]$ on the left-hand side of \cref{eq:appRMTG}. This is the formulation in \cref{eq:GramMatrix}, equivalent to unitary Weingarten calculus. In the following we will need only $k \le 4$. For $d\ge k$, which for us is always satisfied, the matrix $G$ is invertible and the system \eqref{eq:appRMTG} has unique coefficients $c_\tau(X)$ as solutions. From \cref{eq:twirlexpApp} these coefficients uniquely fix the twirl $\mathcal T_k(X)$, whose matrix elements in the computational basis are the Haar averages we are interested in, as explained in \cref{sec:RMTbriefresults}.

\subsection{Single-replica averages}
\label{sec:appRMTsingle}

The results in this subsection are standard. We report them as a warm-up, and to make the similarity with the following results for the subsystem swap operator explicit. 

For $k=1$, the permutation space consists only of the identity. In this case \cref{eq:appRMTG} reduces to
\begin{equation}
\label{eq:appRMTsingleMean}
    \mathcal T_1(O)=\overline O\,\Id\ ,
    \qquad \overline O=\frac{\Tr[O]}{d}\ ,
    \qquad
    \mathbb E_{\rm Haar}[\braket{m|O|n}]=\delta_{mn}\overline O\ ,
\end{equation}
which proves \cref{eq:diagHaarO}. 
Turning to $k=2$, the Haar fluctuations of the matrix elements of $O$ and the single-replica quench spectral weight both follow from the twirl of a 2-replica operator of the form $O_1\otimes O_2$, where $O_1,O_2$ are single-replica operators. Indeed,
\begin{align}
\label{eq:appRMTtwoMoment}
    \mathbb E_{\rm Haar}\big[\braket{m|O_1|n}\braket{n|O_2|m}\big]
    =\bra{e_m,e_n}\mathcal T_2(O_1\otimes O_2)\ket{e_n,e_m} \equiv \big(\mathcal T_2(O_1 \otimes O_2)\big)_{mn,nm} \ .
\end{align}
Writing
\begin{equation}
\label{eq:rogngv-srfbs}
    \mathcal T_2(O_1\otimes O_2)=a\,\Id^{\otimes2}+b\,P_{(12)}\ ,
\end{equation}
the two trace constraints in \cref{eq:appRMTG} read
\begin{equation}
    \begin{pmatrix}
    d^2 & d \\
    d & d^2
    \end{pmatrix}
    \begin{pmatrix}
    a\\
    b
    \end{pmatrix}
    =
    \begin{pmatrix}
    \Tr[O_1]\Tr[O_2]\\
    \Tr[O_1O_2]
    \end{pmatrix}\ , 
\end{equation}
where in the second row of the right-hand side we used property \eqref{eq:intertwofS} of the \emph{global} swap operator $P_{(12)}$. We therefore obtain the exact expressions
\begin{align}
\label{eq:appRMTtwoCoefficients}
    a &=\frac{d\Tr[O_1]\Tr[O_2]-\Tr[O_1O_2]}{d(d^2-1)}\ ,\\
    b &=\frac{d\Tr[O_1O_2]-\Tr[O_1]\Tr[O_2]}{d(d^2-1)}\ . \nonumber
\end{align}
The crossed matrix element in \cref{eq:appRMTtwoMoment} is particularly simple:
\begin{equation}
    \bra{e_m,e_n}\Id^{\otimes2}\ket{e_n,e_m}=\delta_{mn}\ ,\qquad
    \bra{e_m,e_n}P_{(12)}\ket{e_n,e_m}=1\ .
\end{equation}
Consequently,
\begin{equation}
\label{eq:appRMTsingleMaster}
    \mathbb E_{\rm Haar}\big[\braket{m|O_1|n}\braket{n|O_2|m}\big]
    =a\, \delta_{mn}+b\ .
\end{equation}
For a Hermitian $O$, setting $O_1=O_2=O$ and subtracting the squared mean gives
\begin{equation}
\label{eq:appRMTsingleVariance}
    \mathbb E_{\rm Haar}[|\braket{m|O|n}|^2]
    -\big|\mathbb E_{\rm Haar}[\braket{m|O|n}]\big|^2
    =\frac{d-\delta_{mn}}{d^2-1}
    \left(\overline{O^2}-\overline O^{2}\right)\ ,
\end{equation}
where $\overline{O^2}=\Tr[O^2]/d$. At leading order this is \cref{eq:HaarETHoff}, provided both normalized traces are of order one.

For the quench weight, set $O_1=O$, $O_2=\rho(0)$ and use $\Tr[\rho(0)]=1$. Defining $O_0=\Tr[\rho(0)O]$, the same matrix element gives
\begin{align}
\label{eq:appRMTsingleQuench}
    \mathbb E_{\rm Haar}\big[\braket{m|O|n}\braket{n|\rho(0)|m}\big]&=\delta_{mn}\frac{d^2\overline O-O_0}{d(d^2-1)}
    +\frac{O_0-\overline O}{d^2-1}\nonumber\\
    & =\delta_{mn}\left[\frac{\overline O}{d}+\mathcal O(d^{-3})\right]
    +\frac{O_0-\overline O}{d^2}+\mathcal O(d^{-4})\ ,
\end{align}
assuming $O_0$ and $\overline O$ to be of order one. This proves \cref{eq:HaarAVspectralquench}. For a pure initial state, the choice $O_1=O_2=\rho(0)$ also gives
\begin{equation}
\label{eq:appRMToverlaps}
    \mathbb E_{\rm Haar}\big[|\braket{m|\psi(0)}|^2
    |\braket{n|\psi(0)}|^2\big]
    =\frac{1+\delta_{mn}}{d(d+1)}\ ,
\end{equation}
where we used $\rho(0)^2=\rho(0)$. This establishes the validity of \cref{eq:jagkjgV2} at the level of Haar averages. Setting $m=n$ gives the fourth absolute moment used in the main text, establishing \cref{eq:jagkjgV1} at the level of Haar averages. Summing over $n$ and using normalization gives the second moment, $\mathbb E_{\rm Haar}[|\braket{m|\psi(0)}|^2]=1/d$, which is compatible with our definition in \cref{eq:zmDef}. 

\subsection{Mean matrix elements of the subsystem swap}
\label{sec:appRMTmean}

For a single-replica operator, the first step was $\mathcal T_1(O)=\overline O\,\Id$: the averaged operator is proportional to the identity. For $S_A$, the relevant twirl is instead $\mathcal T_2(S_A)$, which contains both $\Id^{\otimes2}$ and the global swap $P_{(12)}$. We choose to work with (anti)symmetrized eigenstates so that, within either sector, these two operators are proportional. This recovers precisely the structure of the single-replica mean.

We use the (anti)symmetrized 2-replica eigenstates from \cref{eq:defmsigma}
\begin{equation}
\label{eq:appRMTstates}
    \ket{\bm m_\sigma}
    =\frac{\ket{m,m'}+\sigma\ket{m',m}}
    {\sqrt{2(1+\delta_{mm'})}}\ , \quad \sigma = \pm 1 \, , \ m \le m' \ .
    \end{equation}
As in the main text we define the sector projectors and their dimensions by
\begin{equation}
\label{eq:appRMTsector}
    \mathsf P_\sigma=\frac{\Id^{\otimes2}+\sigma P_{(12)}}2\ ,
    \qquad d_\sigma=\Tr[\mathsf P_\sigma]=\frac{d(d+\sigma)}2\ .
\end{equation}
They act as
\begin{equation}
\label{eq:appRMTalpha}
    \mathsf P_\sigma\ket{m,m'}=\alpha_{\bm m}\ket{\bm m_\sigma}\ ,
    \qquad \alpha_{\bm m}=\sqrt{\frac{1+\delta_{mm'}}2}\ .
\end{equation}
With the ordering convention of \cref{eq:appRMTstates} we have
\begin{equation}
\label{eq:appRMTorthogonality}
\delta_{\bm m,\bm n}\equiv \delta_{mn}\delta_{m'n'} \, ,  \qquad \sum_{m\leq m'}\ket{\bm m_\sigma}\bra{\bm m_\sigma}
    =\mathsf P_\sigma \ .
\end{equation}
To compute the mean, write $\mathcal T_2(S_A)=a_S\Id^{\otimes2}+b_SP_{(12)}$. The advantage of the sector decomposition becomes then explicit
\begin{equation}
    \mathsf P_\sigma\mathcal T_2(S_A)\mathsf P_\sigma
    =(a_S+\sigma b_S)\mathsf P_\sigma\ .
\end{equation}
The coefficient $(a_S + \sigma b_S)$ is fixed by taking the trace. Since $\mathsf P_\sigma$ is itself a linear combination of permutations, \cref{eq:appRMTprojection} gives
\begin{equation}
    (a_S+\sigma b_S)d_\sigma
    =\Tr[\mathsf P_\sigma\mathcal T_2(S_A)]
    =\Tr[\mathsf P_\sigma S_A]\ .
\end{equation}
Hence
\begin{equation}
\label{eq:appRMTsectorTwirl}
    \mathsf P_\sigma\mathcal T_2(S_A)\mathsf P_\sigma
    =\mu_\sigma\mathsf P_\sigma\ ,\qquad
    \mu_\sigma=\frac{\Tr[\mathsf P_\sigma S_A]}{d_\sigma}\ .
\end{equation}
This is the replica-sector analogue of $\mathcal T_1(O)=(\Tr[O]/d)\Id$. The required traces factorized over the two subsystems, allowing us to use \cref{eq:TrPpiAPPRMT} separately in $A$ and $B$
\begin{align}
    \Tr[S_A]&=d_A d_B^2\ ,\qquad
    \Tr[P_{(12)} S_A]=\Tr[S_B]=d_A^2d_B\ ,\\
    \Tr[\mathsf P_\sigma S_A]&=\frac d2(d_B+\sigma d_A)\ ,\qquad
    \mu_\sigma=\frac{d_B+\sigma d_A}{d+\sigma}\ .\nonumber
\end{align}
As in the main text, to relate the (anti)symmetrized twirl to replica-eigenstate matrix elements we use $[U^{\otimes2},\mathsf P_\sigma]=0$ and \cref{eq:appRMTalpha}
\begin{align}
\label{eq:appRMTmeanContraction}
    &U^{\otimes2}\mathsf P_\sigma\ket{e_m,e_{m'}}
      =\alpha_{\bm m}\ket{\bm m_\sigma}\ ,\\
    &\alpha_{\bm m}\alpha_{\bm n}
    \mathbb E_{\rm Haar}[\braket{\bm m_\sigma|S_A|\bm n_\sigma}]
      =\big(\mathsf P_\sigma\mathcal T_2(S_A)\mathsf P_\sigma\big)_{mm',nn'} = \mu_\sigma  \big(\mathsf P_\sigma\big)_{mm',nn'}\ .
\end{align}
\emph{Excluding} the case $\{\sigma = -1$, $m =m'$ and/or $n = n'\}$, for which the right-hand side of the last equation correctly vanishes, we get
\begin{equation}
\label{eq:appRMTswapMean}
    \mathbb E_{\rm Haar}[\braket{\bm m_\sigma|S_A|\bm n_\sigma}]
    =\mu_\sigma\delta_{\bm m,\bm n}\ ,
\end{equation}
where we used the Kronecker delta defined in \eqref{eq:appRMTorthogonality} based on the ordering in \eqref{eq:appRMTstates}, and the fact that $\braket{e_m,e_{m'}|\mathsf P_\sigma|e_n,e_{n'}}=\alpha_{\bm m}\alpha_{\bm n}\delta_{\bm m, \bm n}$. The previous equation shares the same structure of the single-replica result in \cref{eq:appRMTsingleMean}, and proves \cref{eq:diagHaarSA}. Matrix elements between different $\sigma$ sectors vanish identically because $[S_A,P_{(12)}]=0$.

\subsection{A common calculation for four replicas}
\label{sec:appRMTfour}

Working in the (anti)symmetric sectors also simplifies the second moments. A four-replica twirl involves $4!=24$ permutations, but after projection onto these sectors they fall into only three classes. We derive this reduction once and then use it for both the swap fluctuations and the quench spectral weight.

Let $X_1,X_2$ be fixed \emph{two-replica} operators and set $X=X_1\otimes X_2$. The quantity to be averaged is
\begin{equation}
    \braket{\bm m_\sigma|X_1|\bm n_\sigma}
    \braket{\bm n_\sigma|X_2|\bm m_\sigma}\ .
\end{equation}
Its average is contained in the matrix elements of the 4-replica twirl $\mathcal T_4(X)$, properly (anti)symmetrized. The relevant projector is
\begin{equation}
\label{eq:appRMTQ}
    Q_\sigma\equiv\mathsf P_\sigma^{\otimes2}
    =\frac14\left(\Id^{\otimes4}+\sigma P_{(12)}
    +\sigma P_{(34)}+P_{(12)(34)}\right)\ .
\end{equation}
Applying \cref{eq:appRMTmeanContraction} to the four 2-replica states gives
\begin{align}
\label{eq:appRMTfourContraction}
    \alpha_{\bm m}^2\alpha_{\bm n}^2
    \mathbb E_{\rm Haar}\big[
    \braket{\bm m_\sigma|X_1|\bm n_\sigma}
    \braket{\bm n_\sigma|X_2|\bm m_\sigma}\big]=\big(Q_\sigma\mathcal T_4(X)Q_\sigma\big)_{mm'nn',nn'mm'}\ .
\end{align}
The subscripts denote matrix elements between $\ket{e_m,e_{m'},e_n,e_{n'}}$ and $\ket{e_n,e_{n'},e_m,e_{m'}}$. The two choices needed below are
\begin{equation}
\label{eq:appRMTchoicesofX}
    X=S_A\otimes S_A\qquad\text{and}\qquad
    X=S_A\otimes\rho(0)^{\otimes2}\ .
\end{equation}

\subsubsection{From twenty-four permutations to three operators}
\label{sec:from24to3}

Expanding the twirl in permutations gives
\begin{equation}
    Q_\sigma\mathcal T_4(X)Q_\sigma
    =\sum_{\pi\in S_4}c_\pi(X)Q_\sigma P_\pi Q_\sigma\ .
\end{equation}
The crucial simplification is that most of these twenty-four projected operators $Q_\sigma P_\pi Q_\sigma$ coincide up to a sign.
To see this, consider the following Abelian subgroup of $S_4$, formed by the permutations acting solely \emph{within} the two pairs
\begin{equation}
\label{eq:defHsubgroup}
    H=\{e,(12),(34),(12)(34)\}\ .
\end{equation}
These are the permutations entering the definition \eqref{eq:appRMTQ} of $Q_\sigma$, which absorbs them as
\begin{align}
\label{eq:appRMTH}
    Q_\sigma P_h&=P_hQ_\sigma=\chi_\sigma(h)Q_\sigma\ ,\qquad h\in H\ ,\\
\label{eq:appRMTabsrule}
    \chi_\sigma(e)&=\chi_\sigma((12)(34))=1\ ,\qquad
    \chi_\sigma((12))=\chi_\sigma((34))=\sigma\ .
\end{align}
It follows immediately that
\begin{equation}
    Q_\sigma P_{h\pi h'}Q_\sigma
    =\chi_\sigma(h)\chi_\sigma(h')\,Q_\sigma P_\pi Q_\sigma\ ,
    \qquad h,h'\in H\ .
\end{equation}
Thus we need only distinguish permutations up to rearrangements within each pair, before and after the permutation. The remaining distinction is how many labels are sent from $\{1,2\}$ to $\{3,4\}$:
\begin{equation}
\label{eq:appRMTclasses}
    \begin{array}{c|ccc}
        \#\text{ of labels sent to other pair}&0&1&2\\ \hline
        \text{representative}&\pi_0=e&\pi_1=(13)&\pi_2=(13)(24)\\
        \# \text{ of permutations}&4&16&4
    \end{array}\ .
\end{equation}
For zero or two exchanged labels there are $2!\,2!=4$ possibilities: respectively the elements of the subgroup $H$ itself and $(13)(24)$, $(14)(23)$, $(1324)$, $(1423)$. The other $16$ permutations in $S_4$ exchange one label. Rearranging the input and output labels within their pairs brings each permutation to the corresponding representative. In group-theoretic terms, these are the three double cosets $H\pi_iH$, with $i = 0,1,2$, which include all the $4!$ permutations in $S_4$.

We therefore need only the following three Hermitian operators
\begin{equation}
\label{eq:appRMTbasis}
    B_{0,\sigma}=Q_\sigma\ ,\qquad
    B_{1,\sigma}=Q_\sigma P_{(13)}Q_\sigma\ ,\qquad
    B_{2,\sigma}=Q_\sigma P_{(13)(24)}\ .
\end{equation}
Note that a second $Q_\sigma$ in $B_{2,\sigma}$ is unnecessary because $[Q_\sigma,P_{(13)(24)}]=0$. As a simple example of the reduction, $(123)=(13)(12)$ implies
\begin{equation}
    Q_\sigma P_{(123)}Q_\sigma
    =Q_\sigma P_{(13)}P_{(12)}Q_\sigma
    =\sigma B_{1,\sigma}\ ,
\end{equation}
where we used the absorbtion rules \eqref{eq:appRMTabsrule}.
The projected twirl therefore has the form
\begin{equation}
\label{eq:appRMTthree}
    Q_\sigma\mathcal T_4(X)Q_\sigma
    =\sum_{i=0}^2C_{i,\sigma}(X)B_{i,\sigma}\ .
\end{equation}
The coefficients are determined by exactly the same projection argument as in \cref{sec:appRMTtools}. Since $B_{i,\sigma}$ is a linear combination of permutations and $Q_\sigma B_{i,\sigma}Q_\sigma=B_{i,\sigma}$, using \cref{eq:appRMTprojection} we obtain
\begin{equation}
    \Tr[B_{i,\sigma}Q_\sigma\mathcal T_4(X)Q_\sigma]
    =\Tr[B_{i,\sigma}\mathcal T_4(X)]
    =\Tr[B_{i,\sigma}X]\ .
\end{equation}
Inserting \cref{eq:appRMTthree} gives the $3\times3$ system
\begin{align}
\label{eq:appRMTreducedSystem}
    &v_{i,\sigma}(X)=\sum_{j=0}^2G^{(\sigma)}_{ij}C_{j,\sigma}(X)\ ,\\
\label{eq:ldavdvrld-dh}
    & G^{(\sigma)}_{ij}=\Tr[B_{i,\sigma}B_{j,\sigma}]\ ,\qquad
    v_{i,\sigma}(X)=\Tr[B_{i,\sigma}X]\ .
\end{align}
Only $ v_{i,\sigma}(X)$ depends on the operator being averaged.
Using \cref{eq:TrPpiAPPRMT,eq:appRMTabsrule}, the expansion \eqref{eq:appRMTQ} of $Q_\sigma$ and the cyclicity of the trace, the entries of the symmetric $G^{(\sigma)}$ matrix can be compactly written as
\begin{equation}
    G^{(\sigma)}_{ij}
    =\frac1{16}\sum_{h,h'\in H}\chi_\sigma(h)\chi_\sigma(h')
        d^{{\rm cycles}(h\pi_i h'\pi_j)}\ .
\end{equation}
Some of them can be calculated more easily using $Q_\sigma^2 = Q_\sigma$, the cyclicity of the trace and $[Q_\sigma,P_{(13)(24)}]=0$. For example
\begin{align}
    &G^{(\sigma)}_{00}=\Tr[Q_\sigma]=\Tr[\mathsf P_\sigma^{\otimes 2}]=d_\sigma^2 \ , \\
    &G^{(\sigma)}_{01}= \Tr[Q_\sigma P_{(13)}] = \frac{1}{4}\Tr[P_{(13)}+\sigma (P_{(132)} + P_{(143)})+P_{(1432)}] = \frac{1}{4}(d^3 + 2 \sigma d^2 + d) \ .
\end{align}
The complete matrix is
\begin{equation}
\label{eq:appRMTreducedGram}
    G^{(\sigma)}=\frac{d(d+\sigma)}{16}
    \begin{pmatrix}
        4d(d+\sigma)&4(d+\sigma)&8\\
        4(d+\sigma)&d^2+3\sigma d+4&4(d+\sigma)\\
        8&4(d+\sigma)&4d(d+\sigma)
    \end{pmatrix}\ .
\end{equation}
It is invertible for $d\geq4$ in both sectors.

Both choices of $X$ in \cref{eq:appRMTchoicesofX} commute with $Q_\sigma$. Using $Q_\sigma^2=Q_\sigma$ and cyclicity of the trace, each component of the source vector $ v_{i,\sigma}(X)$ for $X$ such that $[X,Q_\sigma]=0$ reduces to four permutation traces
\begin{align}
\label{eq:appRMTsourceSum}
    v_{i,\sigma}(X)
    =\Tr[XQ_\sigma P_{\pi_i}Q_\sigma]
      =\Tr[XQ_\sigma P_{\pi_i}]=\frac14\sum_{h\in H}\chi_\sigma(h)\Tr[XP_{h\pi_i}]\ \quad\qquad {\rm if} \ [X,Q_\sigma ]=0 \ .
\end{align}

\subsubsection{Matrix elements of (anti)symmetrized twirl}

Before computing $v_{i,\sigma}(X)$ for the two cases \eqref{eq:appRMTchoicesofX} of interest, we evaluate the three permutation matrix elements needed in \cref{eq:appRMTfourContraction} (only three are needed thanks to \cref{eq:appRMTthree}). As in \cref{eq:appRMTswapMean}, we restrict the explicit matrix-element formulas in this and the following subsections to nonzero replica states, i.e.
\begin{align}
\label{eq:appRMTnonzeroStates}
        \sigma=+1:&\qquad m\leq m'\ ,\quad n\leq n'\ ,\\
        \sigma=-1:&\qquad m<m'\ ,\quad n<n'\ . \nonumber
\end{align}
The excluded cases give zero identically, as already accounted for by \cref{eq:appRMTfourContraction}.
Given $B_{0,\sigma}=Q_\sigma$, and remembering $Q_\sigma=\mathsf P_\sigma^{\otimes 2}$, the first matrix element we are interested in is
\begin{align}
\label{eq:RMTAppME1}
    &\braket{e_m,e_{m'},e_n,e_{n'}|B_{0,\sigma}|e_n,e_{n'},e_m,e_{m'}} \\
    &\qquad = \alpha_{\bm m}^2\alpha_{\bm n}^2\braket{\bm e_{\bm m,\sigma},\bm e_{\bm n,\sigma}|\bm e_{\bm n,\sigma},\bm e_{\bm m,\sigma}} = \alpha_{\bm m}^2\alpha_{\bm n}^2 \delta_{\bm m, \bm n} \ , \nonumber
\end{align}
where $\delta_{\bm m, \bm n}$ is the one defined in \cref{eq:appRMTorthogonality} and $\ket{\bm e_{\bm m,\sigma}}$ is the normalized (anti)symmetrization of $\ket{e_m,e_{m'}}$ due to the action of $\mathsf P_\sigma$, i.e.~the analogue of \eqref{eq:appRMTstates} but for computational-basis states. Given $B_{2,\sigma}=Q_\sigma P_{(13)(24)}Q_\sigma$, also its matrix element is straightforward
\begin{align}
\label{eq:RMTAppME2}
    &\braket{e_m,e_{m'},e_n,e_{n'}|B_{2,\sigma}|e_n,e_{n'},e_m,e_{m'}} \\
    &\qquad = \alpha_{\bm m}^2\alpha_{\bm n}^2\braket{\bm e_{\bm m,\sigma},\bm e_{\bm n,\sigma}|P_{(13)(24)}|\bm e_{\bm n,\sigma},\bm e_{\bm m,\sigma}} = \alpha_{\bm m}^2\alpha_{\bm n}^2 \ , \nonumber
\end{align}
where we used $P_{(13)(24)}\ket{\bm e_{\bm n,\sigma},\bm e_{\bm m,\sigma}}=\ket{\bm e_{\bm m,\sigma},\bm e_{\bm n,\sigma}}$. For $B_{1,\sigma}=Q_\sigma P_{(13)}Q_\sigma$ we find instead
\begin{align}
\label{eq:ldaknvvprcdd}
&\braket{e_m,e_{m'},e_n,e_{n'}|B_{1,\sigma}|e_n,e_{n'},e_m,e_{m'}} \\ \nonumber
&\qquad = \alpha_{\bm m}^2\alpha_{\bm n}^2\braket{\bm e_{\bm m,\sigma},\bm e_{\bm n,\sigma}|P_{(13)}|\bm e_{\bm n,\sigma},\bm e_{\bm m,\sigma}} = \alpha_{\bm m}^2\alpha_{\bm n}^2 \ \frac{\Delta_{\bm m, \bm n}}{4}\ ,\\
\label{eq:appRMTDelta}
&\Delta_{\bm m, \bm n} = \delta_{mn}+\delta_{mn'}+\delta_{m'n}+\delta_{m'n'} \ .
\end{align}
Substituting these expressions into \cref{eq:appRMTthree,eq:appRMTfourContraction} and cancelling the common nonzero factor $\alpha_{\bm m}^2\alpha_{\bm n}^2$ gives
\begin{align}
\label{eq:appRMTmaster}
    &\mathbb E_{\rm Haar}\big[
    \braket{\bm m_\sigma|X_1|\bm n_\sigma}
    \braket{\bm n_\sigma|X_2|\bm m_\sigma}\big]=C_{0,\sigma}(X)\delta_{\bm m,\bm n}
      +\frac{C_{1,\sigma}(X)}4\Delta_{\bm m,\bm n}
      +C_{2,\sigma}(X)\ .
\end{align}
This is the four-replica counterpart of \cref{eq:appRMTsingleMaster}. All that remains is now determining $C_{i,\sigma}(X)$ for all $X=X_1 \otimes X_2$ of interest.

\subsubsection{Exact swap second moments and their leading fluctuations}
\label{sec:appRMTvariance}

We now take $X=S_A\otimes S_A$ and denote the corresponding coefficients by $C^{(S)}_{i,\sigma}$. Since this operator permutes copies 1 and 2 and copies 3 and 4 within $A$, its permutation contractions factorize as
\begin{equation}
\label{eq:appRMTswapContraction}
    \Tr[(S_A\otimes S_A)P_\pi]
    =d_A^{{\rm cycles}((12)(34)\pi)}d_B^{{\rm cycles}(\pi)}\ ,
\end{equation}
where we used \cref{eq:TrPpiAPPRMT} separately in $A$ and $B$. The four-term sums in \cref{eq:appRMTsourceSum} gives the corresponding source vector $v_{i,\sigma}^{(S)}$. Using \cref{eq:appRMTswapContraction} we find
\begin{equation}
\label{eq:appRMTswapVector}
    \bm v^{(S)}_\sigma=
    \begin{pmatrix}
        d^2(d_A+\sigma d_B)^2/4\\
        d(d_A+\sigma d_B)^2/4\\
        d(d+\sigma)/2
    \end{pmatrix}\ .
\end{equation}
We can now easily solve the $3\times 3$ system in \cref{eq:appRMTreducedSystem}. It is convenient to give $C^{(S)}_{2,\sigma}$ first and express the other two coefficients in terms of it. Using
\begin{equation}
\label{eq:appRMTK}
    (d+\sigma)^2-(d_A+\sigma d_B)^2
    =(d_A^2-1)(d_B^2-1)\ ,
\end{equation}
we obtain
\begin{align}
\label{eq:appRMTswapCoefficients}
    C^{(S)}_{2,\sigma}
    =\frac{2(d_A^2-1)(d_B^2-1)}{d(d^2-1)(d+3\sigma)}\ , \qquad C^{(S)}_{1,\sigma}
    =-\frac{4C^{(S)}_{2,\sigma}}{d+2\sigma}\ , \qquad C^{(S)}_{0,\sigma}
    =\mu_\sigma^2+\frac{2C^{(S)}_{2,\sigma}}{(d+\sigma)(d+2\sigma)}\ .
\end{align}
Here $\mu_\sigma$ is the diagonal Haar result from \cref{eq:appRMTswapMean}. 

For this choice of $X$, the contraction in \cref{eq:appRMTmaster} is the swap second moment appearing in \cref{eq:offdiagHaarSA}. Subtracting the squared mean $\mu_\sigma^2\delta_{\bm m,\bm n}$ from \cref{eq:appRMTswapMean} therefore gives the exact variance, under the same index restriction from \eqref{eq:appRMTnonzeroStates}
\begin{align}
\label{eq:appRMTswapVariance}
    &\mathbb E_{\rm Haar}[|\braket{\bm m_\sigma|S_A|\bm n_\sigma}|^2]
    -\big|\mathbb E_{\rm Haar}[\braket{\bm m_\sigma|S_A|\bm n_\sigma}]\big|^2\\ \nonumber
    &\qquad=C^{(S)}_{2,\sigma}\left[
    1-\frac{\Delta_{\bm m,\bm n}}{d+2\sigma}
    +\frac{2\delta_{\bm m,\bm n}}{(d+\sigma)(d+2\sigma)}
    \right]\ .
\end{align}
This expression includes all coincident-index cases allowed by \cref{eq:appRMTnonzeroStates}.

For $d\gg1$, with no additional scaling assumption on either subsystem dimension,
\begin{align}
\label{eq:appRMTswapOrders}
    C^{(S)}_{2,\sigma}
    =\frac{2}{d^2}\left(1-\frac1{d_A^2}-\frac1{d_B^2}\right)
    +\mathcal O(d^{-3})\ ,
\end{align}
where we used \eqref{eq:appRMTswapCoefficients}. Using that the bracket in \eqref{eq:appRMTswapVariance} equals $1+\mathcal O(d^{-1})$ uniformly over the allowed indices, the leading variance is
\begin{align}
\label{eq:appRMTswapLeading}
    &\mathbb E_{\rm Haar}[|\braket{\bm m_\sigma|S_A|\bm n_\sigma}|^2]
    -\big|\mathbb E_{\rm Haar}[\braket{\bm m_\sigma|S_A|\bm n_\sigma}]\big|^2=\frac{2}{d^2}\left(1-\frac1{d_A^2}-\frac1{d_B^2}\right)
    +\mathcal O(d^{-3})\ .\nonumber
\end{align}
This proves \cref{eq:HaarFluctOff} for all index classes of nonzero replica eigenstates in both sectors. 

\subsubsection{Exact quench spectral weights}
\label{sec:appRMTquench}

Let $\rho(0)=\ket{\psi(0)}\bra{\psi(0)}$ be a fixed pure state. We now choose
\begin{equation}
\label{eq:appRMTquenchX}
    X_\psi=S_A\otimes\rho(0)^{\otimes2}\ ,
\end{equation}
where $S_A$ acts on copies 1 and 2 and the two density matrices act on copies 3 and 4. The initial two-replica state is symmetric
\begin{equation}
    \mathsf P_+\rho(0)^{\otimes2}=\rho(0)^{\otimes2}\ ,\qquad
    \mathsf P_-\rho(0)^{\otimes2}=0\ .
\end{equation}
Therefore only the symmetric sector contributes, and we denote the corresponding coefficients by $C_i^{(\psi)}=C_{i,+}(X_\psi)$. The Gram matrix and the permutation matrix elements are already known from \cref{eq:appRMTreducedGram,eq:RMTAppME1,eq:RMTAppME2,eq:ldaknvvprcdd}. We need only the three traces $v_i^{(\psi)}=\Tr[B_{i,+}X_\psi]$ to determine $C_i^{(\psi)}$ from the solution of the $3\times 3$ system \eqref{eq:appRMTreducedSystem}.

Given a generic permutation trace, it is useful to first trace out copies 1 and 2:
\begin{equation}
\label{eq:gldgk-ngdnprfv}
    \Tr[X_\psi P_\pi]
    =\Tr_{3,4} \left[\rho(0)^{\otimes2}
      \Tr_{1,2}\left[(S_A\otimes\Id^{\otimes2})P_\pi\right]\right]\ .
\end{equation}
The partial trace can be performed using the following identity for a Hilbert space $\mathcal V$ of dimension $d_\mathcal{V}$
\begin{equation}
\label{eq:appRMTpartial}
    \Tr_{1,2}[P_\pi^{(\mathcal V)}]=d_{\mathcal V}^{\ell(\pi)}P^{(\mathcal V)}_{\bar\pi}\ ,
\end{equation}
where $\ell(\pi)$ counts the cycles of $\pi \in S_4$ entirely contained in $\{1,2\}$, and $\bar\pi \in S_2$ is obtained by deleting 1 and 2 from every cycle and it is hence a permutation of the two remaining labels 3 and 4. We will apply this identity separately in both our subsystems $\mathcal V=A,B$. To verify the rule \eqref{eq:appRMTpartial}, the partial trace can be written in components as
\begin{equation}
\label{eq:appRMTpartialIndices}
    \bra{x_3,x_4}\Tr_{1,2}[P_\pi^{(\mathcal V)}]\ket{y_3,y_4}
    =\sum_{x_1,x_2}\prod_{j=1}^4
    \delta_{x_j,\widetilde y_{\pi^{-1}(j)}}\ ,
    \qquad \widetilde{\bm y}=(x_1,x_2,y_3,y_4)\ .
\end{equation}
The simplest case to verify is $\pi = (1)(2)\overline \pi$: the first two components of $\widetilde{\bm y}$ are unchanged under $\pi$, and therefore the sum over $x_1$ and $x_2$ contributes $c_{34}\sum_{x_1,x_2}\delta_{x_1,x_1}\delta_{x_2,x_2}=c_{34}\,d_{\mathcal V}^2$ where $c_{34}=\delta_{x_3,y_{\overline \pi^{-1}(3)}}\delta_{x_4,y_{\overline \pi^{-1}(4)}}$. This exactly coincides with $\braket{x_3,x_4|d_{\mathcal V}^{\ell(\pi)}P_{\overline \pi}^{(\mathcal V)}|y_3,y_4}$. If $\pi = (12)\overline \pi$, the only difference is in $\sum_{x_1,x_2}\delta_{x_1,x_2}\delta_{x_2,x_1}= d_{\mathcal V}$, matching again \eqref{eq:appRMTpartial}. An example with cycles not entirely contained within $\{1,2\}$ or $\{3,4\}$ is $\pi = (134)(2)$, which according to the definition yields $\overline \pi=(34)$. Noting that $\pi^{-1}=(143)(2)$, we have
\begin{align}
    \sum_{x_1,x_2}\prod_{j=1}^4
    \delta_{x_j,\widetilde y_{\pi^{-1}(j)}} &= \sum_{x_1,x_2}\delta_{x_1,y_4}\delta_{x_2,x_2}\delta_{x_3,x_1}\delta_{x_4,y_3}  = d_{\mathcal V} \, \delta_{x_3,y_4}\delta_{x_4,y_3} \ .
\end{align}
Given that $\ell(\pi) = 1$, this exactly recovers the element $\braket{x_3,x_4|d_{\mathcal V}^{\ell(\pi)}P_{\overline \pi}^{(\mathcal V)}|y_3,y_4}$. With a bit of patience, it is straightforward to verify the other cases, or to understand the general mechanism behind these examples: sums over traced indices $\{1,2\}$ connect via Kronecker deltas untraced indices $\{3,4\}$ in a way that is always compatible with our definition of $\overline \pi$. 

Now write $P_\pi=P_\pi^{(A)}\otimes P_\pi^{(B)}$ and let $s=(12)$. Since $S_A$ exchanges copies 1 and 2 only in $A$,
\begin{equation}
    (S_A\otimes\Id^{\otimes2})P_\pi
    =P_{s\pi}^{(A)}\otimes P_\pi^{(B)}\ .
\end{equation}
Applying \cref{eq:appRMTpartial} separately in the two subsystems and using \cref{eq:gldgk-ngdnprfv} gives
\begin{align}
\label{eq:appRMTquenchTrace}
    \Tr[X_\psi P_\pi]
    &=d_A^{\ell(s\pi)}d_B^{\ell(\pi)}
    \bra{\psi(0)}^{\otimes2}
    \big(P_{\overline{s\pi}}^{(A)}\otimes P_{\bar\pi}^{(B)}\big)
    \ket{\psi(0)}^{\otimes2}\\*
    &=d_A^{\ell(s\pi)}d_B^{\ell(\pi)}
    \begin{cases}
        1\ ,&\overline{s\pi}=\bar\pi\ ,\\
        \mathbb P_A(0)\ ,&\overline{s\pi}\neq\bar\pi\ .
    \end{cases}\nonumber
\end{align}
The last equality follows from: (i) the fact that each residual permutation of $\{3,4\}$ is either the identity $e$ or $(34)$; (ii) $P^{(A)}_{z}\otimes P^{(B)}_z\ket{\psi(0)}^{\otimes 2}=P_z\ket{\psi(0)}^{\otimes2}=\ket{\psi(0)}^{\otimes2}$ for any $z \in S_2$; (iii) $\bra{\psi(0)}^{\otimes 2}P^{(A)}_{(34)}\otimes P^{(B)}_{e}\ket{\psi(0)}^{\otimes 2}=\mathbb P_A(0)=\mathbb P_B(0)$, where we used \cref{eq:mainidentity} because $P_{(34)}^{(A)}=S_A$ on copies 3 and 4. This explains why the initial purity is the only property of the fixed pure state that will enter the final result. 

Using \cref{eq:appRMTquenchTrace} to evaluate the four-term sum \eqref{eq:appRMTsourceSum}, we obtain
\begin{align}
\label{eq:appRMTquenchVector}
    v_0^{(\psi)}=\frac d2(d_A+d_B)\ ,\qquad v_1^{(\psi)}=\frac{d_A+d_B}2\ ,\qquad v_2^{(\psi)}=\mathbb P_A(0)\ .
\end{align}
Solving the same $3\times3$ system and again expressing the result in terms of the last coefficient gives
\begin{align}
\label{eq:appRMTquenchCoefficients}
    C_2^{(\psi)}
    =\frac{4[\mathbb P_A(0)-\mu_+]}{d^2(d-1)(d+3)}\ ,\qquad C_1^{(\psi)}
    =-\frac{4C_2^{(\psi)}}{d+2}\ ,\qquad C_0^{(\psi)}
    =\frac{\mu_+}{d_+}+\frac{2C_2^{(\psi)}}{(d+1)(d+2)}\ .
\end{align}
Here $\mu_+=(d_A+d_B)/(d+1)$ and $d_+=d(d+1)/2$, as defined in \cref{sec:appRMTmean}.
The exact averaged quench weight follows from \cref{eq:appRMTmaster} with $\sigma=+1$
\begin{align}
\label{eq:appRMTquenchExact}
    &\mathbb E_{\rm Haar}\big[
    \braket{\bm m_+|S_A|\bm n_+}
    \braket{\bm n_+|\rho(0)^{\otimes2}|\bm m_+}\big]\\
    &\qquad=\frac{\mu_+}{d_+}\delta_{\bm m,\bm n}
    +C_2^{(\psi)}\left[
    1-\frac{\Delta_{\bm m,\bm n}}{d+2}
    +\frac{2\delta_{\bm m,\bm n}}{(d+1)(d+2)}
    \right]\ .\nonumber
\end{align}
The bracket has exactly the same index dependence as the swap variance \eqref{eq:appRMTswapVariance} at $\sigma=+1$. This formula holds for every ordered pair $m\leq m'$, $n\leq n'$, including repeated indices within either pair, and for arbitrary fixed pure initial states (including those highly entangled across the cut $\partial$ between $A$ and $B$).

The two prefactors in \cref{eq:appRMTquenchExact} have the expansions
\begin{align}
\label{eq:appRMTquenchOrders}
    \frac{\mu_+}{d_+}
    &=\frac{2}{d^2}\left(\frac1{d_A}+\frac1{d_B}\right)
    \big[1+\mathcal O(d^{-1})\big]\ ,\\
    C_2^{(\psi)}
    &=\frac4{d^4}\left[\mathbb P_A(0)-\frac1{d_A}-\frac1{d_B}\right]
    +\mathcal O(d^{-5})\ .\nonumber
\end{align}
On the diagonal, $\mu_+/d_+$ supplies the leading term, while the term proportional to $C_2^{(\psi)}$ is $\mathcal O(d^{-4})$. Since $d_A^{-1}+d_B^{-1}\geq2d^{-1/2}$, the latter is smaller than the leading term by $\mathcal O(d^{-3/2})$, and is absorbed in the relative $\mathcal O(d^{-1})$ correction. We therefore obtain
\begin{align}
\label{eq:appRMTquenchDiagonal}
    \mathbb E_{\rm Haar}\big[
    \braket{\bm m_+|S_A|\bm m_+}
    \braket{\bm m_+|\rho(0)^{\otimes2}|\bm m_+}\big]=\frac{2}{d^2}\left(\frac1{d_A}+\frac1{d_B}\right)
    \big[1+\mathcal O(d^{-1})\big]\ ,
\end{align}
which proves \cref{eq:nalgp4ggd11}. 
For $\ket{\bm m_+}\neq\ket{\bm n_+}$, the Kronecker delta vanishes and the bracket in \cref{eq:appRMTquenchExact} is $1-\Delta_{\bm m,\bm n}/(d+2)= 1+\mathcal O(d^{-1})$. Therefore
\begin{align}
\label{eq:appRMTquenchOff}
    \mathbb E_{\rm Haar}\big[
    \braket{\bm m_+|S_A|\bm n_+}
    \braket{\bm n_+|\rho(0)^{\otimes2}|\bm m_+}\big]=\frac4{d^4}\left[\mathbb P_A(0)-\frac1{d_A}-\frac1{d_B}\right]
    +\mathcal O(d^{-5})\ .
\end{align}
This proves \cref{eq:offdiagETHoutHaar}, without any assumption on $\min(d_A,d_B)^{-1}\le \mathbb P_A(0)\le 1$. The dependence on coincident indices is subleading, so the off-diagonal result applies to every off-diagonal class from \cref{tab:classesOfEigenstates}. The $\mathcal O(d^{-4})$ Haar average is smaller than the $d^{-3}$ scale obtained by multiplying the typical magnitudes of the swap matrix element and the overlaps. The joint average therefore captures the phase cancellations discussed in the main text.

\subsection{Purity averaged over subsystem product states}
\label{sec:appRMTproduct}

Finally, we derive \cref{eq:Eprod}. This identity concerns the independent Haar averages of the initial subsystem states, and holds for any \emph{fixed deterministic} unitary evolution $U(t)$. It is closely related to the product-state average defining the entangling power of a unitary in \cite{zanardi_entangling_2000}.

For $\ket{\psi(0)}=\ket a_A\ket b_B$, assume the the two subsystem states are independently Haar distributed. Applying \cref{eq:rogngv-srfbs,eq:appRMTtwoCoefficients} to a pure state $\ket{a}\in{\rm Haar}(A)$ gives
\begin{equation}
    \mathbb E_{\rm Haar(A)} \left[(\ket a\bra a)^{\otimes2}\right]
    =\frac{\Id_A^{\otimes2}+P_{(12)}^{(A)}}{d_A(d_A+1)}\ ,
\end{equation}
where $S_A = P_{(12)}^{(A)}\otimes \Id_B^{\otimes 2}$. The analogous identity holds in $B$. Independence of the two averages then gives
\begin{align}
\label{eq:appRMTproductMoment}
    M\equiv\mathbb E_{\ket{\psi}}[\rho(0)^{\otimes2}]
    =\frac{(\Id^{\otimes2}+S_A)(\Id^{\otimes2}+S_B)}
    {d(d_A+1)(d_B+1)}=\frac{2(\Id^{\otimes2}+S_A)\mathsf P_+}
    {d(d_A+1)(d_B+1)}\ .
\end{align}
In the last equality we used $S_AS_B=P_{(12)}$ and $S_A^2=\Id^{\otimes2}$. The swap representation of purity in \cref{eq:mainidentity} therefore gives
\begin{equation}
\label{eq:appRMTproductTrace}
    \mathbb E_{\ket{\psi}}[\mathbb P_A(t)]
    =\Tr[M S_A(t)]
    =\frac{2\Tr[\mathsf P_+S_A(t)]
    +2\Tr[\mathsf P_+S_A(t)S_A]}
    {d(d_A+1)(d_B+1)}\ .
\end{equation}
Because $[\mathsf P_+,U(t)^{\otimes2}]=0$, the first trace is independent of time. The second is (up to normalization $d_+$) the correlator defined in \cref{eq:Cplus}
\begin{align}
    \Tr[\mathsf P_+S_A(t)]&=\Tr[\mathsf P_+S_A]=\frac d2(d_A+d_B)\ ,\\
    \Tr[\mathsf P_+S_A(t)S_A]&=d_+C_A^{(+)}(t)\ .\nonumber
\end{align}
Substituting these two expressions into \cref{eq:appRMTproductTrace} yields
\begin{equation}
\label{eq:appRMTproductFinal}
    \mathbb E_{\ket{\psi}}[\mathbb P_A(t)]
    =\frac{d+1}{(d_A+1)(d_B+1)}C_A^{(+)}(t)
    +\frac{d_A+d_B}{(d_A+1)(d_B+1)}\ ,
\end{equation}
which is exactly \cref{eq:Eprod}. 
At $t=0$, $S_A^2=\Id^{\otimes2}$ gives $C_A^{(+)}(0)=1$, and the right-hand side is one, as required for an initial product state. No assumption about the unitary evolution $U(t)$, or about its spectrum and eigenstates, is needed for this identity.

\section{Coincident indices and full-ETH}
\label{app:coincidentindices}

The solution of the apparent difficulty with the subleading off-diagonal classes identified in \cref{sec:conn2&4ETH} stems from the following convenient choice of the basis $\{O_\mu\}$
\begin{equation}
\label{eq:gglkrgj5333}
    O_1 = \frac{\Id_A}{\sqrt{d_A}} \quad \longrightarrow \quad \Tr_A[O_\mu] = 0 \ \ \forall  \ \mu > 1 \ ,
\end{equation}
where we have used the orthonormality in \eqref{eq:SAexpanded2}. From this we get the following consequences for each class (again, we assume for simplicity that $d_A$ is finite while $d \to \infty$).\\

\emph{Class II $\to \ \{(a,a),(a,b)\}$} . \ Using the full-ETH factorization rules (in this case corresponding to a non-crossing partition in full-ETH language \cite{foini_eigenstate_2019, pappalardi_eigenstate_2022}), \cref{eq:connectingFullETH1} amounts to leading order to $\overline{(O_\mu)_{11}}\cdot\overline{(O_\nu)_{11}} \cdot \overline{(O_\mu)_{14} (O_\nu)_{41}}$. Using the Floquet-ETH prediction $\overline{(O_\mu)_{mm}}\sim\Tr[O_\mu]/d=\Tr_A[O_\mu]/d_A$, we have that the leading diagonal order of $\overline{(O_\mu)_{11}}$ vanishes for any $\mu > 1$ because of \cref{eq:gglkrgj5333}. This means that the only surviving term at leading diagonal order comes from $\mu=\nu=1$, for which however $(O_1)_{14}=(O_1)_{41}=0$ because of $O_1 \propto \Id_A$ and orthogonality between $\ket{1}$ and $\ket{4}$. The other term $[3 \leftrightarrow4]$ in \cref{eq:SAmatrixelFromFullETH} is identical to the first one for class II. The next-to-leading order terms in the full-ETH factorization, which do not vanish, are in this case $\mathcal O(d^{-2})$ \cite{vallini_refinements_2026}. This resolves the paradox because no term of overall order $d^{-1}$ survives, leaving space for the $d^{-2}$ scaling we expect from RMT. \\

\emph{Class III} $\to \ \{(a,a),(b,b)\}$ . \ Here the terms in \cref{eq:SAmatrixelFromFullETH} factorize similarly to \cref{eq:connectingFullETH1} (the factorization features an additional order-one multipliticity prefactor, see Ref.~\cite{vallini_refinements_2026}). Therefore this class yields automatically the order $d^{-2}$.\\

\emph{Class V} $\to \ \{(a,a),(b,c)\}$ . \ Here the terms in \cref{eq:SAmatrixelFromFullETH} factorize to leading order exactly as in \cref{eq:connectingFullETH1} by full-ETH factorization rules (these is a standard non-crossing partition \cite{pappalardi_eigenstate_2022}).\\

\emph{Class VI} $\to \ \{(a,b),(a,c)\}$ . \ Similarly to class II, also in this case the leading diagonal order in \cref{eq:connectingFullETH1} vanishes because of $O_1\propto \Id_A$, and contributes as $\sim d^{-2}$ to the next leading order. On the other hand, the term coming from $[3 \leftrightarrow 4]$ in \eqref{eq:SAmatrixelFromFullETH} factorizes exactly as in \cref{eq:connectingFullETH1} to $\overline{(O_\mu)_{14}(O_\nu)_{41}} \cdot \overline{(O_\nu)_{12}(O_\mu)_{21}}$, leading automatically to the $\sim d^{-2}$ scaling. \\

One can verify that the leading \emph{diagonal} scalings that appear in the 4-ETH terms for the classes II, III, V, VI also vanish by analogous reasoning based on the choice \eqref{eq:gglkrgj5333}. However, for some of these classes the 4-ETH terms might factorize into lower-order terms that do not involve diagonal factors, and hence contribute at order $d^{-2}$ themselves (see, e.g., the factorization of \cref{eq:connectingFullETH2} for the class III). This fact is at the basis of the cancellation expected for some of these classes in the sector $\sigma = -1$, which must yield zero in \cref{eq:SAmatrixelFromFullETH} whenever $m = m'$ and/or $n = n'$.

What can we say instead, assuming full-ETH, about the diagonal classes I and IV from \cref{tab:classesOfEigenstates}? \\

 \emph{Class I} $\to \ \{(a,a),(a,a)\}$ . \ In this case there is no difference anymore between 2- and 4-ETH terms in \cref{eq:SAmatrixelFromFullETH}. Due to the choice \eqref{eq:gglkrgj5333}, the leading diagonal contribution comes from $\mu = \nu = 1$, and we can restrict our analysis to $\sigma = +1$ because class I is not compatible with $\sigma = -1$ (indeed one trivially gets zero in \cref{eq:SAmatrixelFromFullETH} for $\sigma = -1$). We then obtain
 \begin{equation}
 \label{eq:lkgòdèe4}
     |\braket{m,m|S_A|m,m}|^2 \sim \left(\frac{(\Id_A)_{mm}}{\sqrt{d_A}}\right)^4 = \frac{1}{d_A^2} \ .
 \end{equation}
This matches exactly the limit $d\to \infty, d_A$ finite of our diagonal ETH ansatz \eqref{eq:ETHansatz1} for the subsystem swap operator. \\

 \emph{Class IV} $\to \ \{(a,b),(a,b)\}$ . \ The dominant term, thanks to its diagonal $\mu = \nu = 1$ contribution, is $(O_\mu)_{13}(O_\nu)_{31} \cdot (O_\mu)_{24} (O_\nu)_{42}$ from \cref{eq:SAmatrixelFromFullETH}, which leads at leading order to $\overline{(O_1)_{11}} \cdot \overline{(O_1)_{11}} \cdot \overline{(O_1)_{22}} \cdot \overline{(O_1)_{22}}$. This yields, similarly to \eqref{eq:lkgòdèe4}
 \begin{equation}
     |\braket{\bm m_+|S_A|\bm m_+}|^2 \sim \left(\frac{(\Id_A)_{mm}}{\sqrt{d_A}}\right)^2\left(\frac{(\Id_A)_{m'm'}}{\sqrt{d_A}}\right)^2 = \frac{1}{d_A^2} \ .
 \end{equation}
The other three terms appearing in \cref{eq:SAmatrixelFromFullETH} are subleading by the full-ETH factorization \eqref{eq:factorizationFullETH}. Therefore, also in this case we recover our diagonal ETH ansatz \eqref{eq:ETHansatz1} in the limit of $d \to \infty$ and $d_A$ finite. In particular, this derivation makes explicit the \emph{absence} at leading order, for the diagonal matrix elements, of a dressing with a function that depends on $\omega_{mn}=\varphi_m - \varphi_n$ (the only frequency present in this class). \\

As mentioned in \cref{sec:conn1repforQuench}, similar questions on coincident-indices classes emerge in relation to the out-of-equilibrium ansätze in \cref{eq:outEqETH1,eq:outEqETH2}. Also here they can be resolved thanks to the basis choice \eqref{eq:gglkrgj5333}. The simplest example is class I $\to \ \{(a,a),(a,a)\}$, where from \cref{eq:bfdvlda59bw,eq:dgkal4gglv} we get
\begin{align}
\label{eq:vlrl39vn}
    \sum_{\mu=1}^{d_A^2}\overline{|(O_\mu)_{mm}|^2 |\Psi_{mm}|^2 } \sim \overline{|(O_1)_{mm}|^2 |\Psi_{mm}|^2 }  \sim \frac{1}{d_A} \overline{|\Psi_{mm}|^2} \sim \frac{2}{d^2}\frac{1}{d_A} \ .
\end{align}
Here we have used that $\mu=1$ is the only term to have non-vanishing diagonal elements for $d \to \infty$, together with the interpretation of $z_m$ from \cref{eq:zmDef} as a complex Gaussian random variable for what concerns its absolute moments (we note that indeed the related RMT statement gives $\mathbb E_{\rm Haar}[|\Psi_{mm}|^2] \sim 2/d^2$). The result \eqref{eq:vlrl39vn} exactly matches the diagonal ansatz \eqref{eq:outEqETH1} in the limit $d \to \infty$, $d_A$ finite.

\section{Further numerical results}
\label{sec:furthernumerics}

\subsection{Swap matrix elements for subleading off-diagonal classes}
\label{app:swapforsubleadingclasses}
\begin{figure*}[b]
    \centering
    \includegraphics[width=0.85\linewidth]{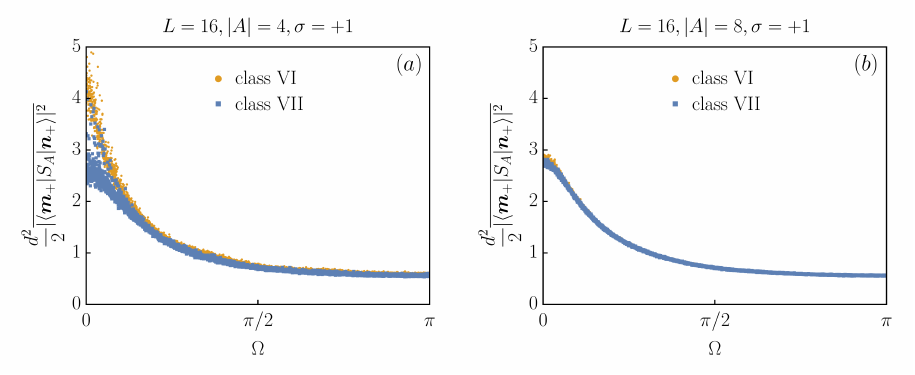}
    \caption{Comparison of the $L=16$ rescaled off-diagonal matrix elements $d^2/2 \, \overline{M_{\bm q}^{(+)}}$ from classes VI and VII (\cref{tab:classesOfEigenstates}), as functions of $\Omega$. The local-spectrum radius is $\mathcal R = 11$ ($\mathcal M = 3$) for class VI and $\mathcal R = 5$ ($\mathcal M = 4$) for class VII. We folded the $\Omega$-support $[0,2\pi]$ onto $[0,\pi]$ thanks to symmetry around $\pi$. \textbf{(a)} $|A| = 4$. \textbf{(b)} $|A|=8$.}
    \label{fig:class_VI_eth_comparison}
\end{figure*}
\begin{figure}[t]
    \centering
    \includegraphics[width=0.85\linewidth]{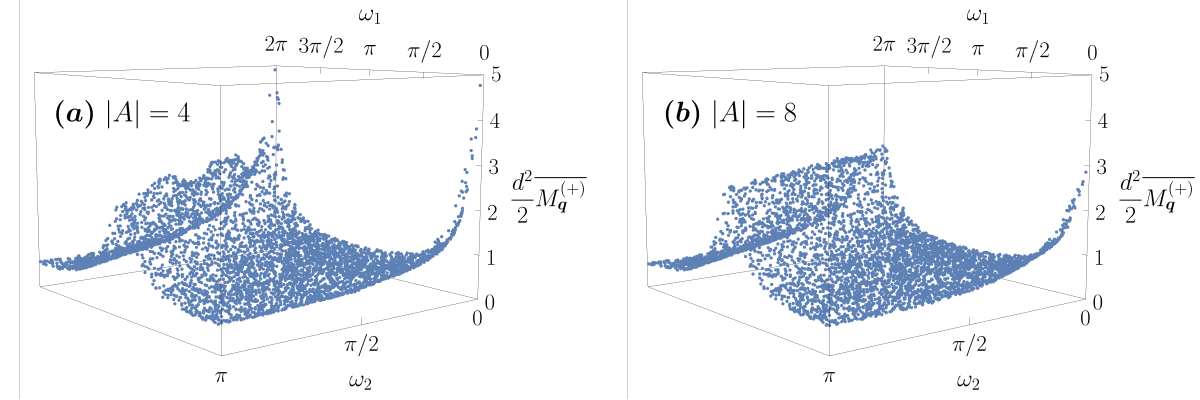}
    \caption{3D plots for rescaled off-diagonal elements $d^2/2\, \overline{M_{\bm q}^{(+)}}$ from class V (\cref{tab:classesOfEigenstates}), as functions of $\omega_1$ and $\omega_2 $ (these capture the entire frequency structure for this class). Here $L = 16$, with local-spectrum radius $\mathcal R = 11$ ($\mathcal M = 3$). We folded $\omega_2$ to $ [0,\pi]$ thanks to $(\omega_1, \omega_2)\to(-\omega_1 \ {\rm mod}\ 2\pi, 2\pi - \omega_2)$ for $\omega_2 > \pi$. \textbf{(a)} $|A|=4$. \textbf{(b)} $|A|=8$.}
    \label{fig:class_V_comparison}
\end{figure}
Although in the thermodynamic limit class VII from \cref{tab:classesOfEigenstates} is the only one relevant for the dynamics of $\mathbb E_{\ket \psi}[\mathbb P_A(t)]$ from its initial value of $1$ to the relaxation at late times, the other subleading off-diagonal classes give us the opportunity to check: (i) that our off-diagonal ETH ansatz \eqref{eq:ETHansatz2} is indeed correct irrespective of coincidence of indices; (ii) what happens to the function $f_A(\Omega, \omega_1,\omega_2)$ when one of its fundamental frequencies is exactly zero due to coincident indices, e.g.~$\omega_1 = 0$ when $m = n$ in class VI from \cref{tab:classesOfEigenstates} (a priori this could be quite different than having simply $\omega_1 \approx 0$). For point (i), we have verified that indeed all the four subleading off-diagonal classes satisfy an ETH ansatz of the form \eqref{eq:ETHansatz2}. We show an example in \cref{fig:class_VI_eth_comparison} for class VI. We see that the rescaling of the matrix elements $\overline{M_{\bm q}^{(+)}}$ by the entropic factor $d^2/2$ outputs order-one data for both classes VI and VII, in agreement with \eqref{eq:ETHansatz2}. We also notice that the emergent $f_A$ for $\Omega$ far from $0$ is essentially identical for classes VI and VII. For the larger subsystem size $|A| =8$ in \cref{fig:class_VI_eth_comparison}(b), where $f_A$ is to a very good degree a function of $\Omega$ alone in class VII, we obtain quite good agreement between classes VI and VII all the way up to $\Omega \approx 0$. Things are different for the smaller subsystem size $|A|=4$ in  \cref{fig:class_VI_eth_comparison}(a), where class VI deviates from the main signal of class VII, and instead follows the few sampled points from VII that show an enhancement around $\Omega \approx 0$. Thanks to the evidence in \cref{fig:classVII_eth_3D}, this appears to be perfectly consistent with $f_A$ possessing no discontinuity when one of its fundamental frequencies passes from being $\approx 0$ to exactly zero due to index coincidence. Indeed, we saw in \cref{fig:classVII_eth_3D} that the enhancement in the rescaled matrix elements happens in proximity of $\Omega \approx 0$, $\omega_1 \approx 0$ (and/or $\Omega \approx 0$, $\omega_2 \approx 0$, although this has not been plotted explicitly). These approximate equalities also imply $\omega_{m'n'}\approx 0$ (and/or $\omega_{m'n}\approx 0$). Given that for class VI one of these frequencies $\omega$ is always exactly zero due to coincident indices, a continuous $f_A$ would imply exactly what we see in \cref{fig:class_VI_eth_comparison}(a). This evidence directly addresses point (ii) above.

Although we do not report plots for all the subleading off-diagonal classes, we stress that these are very convenient for characterizing the enhancement of matrix elements at small values of the fundamental frequencies $\Omega, \omega_1, \omega_2$ for $|A|$ small, and to capture additional frequency dependence that is absent for large $|A|$. Indeed, their full frequency structure never requires three independent frequencies and hence it can be fully captured via 3D plots. We report an example in \cref{fig:class_V_comparison}, for class V from \cref{tab:classesOfEigenstates}. Here $m = m'$, and therefore $\omega_1$ and $\omega_2$ are a complete set of frequencies for this class. At large $|A|$, where we expect dependence only on $\Omega$, the 3D should appear constant along the lines $\omega_1 + \omega_2 \ ({\rm mod} \ 2\pi)= {\rm const}$, which indeed is what we observe in \cref{fig:class_V_comparison}(b) for $|A|=8$. On the other hand, in \cref{fig:class_V_comparison}(a) for $|A| = 4$ we can clearly observe the expected enhancement close to $\omega_1\approx \omega_2\approx 0 \ ({\rm mod} \ 2\pi)$ and also resolve additional structure at intermediate values of the frequencies along the lines $\omega_1 + \omega_2 \ ({\rm mod} \ 2\pi)= {\rm const}$.

\subsection{Additional figures}
\label{app:addFigures}
\begin{figure}[b!]
    \centering
    \includegraphics[width=0.4\linewidth]{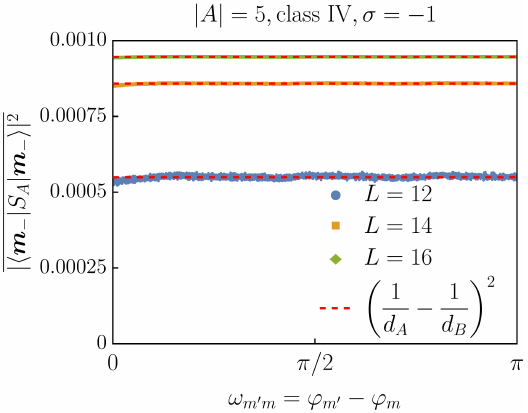}
    \caption{Neighbourhood-averaged diagonal elements $\overline{|\braket{\bm m_- |S_A|\bm m_-}|^2}$ from class IV, as functions of $\omega_{m'm}$ for $|A| = 5$ and $L=12,14,16$ (with neighbourhood radii, respectively, $\mathcal R = 8, 8,16$). The red dashed line is the leading prediction $(d_A^{-1}-d_B^{-1})^2$.}
    \label{fig:class_IV_eth_sigmaMinus}
\end{figure}
\begin{figure}[h!]
    \centering
    \includegraphics[width=0.35\linewidth]{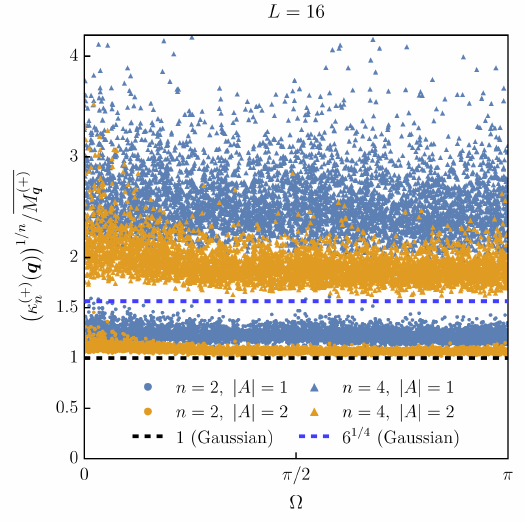}
    \caption{Analogue of \cref{fig:classVII_eth_R} but for the smaller subsystems $|A|= 1,2$. The plot shows the rescaled 2nd and 4th cumulants from \cref{eq:rescaledcums} as functions of $\Omega$. The dashed lines with values $1$ (for $\kappa_2$) and $6^{1/4}$ (for $\kappa_4$) are the predictions for $R_{\bm m, \bm n}^{(A)}$ modelled as a complex Gaussian random variable with unit variance.}
    \label{fig:classVII_eth_R_small_subsystem}
\end{figure}
\begin{figure}[h!]
    \centering
    \includegraphics[width=0.7\linewidth]{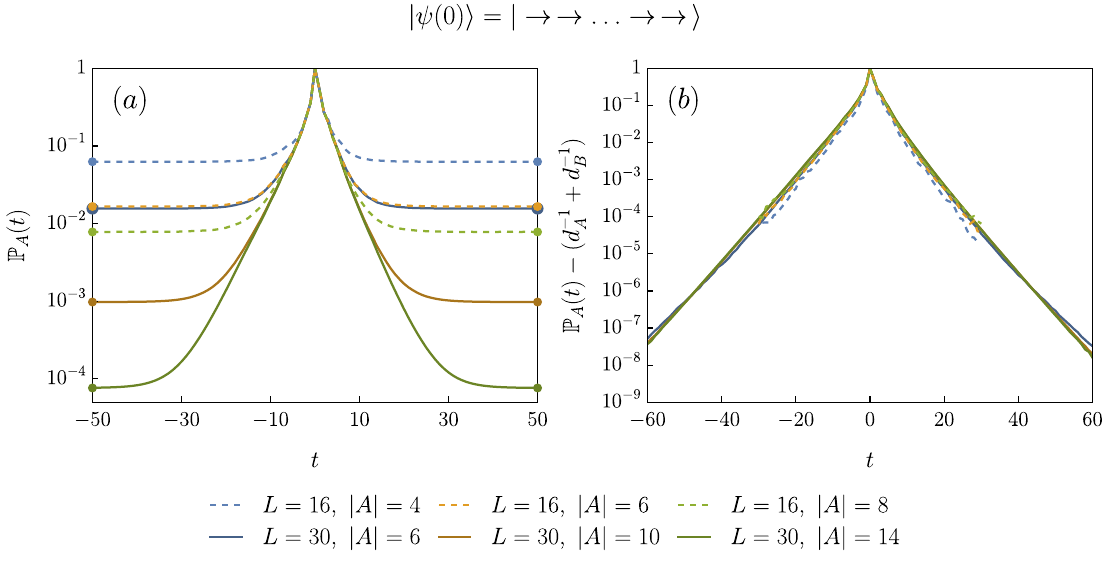}
    \caption{Same as \cref{fig:purity_dynamics_all_up}, but for the fully polarized initial state along the $x$ direction $\ket{\psi(0)}=\ket{\rightarrow \rightarrow \ldots \rightarrow \rightarrow}$.}
    \label{fig:purity_dynamics_plus_x}
\end{figure}
\begin{figure}[h!]
    \centering
    \includegraphics[width=0.4\linewidth]{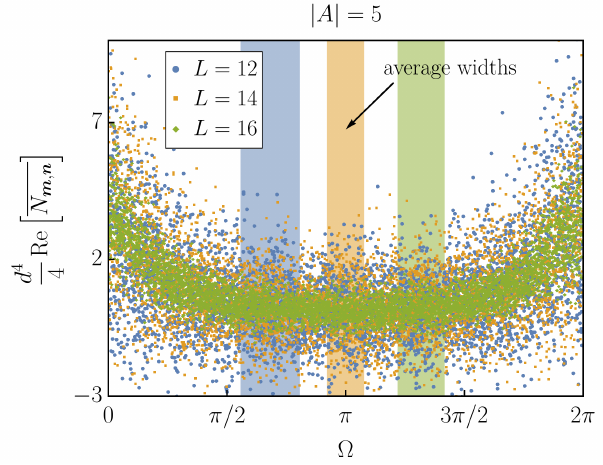}
    \caption{Plot analogous to \cref{fig:offdiagonal_outETH_all_up}(a), but for class V (\cref{tab:classesOfEigenstates}). The vertical strips have widths equal to $2\pi/d \cdot 8\mathcal R$, i.e.~the size of local-spectrum $\Omega$-windows for class V. For the three sizes $L = 12,14,16$ the radii $\mathcal R$ are, respectively, $\mathcal R = 64, 161, 813$, yielding $(2\mathcal R+1)^{3}\approx 2.15, 33.7,4307 \times 10^6$ terms per local-spectrum average.}
    \label{fig:off_diagonal_outETH_class_V}
\end{figure}
\begin{figure*}[h!]
    \centering
    \includegraphics[width=0.8\linewidth]{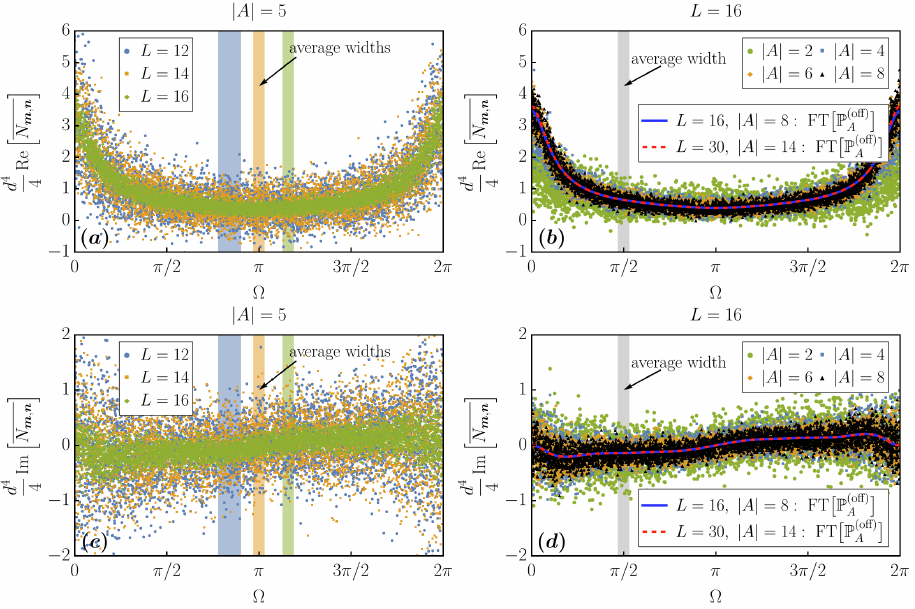}
    \caption{Same as \cref{fig:offdiagonal_outETH_all_up}, but for the fully polarized initial state along $x$: $\ket{\psi(0)}=\ket{\rightarrow \rightarrow \ldots \rightarrow \rightarrow}$. The FT come from the data in \cref{fig:purity_dynamics_plus_x}.}
    \label{fig:offdiagonal_outETH_plus_x}
\end{figure*}
\begin{figure}[h!]
    \centering
    \includegraphics[width=0.4\linewidth]{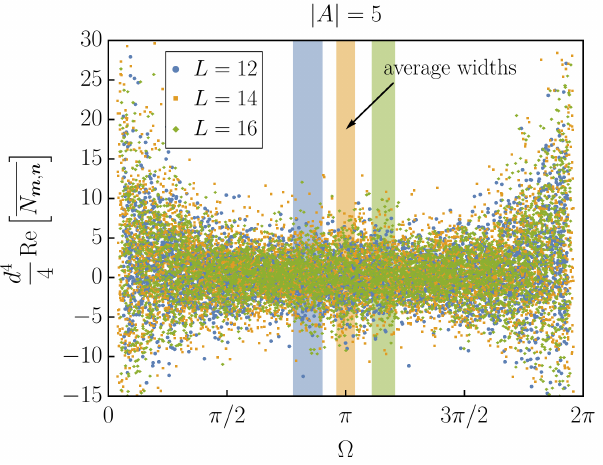}
    \caption{Plot analogous to \cref{fig:offdiagonal_outETH_all_up}(a) and \cref{fig:off_diagonal_outETH_class_V}, but for class VI (\cref{tab:classesOfEigenstates}). The vertical strips have widths equal to $2\pi/d \cdot 4\mathcal R$, i.e.~the size of local-spectrum $\Omega$-windows for class VI.}
    \label{fig:off_diagonal_outETH_class_VI}
\end{figure}

\clearpage

\bibliography{references}

\end{document}